\documentclass[12pt]{JHEP3}
\usepackage[centertags]{amsmath}
\usepackage{amssymb}
\usepackage{epsfig} 
\usepackage{amsmath} 
\usepackage{graphicx} 
\usepackage{subfig}
\usepackage{multirow}
\newcommand{\spp}{\vphantom{\bigg(}}
\def\be{\begin{equation}}
\def\ee{\end{equation}}
\def\bea{\begin{eqnarray}}
\def\eea{\end{eqnarray}}

\title{\boldmath Exploring New Physics in the {\boldmath$\text{C}_\text{7}$}-{\boldmath$\text{C}_\text{7}^\prime$} 
plane}
\author{S\'ebastien Descotes-Genon$^a$, Diptimoy Ghosh$^b$, Joaquim Matias$^c$, Marc Ramon$ ^c$ 
\\
$^a$ Laboratoire de Physique Th\'eorique, CNRS/Univ. Paris-Sud 11 (UMR 8627) \\
91405 Orsay Cedex, France \\
$^b$ Tata Institute of Fundamental Research, Homi Bhabha Road, \\ Mumbai 400005, India \\
$^c$ Universitat Autonoma de Barcelona, 08193 Bellaterra, Barcelona, Spain \\
E-mail: 
\email{descotes@th.u-psud.fr},
\email{diptimoyghosh@theory.tifr.res.in},
\email{matias@ifae.es},
\email{mramon@ifae.es}
}
\preprint{TIFR/TH/11-13\\
UAB-FT/693\\
LPT-ORSAY/11-31}
\abstract{
The Wilson coefficient $C_7$ governing the radiative electromagnetic decays of B meson has been 
calculated to a very high accuracy in the Standard Model, but experimental bounds on either the magnitude or the sign of $C_7$ are often model-dependent. In the present paper, we attempt at constraining both the magnitude and sign of $C_7$ using a systematic approach.
We consider already measured observables like the branching ratios of $B \to X_s \mu^+ \mu^-$ and
$B \to X_s \gamma$, the isospin and CP asymmetries in $ B\to K^{*}\gamma $, as well as $A_{\mathrm{FB}}$ and $F_{\mathrm{L}}$ in $B \to K^* \ell^+\ell^-$. We also discuss the transverse observable $A_{\mathrm{T}}^{(2)}$ which, once measured, may help to disentangle some of the scenarios considered.
We 
explore the constraints on $C_7,C_9,C_{10}$ as well as their chirality-flipped counterparts. Within our framework, we find that we need to extend the constraints up to $1.6\, \sigma$ to allow for the "flipped-sign solution" of $C_7$. The SM  solution  for $C_7$ exhibits a very mild tension  if New Physics is allowed in dipole operators only. 
 We provide semi-numerical expressions for all these observables as functions of the relevant Wilson coefficients at the low scale.
}

\begin{document}
\section{Introduction}
In the last decade, one of the main avenues to search for New Physics signals in $B$ and $K$ decays has consisted in overdetermining the parameters of the Cabibbo-Kobayashi-Maskawa matrix (which encodes charged weak transitions in the Standard Model (SM)) 
and its representation as a unitarity triangle embedding CP-violation. The resulting picture has shown a very good overall agreement of all the constraints, apart from some discrepancies (direct CP asymmetries difference between $B^- \to K^- \pi^0$ and ${\bar B}^0 \to K^- \pi^+$,
$B\to\tau\nu$ versus $\sin 2\beta$, $B_s$ meson mixing from $J/\Psi \phi$ channel, and the dimuon asymmetry), which are still under experimental scrutiny but  may be understood in terms of New Physics contributions \cite{Lenz:2010gu,Lunghi:2010gv,Bevan:2010gi,Bhattacherjee:2010ju,Dighe:2011du}.
 
In the meanwhile, a long list of rare $B$ decays has been determined at present with high theoretical and experimental accuracy. A tool of choice for these analysis 
is the effective Hamiltonian describing flavour transitions, allowing an elegant
separation between long-distance operators ${\cal O}_i$ (leading to contributions governed by strong and electromagnetic SM interactions) and short-distance Wilson coefficients $C_i$ (summing up
all the details of the fundamental theory lying beyond the SM at higher energies).
Once expressed in this language, the analysis of rare $B$ decays corresponds to constraining
the allowed range of Wilson coefficients (WC), taking into account several observables. One must be careful that New Physics (NP) can not only change the value of the SM Wilson coefficients, but also introduce new operators with a Dirac structure that is different from the SM ones.
We hope that overconstraining these Wilson coefficients will push them into regions
incompatible with the Standard Model, providing hints of the structure of the underlying theory responsible for these New Physics effects (right-handed currents, scalar or tensor contributions, etc.). 

This program turns out to be quite challenging as many
observables depend not on a single WC but a combination of many of them. 
Hence the constraint on a particular WC depends very much on the assumptions made on the 
type of New Physics  present and its impact on different WCs. Many model-independent 
analyses with the aim of avoiding fine tunning assume that only the Wilson coefficient analysed receives a contribution from New Physics (all the other ones being set to their SM values). The limits  of such an approach
are quite obvious, and  the conclusions that can be extracted are rather limited, specially when the framework is not clearly defined. As an illustration, it was proposed sometime ago to consider a NP contribution to the WC of the electromagnetic operator ${\cal O}_7$
approximately twice as large as the SM one but of the opposite sign, so that
the prediction for ${\cal B}(B\to X_s\gamma)$ would be similar to that of the Standard Model, which was in good 
agreement with the current experimental value. This solution attracted some interest recently, as it could explain the Belle measurements~\cite{Belle:2009zv} for the
exclusive decay $B\to K^*\ell^+\ell^-$ suggesting that the forward-backward asymmetry did not exhibit any zero at low energies.
In ref.~\cite{Gambino:2004mv}, this so-called `flipped-sign solution" was shown to be at odds
with the prediction of ${\cal B}(B\to X_s\ell^+\ell^-)$. More generally, this question can be answered only once we fix the values
of the other operators that can contribute to the observables: the
conclusions may change if NP is allowed to contribute also to the
semileptonic operators $O_{9,10}$, or if relevant operators with a non-SM
structure are included. Other solutions to this forward-backward asymmetry issue were also discussed in ref. \cite{Alok:2009tz}.

Fortunately, the rich phenomenology of $B$ decays together with the increasingly large amount of data from $B$ factories and hadron machines open new perspectives to deal with larger sets of operators. In this article, we propose to focus on the two Wilson coefficients associated with the electromagnetic operator ${\cal O}_7$ and its chirally-flipped counterpart ${\cal O}_{7^\prime}$ as tools to search for New Physics in a systematic approach. Our goal is that these coefficients
play here a similar role to the $\bar\rho$ and $\bar\eta$ parameters in the studies of the unitarity triangle. $C_7$ and $C_{7^\prime}$ do not exhaust all the information that can be obtained concerning New Physics, exactly as $\bar\rho$ and $\bar\eta$ are not sufficient to describe the full structure of the CKM matrix, but they provide an interesting summary of the situation and a good starting point to investigate NP contributions with other structures.

We will focus on the allowed regions for this pair of Wilson coefficients under different scenarios defined later on and corresponding to letting more and more  Wilson coefficients  
receive New Physics contributions. Each scenario will be more general than the previous one. The basic idea is that different choices of NP scenarios  may in principle  
lead to different solutions or allowed regions for each Wilson coefficient 
in agreement with all present constraints. The non-overlapping regions may be 
distinguished thanks to additional observables, yet to be measured, providing a criterion 
to distinguish between the different NP scenarios.

We will consider seven observables in our analysis. Six of them  are believed to exhibit a limited sensitivity to hadronic uncertainties\footnote{Notice that even though we analysed the branching ratios for  ${\cal B}(B \to K^* \gamma)$ and  ${\cal B}(B \to K^* \ell^+ \ell^-)$ we decided not to include them in the list, mainly due to the presence of significant hadronic uncertainties in form factors (see Fig.~\ref{dGammaPlot}). }:
\begin{enumerate}
\item for inclusive decays, the branching ratios  ${\cal B}(B \to X_s \gamma)$ and ${\cal B}(B \to X_s \ell^+ \ell^-)$,
\item for  $B \to K^* \ell^+ \ell^-$, the polarization fraction $F_{\mathrm{L}}$, the forward-backward asymmetry $A_{\mathrm{FB}}$ and the transverse asymmetry $A_{\mathrm{T}}^{(2)}$. 
\item for $B\to K^*\gamma$,  the exclusive CP asymmetry $S_{K^* \gamma}$. This observable is not in the same footing of robustness as the previous observables, however its main theoretical uncertainties are reasonably under control.
\end{enumerate}
The list could be extended to include other future and theoretically-clean  observables like $A_{\mathrm{T}}^{(i)}$ ($i=3,4,5$) proposed in ref.~\cite{Egede:2010zc}. However for the sake of simplicity we will not include them in this paper. 
All of the observables above are measured with different levels of accuracy except for  $A_{\mathrm{T}}^{(2)}$, which will be measured in the near future and can be used  as an efficient probe to constrain the dipole operators in a different way from
current observables. The seventh observable in our analysis, not included in this list, is the isospin asymmetry  $A_I(B \to K^* \gamma)$. Even though it is strongly sensitive to hadronic uncertainties,  we include this asymmetry 
 because of its discriminating power in our discussion of NP solutions.

Our New Physics {\it framework}  is defined by considering that NP enters in ${\cal O}_i$ with $i=7,9,10$ (electromagnetic and semileptonic operators), together with the chirally-flipped operators ${\cal O}_{i^\prime}$ with $i=7,9,10$. The precise definition and conventions for all those operators is presented in Section 2.
We will split\footnote{This splitting is not unique and different choices are possible. The only condition is to start from a restrictive NP scenario, where only dipole operators are affected by NP, and end up with the most general scenario.} this framework in three different scenarios corresponding to switching on NP step by step, starting from dipole operators and finishing with the full set of operators in the framework:

\begin{itemize}
\item Scenario A.  In this scenario the main New Physics contributions affect the electromagnetic dipole operators ${\cal O}_7$, ${\cal O}_{7^\prime}$.
\item Scenario B. Here New Physics affects not only ${\cal O}_7$, ${\cal O}_{7^\prime}$, but also the SM-like semileptonic operators ${\cal O}_9$ and ${\cal O}_{10}$.
\item Scenario C. This is the most general case in the framework we have defined, where all operators ${\cal O}_{7,9,10}$ and ${\cal O}_{7^\prime,9^\prime,10^\prime}$ can receive NP contributions.
\end{itemize}

This will allow us to have a better control, once confronted with data, on the impact of enlarging, step by step, the set of operators, as well as
providing information on the effects from right-handed currents~\cite{Pati:1974vw,Mohapatra:1974gc,Mohapatra:1974hk,Bernard:2006gy,Bernard:2007cf,Crivellin:2009sd,Buras:2010pz,Alok:2011gv}. 
Our guideline in splitting the framework in scenarios will be to try to find in a systematic way the minimal set of operators compatible with data inside a framework (and extend it if necessary). Once this is done, a future step would be to find which theories can contribute to the selected operators.

We will
assume that NP enters only these operators, and that their Wilson coefficients are real. If no solution compatible with all constraints is found at the end of our analysis, within our defined framework, the next step will consist in generalizing the framework to other operators (like scalars, tensors, the chromomagnetic operator\footnote{This generalization maybe particularly interesting because it would affect most of the observables described here, and only weak bounds on this operator are available till now.} or further chirally-flipped operators). The generalization is systematic and straightforward and will be presented elsewhere, but some details will be given here. We classify our observables in three {\it categories}:
\begin{enumerate}
\item Class-I observables mainly sensitive to ${\cal O}_7$ and ${\cal O}_{7^\prime}$, but not to ${\cal O}_{i=9,10,9^\prime,10^\prime}$. 
\item Class-II observables exclusively sensitive 
 to ${\cal O}_7$ and ${\cal O}_{7^\prime}$,  to semileptonic operators (${\cal O}_9$ and ${\cal O}_{10}$) and their chiral counterparts  ${\cal O}_{9^\prime}$, ${\cal O}_{{10}^\prime}$. Only these operators intervene, even within more general frameworks than the one considered here. 
\item Class-III observables that are also sensitive to all the previous operators ${\cal O}_i$ with $i=7...10'$, and in addition have the potential of exhibiting a sensitivity to NP contributions from other operators like scalars, tensors, chromomagnetic operator...\footnote{There is an important distinction in our way of treating Class-I observables with respect to the other classes: 
the definition of Class-I observables involves only their sensitivity to dipole operators and their lack of contributions from semileptonic operators. Other potential sensitivities beyond the defined framework are not relevant at this stage. This is essential to be able to define primary regions in a systematic way for each framework. 
On the contrary, we prefer to split Class II from Class III, to identify more easily the observables that will change  if new sources beyond the framework are included.
} including all the previous operators ${\cal O}_i$ with $i=7,7',9,9',10,10'$ but also scalar, tensor, chromomagnetic, etc., operators. 
\end{enumerate}
Notice that, strictly speaking, within our defined framework, Class-II and Class-III observables coincide. However, having in mind a systematic forthcoming generalization of this work  we need to split them as a function of their potential NP sensitivity beyond our present framework. 
We will also discuss how this classification  would change if we extend  the framework to include  additional Dirac structures. It would basically require to re-classify some  observables (mostly in Class I)  
and introduce more Class II subdivisions (even if not required here, one could also add intermediate stages between Class II and Class III at will).

In this paper we will illustrate the method on the practical example of determining
the sign of $C_7$, already discussed in ref.~\cite{Gambino:2004mv}, using a subset of our observables\footnote{An interesting analysis was also presented in ref.~\cite{Bobeth:2008ij}, considering another subset of our observables but adding NP to one Wilson coefficient at a time. See Sect. 4 for further details.}. We will focus not only on the restrictive ``flipped-sign'' solution, but allow also for deviations in the modulus of $C_7$. As it is well known, the sign of $C_7$ has an important impact on 
observables like the forward-backward asymmetry ($A_{\mathrm{FB}}$) in the rare exclusive semileptonic 
decay $B\to K^* \mu^+ \mu^-$, that is at present slightly at odds with the SM prediction.

In section 2, we present in detail the operators entering our framework and the observables of interest, with their current
experimental accuracy as well as numerical expressions for the implementation of their theoretical determination. In section 3, we discuss the three different scenarios and  combine the present constraints for each of those scenarios to look for different solutions or allowed regions in the WC planes. In section 4, we summarize the elements learned concerning the sign of $C_7$ and the values of the WCs. Most technical details concerning the inputs and the computation of the observables are collected in the appendices. 

\section{Operators, method and observables} \label{sec:observables}

\subsection{$b\to s$ effective Hamiltonian}

We consider the effective Hamiltonian for radiative $b \rightarrow s$ transitions~\cite{Misiak:2006zs,Lunghi:2006hc}
\begin{equation} \label{eq:heff}
{\cal H}_{\mathrm{eff}} = -\frac{4G_F}{\sqrt{2}} \left(\lambda_t^{(s)} {\cal{H}}_{\mathrm{eff}}^{(t)}+ \lambda_u^{\mathrm{(s)}} {\cal{H}}_{\mathrm{eff}}^{(u)}\right) + h.c.,
\end{equation}
with the CKM matrix combinations $\lambda_q^{(s)}=V_{qb} V_{qs}^{*}$, and
\begin{eqnarray}
{\cal{H}}_{\mathrm{eff}}^{(t)} &=& C_1 {\cal{O}}_1^c + C_2 {\cal{O}}_2^c + \sum_{i=3}^{6} C_i {\cal{O}}_i + \sum_{i=7}^{10} (C_i {\cal{O}}_i + C_{i^\prime} {\cal{O}}_{i^\prime}), \nonumber \\
 {\cal{H}}_{\mathrm{eff}}^{(u)} &=& C_1 ({\cal{O}}_1^c - {\cal{O}}_1^u) + C_2 ({\cal{O}}_2^c - {\cal{O}}_2^u). 
 \end{eqnarray}
$C_{i^{(\prime)}} \equiv C_{i^{(\prime)}} (\mu_b)$ and ${\cal{O}}_{i^{(\prime)}} \equiv {\cal{O}}_{i^{(\prime)}} (\mu_b)$ are the Wilson coefficients and the local effective operators respectively. The contribution of ${\cal{H}}_{\mathrm{eff}}^{(u)}$ is usually dropped for being doubly Cabibbo-suppressed with respect to that of ${\cal{H}}_{\mathrm{eff}}^{(t)}$, but we will keep it for the observables of interest. In eq.~(\ref{eq:heff}) we use the same operator basis as ref.~\cite{Kruger:2005ep}. We focus  our attention on the operators
\begin{align}
{\mathcal{O}}_{7} &= \frac{e}{16 \pi^2} m_b
(\bar{s} \sigma_{\mu \nu} P_R b) F^{\mu \nu} ,&
{\mathcal{O}}_{{7}^\prime} &= \frac{e}{16 \pi^2} m_b
(\bar{s} \sigma_{\mu \nu} P_L b) F^{\mu \nu} , \nonumber
\\
{\mathcal{O}}_{9} &= \frac{e^2}{16 \pi^2} 
(\bar{s} \gamma_{\mu} P_L b)(\bar{\ell} \gamma^\mu \ell) ,&
{\mathcal{O}}_{{9}^\prime} &= \frac{e^2}{16 \pi^2} 
(\bar{s} \gamma_{\mu} P_R b)(\bar{\ell} \gamma^\mu \ell) , \nonumber
\\
\label{eq:O10}
{\mathcal{O}}_{10} &=\frac{e^2}{16 \pi^2}
(\bar{s}  \gamma_{\mu} P_L b)(  \bar{\ell} \gamma^\mu \gamma_5 \ell) ,&
{\mathcal{O}}_{{10}^\prime} &=\frac{e^2}{16\pi^2}
(\bar{s}  \gamma_{\mu} P_R b)(  \bar{\ell} \gamma^\mu \gamma_5 \ell) ,
\end{align}
where $P_{L,R}=(1 \mp \gamma_5)/2$ and $m_b \equiv m_b(\mu_b)$ denotes the running $b$ quark mass in the $\overline{\mathrm{MS}}$ scheme. The primed operators, with flipped chirality with respect to the unprimed ones, are either highly suppressed or vanish in the SM. Hence, 
\begin{equation} 
C^{SM}_{7'}=\frac{m_s}{m_b} C^{SM}_{7}, \qquad C^{SM}_{9',10'}=0
\label{C7pSM}
\end{equation}
In the following, we will assume that only the Wilson coefficients of the operators in eq.~(\ref{eq:O10}) are potentially affected by NP according to our framework.

The determination of the Wilson coefficients in the Standard Model follows the discussion in refs.~\cite{Misiak:2006zs,Lunghi:2006hc} to perform the matching at the high scale $\mu_0$ (potentially affected by short-distance NP) and the running of the Wilson coefficients from the high-scale down to $\mu_b$, leading to SM Wilson coefficients at NNLO accuracy. The error budget of the observables includes a  variation of  $\mu_b$  from twice to half its central value (we take $\mu_b=4.8$ GeV). We have also checked that the variation of the high scale $\mu_0$ yields only a tiny uncertainty on the observables. 
We follow refs.~\cite{Huber:2005ig,Gambino:2003zm,Gorbahn:2004my,Bobeth:2003at} and include QED corrections through five additional operators (${\cal O}_{3,4,5,6Q}$ and ${\cal O}_b$) mixing with the ones displayed in eq.~(\ref{eq:heff}). The values of the Wilson coefficients at the low-scale $\mu_b=4.8$ GeV are given in table~\ref{wilson}, where the definitions \cite{Chetyrkin:1996vx}
\begin{eqnarray}
C_7^{\mathrm{eff}}  & \equiv & C_7 -\frac{1}{3}\, C_3 - 
\frac{4}{9}\, C_4 - \frac{20}{3}\, C_5\, -\frac{80}{9}\,C_6\,, \nonumber\\
C_8^{\mathrm{eff}}  & \equiv & C_8 + C_3 -
\frac{1}{6}\, C_4 + 20 C_5\, -\frac{10}{3}\,C_6 \,             \nonumber
\label{C7effC8eff}
\end{eqnarray}
have been used, since $C_7$ and $C_8$ always appear in these particular combinations with other $C_i$ in matrix elements.
 
In tables~\ref{wilson} and \ref{tab:inputs} we present the most important inputs used in our observables including the values of the Wilson coefficients in the SM. 

\begin{table}
\renewcommand{\arraystretch}{1.4}
\addtolength{\arraycolsep}{1pt}
\footnotesize{$$
\begin{array}{|c|c|c|c|c|c|c|c|c|c|}
\hline
C_1(\mu_b) &   C_2(\mu_b) &  C_3(\mu_b) &  C_4(\mu_b) & C_5(\mu_b) &  C_6(\mu_b)
& C_7^{\rm eff}(\mu_b) & C_8^{\rm eff}(\mu_b) & C_9(\mu_b) & 
C_{10}(\mu_b)\\ \hline
-0.2632 & 1.0111 & -0.0055 & -0.0806 & 0.0004 &
0.0009 &  -0.2923 & -0.1663 & 4.0749 & -4.3085
 \\
\hline
\end{array}
$$}
\renewcommand{\arraystretch}{1}
\addtolength{\arraycolsep}{-1pt}
\caption[]{NNLO Wilson coefficients in the Standard Model at the scale $\mu_b$=4.8 GeV, obtained from the inputs in table~\ref{tab:inputs}.
For the computation of the observables, we considered a variation of $\mu_b$ from half to twice its value.   
}
\label{wilson}
\end{table}

\TABLE{
\centering
\footnotesize{\begin{tabular}{|lc|lc|}
\hline
$\mu_b=4.8 \ {\rm GeV}$ &   & $\mu_0=2M_W$ &  \cite{Misiak:2006zs} \\
\hline
$m_B=5.27950   \ {\rm GeV}$ & \cite{Nakamura:2010zzi} & $m_{K^*}=0.89594   \ {\rm GeV}$  & \cite{Nakamura:2010zzi} \\\hline
$m_{B_s} = 5.3663  \ {\rm{GeV}}$ & \cite{Nakamura:2010zzi} & $m_{\mu} = 0.105658367  \ {\rm GeV}$ & \cite{Nakamura:2010zzi} \\\hline
$ \sin^2 \theta_W                      = 0.2313$ & \cite{Nakamura:2010zzi}         &  & \\
$M_W=80.399\pm 0.023  \ {\rm GeV}$ &  \cite{Nakamura:2010zzi} &
$M_Z=91.1876 \ {\rm GeV}$ &  \cite{Nakamura:2010zzi} \\\hline
$ \alpha_{em}(M_Z)                     =1/128.940 $ & \cite{Misiak:2006zs}&
$ \alpha_s(M_Z)                        = 0.1184 \pm 0.0007  $ & \cite{Nakamura:2010zzi} \\\hline
$ m_t^{\rm pole} = 173.3\pm 1.1  \ {\rm GeV}      $ & \cite{Alcaraz:2009jr}
 & $ m_b^{1S}                   = 4.68 \pm 0.03   \ {\rm GeV}   $  &\cite{Bauer:2004ve} \\
$ m_c^{\overline{MS}}(m_c)                   = 1.27 \pm 0.09  \ {\rm GeV}    $ & \cite{Nakamura:2010zzi} &
$ m_s^{\overline{MS}}(2\ {\rm GeV})=0.101 \pm 0.029 \ {\rm GeV}$ &   \cite{Nakamura:2010zzi} \\\hline
$\lambda_{CKM}=0.22543\pm 0.0008$ &  \cite{ckmfitter} &
$A_{CKM}=0.805\pm 0.020$ & \cite{ckmfitter}  \\
$\bar\rho=0.144\pm 0.025$ & \cite{ckmfitter} &
$\bar\eta=0.342\pm 0.016$ & \cite{ckmfitter}  \\\hline
${\cal B}(B \to X_c e \bar\nu) = 0.1061 \pm 0.00017 $ & \cite{Misiak:2006zs} & $ C                                  = 0.58 \pm 0.016     $  & \cite{Misiak:2006zs} \\
$ \lambda_2                            = 0.12      \ {\rm GeV}^2         $ & \cite{Misiak:2006zs} & &\\\hline
 $ \Lambda_h=0.5\ {\rm GeV}$ & \cite{Kagan:2001zk} & $ f_B = 0.200 \pm 0.025\ {\rm GeV}    $ & \cite{Altmannshofer:2008dz}\\
 $f_{K^*,||}=0.220 \pm 0.005$\ {\rm GeV} & \cite{Altmannshofer:2008dz}
 & $f_{K^*,\perp}(2\ {\rm GeV})=0.163\pm 0.008\ {\rm GeV}$ & \cite{Altmannshofer:2008dz}
\\
$ \xi_\perp(0)=    0.31^{+0.20}_{-0.10} $ &  \cite{Khodjamirian:2010vf} &$  \xi_{||}(0)=    0.10\pm 0.03  $ & \cite{Khodjamirian:2010vf}\\
$ a_{1,||,\perp}(2\ {\rm GeV})=0.03\pm 0.03$ & \cite{Altmannshofer:2008dz}& $ a_{2,||,\perp}(2\ {\rm GeV})=0.08\pm 0.06$ & \cite{Altmannshofer:2008dz}\\
$\lambda_B(\mu_h)=0.51\pm 0.12\ {\rm GeV}$& \cite{Altmannshofer:2008dz}
&&\\
\hline
 $ f_{B_s} = 0.2358 \pm 0.0089\ {\rm GeV}$ & \cite{ckmfitter} & $ \tau_{B_s} = 1.472 \pm 0.026\ {\rm ps}    $ & \cite{Nakamura:2010zzi} \\
 \hline
\end{tabular}}
\caption{Input parameters, based on refs.~\cite{Misiak:2006zs}, \cite{Nakamura:2010zzi}, \cite{Alcaraz:2009jr}, \cite{ckmfitter}, \cite{Altmannshofer:2008dz}, \cite{Kagan:2001zk}.}
\label{tab:inputs}
}

\subsection{Method} \label{Method}

We start by describing in full detail  how the method applies to our previously defined framework (New Physics allowed only in electromagnetic dipole, semileptonic operators, with SM and flipped-chirality structures).  We will proceed in the following way:

\begin{enumerate}
\item We start by classifying observables in three classes, as already mentioned:
Class-I observables sensitive only to ${\cal O}_7$ and ${\cal O}_{7'}$ contributions, Class-II observables  exclusively sensitive to the full set of operators that we consider may be affected by New Physics (${\cal O}_7$, ${\cal O}_9$, ${\cal O}_{10}$ as well as their flipped chirality counterparts)
and Class-III observables, not only sensitive to all these operators, but also to further  new operators (scalars, tensors, etc).
\item We define a reference frame of allowed regions using observables sensitive to NP only through a pair of Wilson coefficients, in our case $C_7$, $C_{7'}$. These reference
regions, that we will call {\it primary regions}, are determined by Class-I observables, and are the maximally 
allowed regions  inside our defined framework. They can only shrink when new observables are added. In principle, one could define a different reference frame, where NP enters only in $C_7$, but this can be inferred directly from the projection of our
reference region along the $C_7$ axis. We are in Scenario A.
\item We add a larger set of observables with sensitivity to larger sets of operators
(Class II and Class III), but still inside Scenario A. Those new observables when restricted to the $(C_7 ,C_{7'})$ plane may cut
further the primary regions, defining a smaller allowed region inside the primary ones.
\item In order to expand again these allowed regions (with the maximal area always 
defined as the primary regions), we will now move to Scenario B and C, allowing new contributions for the extra
coefficients, in our case, $C_9$, $C_{10}$ and the flipped-chirality ones. 
\end{enumerate}

The same procedure should be repeated if other structures are included defining an extended framework. A discussion can be found in Sec. \ref{generalframe}.

We will list the observables of interest for our analysis, providing in each case a semi-numerical expression for the observables with their central values and their
uncertainties in the Standard Model, as well as their dependence on the deviation $\delta C_i=C_i-C_i^{SM}$ at the low scale $\mu_b$. This treatment 
assumes  that the analysis of uncertainties performed in the SM is not significantly affected by the presence of NP.

In many places along this paper we will refer to the correlation between pairs of WCs, sometimes denoted as $(C_i,C_j)$ or $(\delta C_i,\delta C_j)$. The relation between both is linear $C_i(\mu_b)=C^{SM}_{i}(\mu_b)+\delta C_i$. In all cases we will plot only the correlation between $(\delta C_i,\delta C_j)$.

\subsection{Class-I observables}  

Class-I observables receive contributions from ${\cal O}_7$, ${\cal O}_{7^\prime}$ but not from the semileptonic operators ${\cal O}_{9,10}$ or ${\cal O}_{9^\prime,10^\prime}$. Three observables considered here fall into this category: the branching ratio of the inclusive radiative decay $B \to X_s \gamma$, as well as the isospin asymmetry ($A_I$) and the CP-asymmetry ($S_{K^* \gamma}$) of the exclusive decay $B\to K^*\gamma$. 

$\bullet$ {\it ${\cal B}(\bar B \to X_s \gamma)$} is one of the cleanest observables in $B$ physics from the theoretical point of view. Apart from contributions to the chromomagnetic operator, it is only sensitive
to electromagnetic dipole operators, without pollution from other New Physics contributions.
The currently available experimental world average is~\cite{Asner:2010qj}:
\bea 
\label{aver}
{\cal B}(\bar B \to X_s \gamma)_{{}_{E_\gamma > 1.6\,{\rm GeV}}} = 
\left( 3.55 \pm 0.24 \pm 0.09 \right)\times 10^{-4}
\eea
 The following formula updates the expression in ref.~\cite{Lunghi:2006hc}, using  ref.~\cite{Freitas:2008vh}, 
based on the NNLO SM results of \cite{Misiak:2006zs, Misiak:2006ab, Misiak:2010sk} (more details can be found in App.~\ref{app:numerics}).
\bea \label{bsgamma}
{\cal B} (\bar B \to X_s \gamma)_{{}_{E_\gamma > 1.6\,{\rm GeV}}}  =\!\!
&& \Big[
a_{(0,0)} \pm \delta_a
+ a_{(7,7)} \left[(\delta C_7)^2 + (\delta C_{7'})^2\right]  + {\phantom{\Bigg\{}} \nonumber \cr
&& 
+ a_{(0,7)} \; \delta C_7
+ a_{(0,7')} \; \delta C_{7'} 
\Big]\cdot 10^{-4} \, \qquad
\eea
where the scale of the NP contributions to the Wilson coefficients $\delta  C_i (\mu_b)$ (with $i=7,7^\prime$) is taken at $\mu_b=4.8~{\rm GeV}$. The coefficients $a_i$ are collected in table~\ref{tab:coeffBtoXsgamma}, from which one can extract the SM prediction, in good agreement with the experimental measurements:
\begin{equation}
{\cal B} (\bar B \to X_s \gamma)_{{}_{E_\gamma > 1.6\,{\rm GeV}}}^{SM}=(3.15\pm 0.23)\cdot 10^{-4}
\end{equation}

\TABLE[t]{
\centering
\begin{tabular}{|l|l|l|l|}
\hline
\spp $a_{(0,0)}  = 3.15 \quad \delta_a=0.23 $ & $a_{(0,7)}  = -14.81 $ & $a_{(7,7)} = 16.68$ & $a_{(0,7')}  = -0.23 $  \\
\hline
\end{tabular}
\caption{Coefficients describing the dependence of ${\mathcal B}(B\to X_s\gamma)$ on $C_7$ and $C_{7'}$.
\label{tab:coeffBtoXsgamma}}
}

$\bullet$ {\it $A_I(B \to K^* \gamma)$}: 
The measurement of the isospin asymmetry ($A_I$) in $B \to K^{*}\gamma $ 
was reported by BaBar and Belle, with a slightly larger neutral decay rate and hence a positive $ A_I$. 
\begin{eqnarray}
A_I & \equiv &
\frac{\Gamma(\bar{{B}}^0 \to \bar{{K}}^{*0}\gamma)
-\Gamma({B}^- \to {K}^{*-}\gamma)}
{\Gamma(\bar{{B}^0} \to \bar{{K}^{*0}}\gamma)
+\Gamma({B}^- \to {K}^{*-}\gamma)}=  \left( I \cdot R^{+/0}\, {\tau^+}/{\tau_0} -1 \right)/2 \end{eqnarray}
where the two isospin-breaking ratios are $I={\cal B}({\bar B^0} \to {\bar K^{*0}} \gamma)/ {\cal B}(B^{-} \to K^{*-} \gamma)$ 
and 
$R^{+/0}=\Gamma(\Upsilon(4s) \to B^+ B^-)/\Gamma(\Upsilon(4s) \to B^0 {\bar B^0})$\,.

The recent update of BaBar collaboration~\cite{BaBar:2009we} with a five times larger sample than their previous result has moved 
substantially $A_I$ in  the positive direction ($A_{I}=0.066 \pm 0.021 \pm 0.022$),
being now consistent with zero at more than $2\, \sigma$, while previously the consistency was below $1\, \sigma$. The older result from the Belle collaboration~\cite{Nakao:2004th}:
$A_{I}=0.012 \pm 0.044 \pm 0.026$
requires an update to determine whether it follows the same trend as BaBar. The average of these two measurements according to the 
Heavy Flavor Averaging Group is~\cite{Asner:2010qj}:
\begin{equation}A_{I}^{\mathrm{exp}}(B \to K^* \gamma)= 0.052 \pm 0.026\,.\end{equation}

In the Standard Model, $A_I$ vanishes in na\"{\i}ve factorisation, and it gets contribution only 
from non-factorizable graphs where a photon is radiated from the spectator quark. This quantity was first calculated in the SM within the QCD Factorisation (QCDF) framework in ref.~\cite{Kagan:2001zk} 
and confirmed in ref.~\cite{Feldmann:2002iw}, with a result $9.3^{+3.8}_{-3.2}~\%$
\cite{Feldmann:2002iw}. Later on, it was reevaluated adding some (Cabibbo-suppressed) annihilation contributions but changing the factorisation scale from around $2$ GeV to near $4.8$ GeV, due to the fact that, below this scale, the four-quark operators factorise, so that the gluon exchange responsible for the running of these operators does not probe small scales and thus does not induce running below $\mu_b$~\cite{Beneke:2004dp}.

This observable is dominated by $1/m_b$ corrections inducing important hadronic uncertainties, but we include it because of its particular sensitivity to $C_7$ and $C_{7'}$ which will prove very important in our discussions. 
The corresponding numerical expression is:
\begin{equation}
A_I(B \to K^* \gamma)= 
 c \times \frac{\sum_{k} d_{k} (\delta C_7)^k}
   {\sum_{k,l} e_{(k,l)} (\delta C_7)^k (\delta C_{7'})^l}\pm \delta c\,.
\end{equation}
where the non-zero coefficients
are collected in table~\ref{tab:coeffAIKstargamma}, out of which one extracts the SM prediction:
\begin{equation}
A_I(B \to K^* \gamma)^{SM}=0.041\pm 0.025
\end{equation}
once again in good agreement with the experimental value.

\begin{table}\begin{center}
\footnotesize{\begin{tabular}{|c|c|}
\hline
\spp $c= 4.11 \%$& $\delta c= 2.52 \%$ \\
\hline
\spp  $d_{0} =1$& $d_{1} =-2.51757$ \\
\hline
\spp $e_{(0,0)} =1$ & $e_{(1,0)} =-5.0165$ \\
\spp $e_{(0,1)} =-0.0919061$ & $e_{(2,0)} = 6.30856$  \\
\spp $e_{(0,2)} =7.49847$ & \\
\hline
\end{tabular}}
\caption{Coefficients describing the dependence of $A_I(B\to K^*\gamma)$ on $C_{7}$ and $C_{7'}$.\label{tab:coeffAIKstargamma}}
\end{center}\end{table}

$\bullet$ $S_{K^*\gamma}$: The radiative decay $b \to s \gamma$ constitutes a major probe of both the flavour structure of the SM and NP. In the SM, the left-handed structure of the weak interactions makes the emitted photon mainly left-handed in $b$ decays and right-handed in $\bar{b}$ decays, as can be seen  from the structure of the (dominant) electromagnetic dipole operator $\bar{s}_{L(R)} \sigma_{\mu \nu} b_{R(L)}$. The needed helicity flip of one of the external quarks results into a factor $m_b$ for $b_R \to s_L \gamma_L$ and a factor $m_s$ for $b_L \to s_R \gamma_R$. Therefore, at LO in the SM, the emission of right-handed photons is suppressed by a factor $m_s/m_b$. This suppression can be overridden in a large number of NP scenarios where the helicity flip occurs on a internal line, which may cause appearance of a factor much larger than $m_s/m_b$.

The photon helicity is difficult to probe directly, but can be accessed indirectly using the time-dependent CP asymmetry in $B^0 \to K^{*0} \gamma$:
\begin{equation}
A_{CP}= \frac{\Gamma(\bar{B}^0(t) \to \bar{K}^{*0} \gamma) - \Gamma(B^0(t) \to K^{*0} \gamma)}{\Gamma(\bar{B}^0(t) \to \bar{K}^{*0} \gamma) + \Gamma(B^0(t) \to K^{*0} \gamma)}  = S_{K^*\gamma} \sin(\Delta m_B t) - C_{K^*\gamma} \cos(\Delta m_B t),
\label{ACP}
\end{equation}
where $K^{*0}$ and $\bar{K}^{*0}$ are observed through their decay into the CP eigenstate $K_S \pi^0$ and $B^0$ mixing is assumed to be SM-like\footnote{This assumption is compatible with the latest measurements of the CP-violating parameter
$|p/q|=1.0024 \pm 0.0023$~\cite{Asner:2010qj} derived from the data gathered at $B$ factories only.}. The helicity suppresion of right-handed photons make $A_{CP}$ dominated by
$B$-meson mixing in the SM, irrespective of hadronic uncertainties. Since NP can relieve this suppression, eq.~(\ref{ACP}) is a good candidate for null-tests of the SM~\cite{Grinstein:2004uu, Grinstein:2005nu, Ball:2006cva, Ball:2006eu}. In the present article, we will focus on $S_{K^*\gamma}$ in eq.~(\ref{ACP}), as it involves the interference of photons with different polarisation and provide interesting constraints on $C_{7'}$ (see Appendix~\ref{sec:skstargamma} for further details). 

The experimental results available from the $B$ factories for $S_{K^* \gamma}$ are the following:
\begin{equation*}
S_{K^* \gamma}^{\mathrm{exp}}=\left\{ \begin{array}{llll}
&\!\!\!-0.32^{+ 0.36}_{-0.33}\,({\rm stat.}) \pm 0.05\, ({\rm syst.})
& \;\, \mbox{Belle \cite{Ushiroda:2006fi}} 
&  \;\,\mbox{($535\cdot 10^6$ $B\bar B$ pairs),}\\
&\!\!\!-0.03 \pm 0.29\,({\rm stat.}) \pm 0.03\, ({\rm syst.})
& \;\, \mbox{BaBar \cite{Aubert:2008gy}} 
& \;\,\mbox{($467\cdot 10^6$ $B\bar B$ pairs),}
\end{array} \right.
\end{equation*}
with the HFAG average~\cite{Asner:2010qj}
\begin{equation}
S_{K^* \gamma}^{\mathrm{exp}}=-0.16 \pm 0.22. 
\end{equation}
A numerical expression for this observable with our inputs is:
\begin{equation}
S_{K^* \gamma} = f\phantom{.}^{+\delta_f^u}_{-\delta_f^d} + \frac{\sum_{k,l} g_{(k,l)} (\delta C_{7})^k (\delta C_{7^\prime})^l}{\sum_{k,l} h_{(k,l)} (\delta C_{7})^k (\delta C_{7^\prime})^l},
\end{equation}
where  $f$ corresponds to the SM central value and $\delta_f^{u}$, $\delta_f^{d}$ the corresponding error bars.  The non-vanishing $g$ and $h$ coefficients can be found in table \ref{tab:coeffSKsgamma}. One can see that the SM prediction
is:
\begin{equation}
S_{K^* \gamma}^{SM}=-0.03 \pm 0.01 
\end{equation}
\begin{table}
\begin{center}
\footnotesize{\begin{tabular}{|c|c|c|}
\hline
\multirow{3}{*}{$f = - 0.0297336$}{\phantom{\tiny{-}}} 
& \spp $\delta_f^u = 0.0089893$ \\
&  \spp $\delta_f^d = 0.0089767$ \\
\hline
\spp $g_{(0,1)} = + 152.774$ & $h_{(0,0)} = + 39.9999$ \\
\spp $g_{(1,0)} = - 3.17764$ & $h_{(0,1)} = - 4.51218$ \\
\spp $g_{(1,1)} = - 415.441$ & $h_{(1,0)} = - 214.866$ \\
\spp $g_{(0,2)} = + 8.63917$ & $h_{(0,2)} = + 290.553$ \\
\spp $g_{(2,0)} = + 8.63917$ & $h_{(2,0)} = + 290.553$ \\
\hline
\end{tabular}}
\caption{Coefficients describing the dependence of $S_{K^* \gamma}$ on $C_{7}$ and $C_{7'}$.
\label{tab:coeffSKsgamma}}
\end{center}
\end{table}

\subsection{Class II}\label{sec:classII}

In this set, we find some of the observables constructed out of the coefficients of the angular distribution of $B \to K^*(\to K \pi) \ell^+ \ell^-$ for which the hadronic uncertainties due to form factors cancel largely, and which are only dependent on some of the spin amplitudes involved in this decay. The observables called $A_{\mathrm{T}}^{(i)}$ (with $i=2,3,4,5$) fall inside this category. 

$\bullet$ $A_{\mathrm{T}}^{(2)}$: This is the only observable which has not been measured yet and is included in our analysis though. Its unique sensitivity to ${\cal O}_{7^\prime,9^\prime,10^\prime}$ (shown in \cite{Egede:2010zc,Kruger:2005ep,Egede:2008uy}) and the very limited hadronic uncertainties attached to it makes it into a very appealing observable to distinguish between different NP scenarios.

Its definition in terms of spin amplitudes is \cite{Kruger:2005ep}:
\begin{equation}
A_{\mathrm{T}}^{(2)}(q^2)=\frac{|A_\perp|^2-|A_\||^2}{|A_\perp|^2+|A_\||^2}\,,
\end{equation}
where $A_\perp$ and $A_\|$ are the corresponding spin amplitudes of the $K^*$ and $q^2$ (or $s$ in the following) is the lepton-pair invariant mass squared. This asymmetry avoids one of the main sources of uncertainty for observables based on the $B \to K^* \ell^+ \ell^-$ decay, namely the soft form factors $\xi_\perp$ and $\xi_\|$ \cite{Beneke:2000wa}. $A_{\mathrm{T}}^{(2)}$ is constructed to cancel its dependence on $\xi_\perp(q^2)$ exactly at LO and displays only a very mild sensitivity on it at NLO in QCDF. Its extraction from the uniangular distributions is described in Appendix~\ref{ddd}\footnote{The other asymmetries $A_{\mathrm{T}}^{(i)}$ (with $i=2,3,4,5$) require the determination of the full distribution ($A_{\mathrm{T}}^{(5)}$ is particularly sensitive to ${\cal O}_{10^\prime}$, whereas $A_{\mathrm{T}}^{(3,4)}$ probe the longitudinal spin amplitude).}.

$A_{\mathrm{T}}^{(2)}$ has been computed in QCDF at NLO using our inputs in table 2, the soft form factors described in Appendix \ref{sffsection} and an estimate of $\Lambda/m_b$ suppressed corrections of order $10 \%$. A detailed discussion
of its sensitivity to some of the operators in our framework can be found in ref. \cite{Egede:2010zc}.

\TABLE[t]{
\centering
\footnotesize{
\begin{tabular}{|c|c|c|c|c|c|c|c|}
\hline
\spp $ $  & $1$ & $s$ & $s^2$ & $s^3$ & $s^4$ & $s^5$ & $s^6$ \\
\hline
\spp dim & 1 & $\rm GeV^{-2}$ & $\rm GeV^{-4}$ & $\rm GeV^{-6}$ & $\rm GeV^{-8}$ & $\rm GeV^{-10}$ & $\rm GeV^{-12}$ \\
\hline
\spp  $\!\!F_{(0,0)}\!\!$ &  $\!\!+12904.2\!\!$ & $\!\!-17256.7\!\!$ & $\!\!+10543.8\!\!$ & $\!\!-3519.19\!\!$ & $\!\!+667.247\!\!$ & $\!\!-67.3536\!\!$ & $\!\!+2.78209\!\!$  \\
\spp  $\!\!G_{(0,0)}\!\!$ &  $\!\!+402941\!\!$ & $\!\!-533447\!\!$ & $\!\!+329442\!\!$ & $\!\!-111219\!\!$ & $\!\!+21408.6\!\!$ & $\!\!-2184.57\!\!$ & $\!\!+91.6832\!\!$  \\
\spp  $\!\!P_1\!\!$ &  $\!\!-.0398044\!\!$ & $\!\!+.271220\!\!$ & $\!\!-.205904\!\!$ & $\!\!+.072199\!\!$ & $\!\!-.0119735\!\!$ & $\!\!+8.56923\cdot10^{-4}\!\!$ & $\!\!-1.74034\cdot10^{-5}\!\!$  \\
\spp  $\!\!P_2\!\!$ &  $\!\!-.0398265\!\!$ & $\!\!+.0779803\!\!$ & $\!\!-.106152\!\!$ & $\!\!+.0549163\!\!$ & $\!\!-.0132171\!\!$ & $\!\!+1.50452\cdot10^{-3}\!\!$ & $\!\!-6.58489\cdot10^{-5}\!\!$  \\
\hline
\end{tabular}
}
\caption{Coefficients of the polynomial functions 
$F_{(0,0)}$ and $G_{(0,0)}$ entering SM prediction of $A_{\mathrm{T}}^{(2)}$ and those of the 
 polynomials $P_1$ and $P_2$ corresponding to the associated upper and lower error bands respectively. The second row in this table and the following ones indicates the dimension of the coefficients in each column.}
\label{tab:coeffAT2-FG}
}

\TABLE{
\centering
\footnotesize{
\begin{tabular}{|c|c|c|c|c|c|c|c|}
\hline
\spp $\!\!(i,j)\!\!$  & $1$ & $s$ & $s^2$ & $s^3$ & $s^4$ & $s^5$ & $s^6$  \\
\hline
\spp dim & 1 & $\rm GeV^{-2}$ & $\rm GeV^{-4}$ & $\rm GeV^{-6}$ & $\rm GeV^{-8}$ & $\rm GeV^{-10}$ & $\rm GeV^{-12}$ \\
\hline
\spp  $\!\!(0,7)\!\!$ & $\!\!-35566.4\!\!$ & $\!\!+46009.2\!\!$ & $\!\!-27457.0\!\!$ & $\!\!+9232.04\!\!$ & $\!\!-1776.93\!\!$ & $\!\!+181.164\!\!$ & $\!\!-7.59797\!\!$ \\
\spp $\!\!(0,7')\!\!$ &  $\!\!-2260921\!\!$ & $\!\!+2797565\!\!$ & $\!\!-1657496\!\!$ & $\!\!+557358\!\!$ & $\!\!-106615\!\!$ & $\!\!+10798.0\!\!$ & $\!\!-448.880\!\!$  \\
\spp $\!\!(0,9)\!\!$ &  $\!\!-495.374\!\!$ & $\!\!+80.4698\!\!$ & $\!\!+25.6073\!\!$ & $\!\!-1.54246\!\!$ & $\!\!-1.27554\!\!$ & $\!\!+0.20500\!\!$ & $\!\!-0.0141835\!\!$ \\
\spp $\!\!(0,9')\!\!$ & $\!\!-17643.1\!\!$ & $\!\!+2256.36\!\!$ & $\!\!+1655.96\!\!$ & $\!\!-634.239\!\!$  & $\!\!+148.767\!\!$ & $\!\!-18.4135\!\!$  & $\!\!+0.947823\!\!$ \\
\spp $\!\!(0,10)\!\!$ & $\!\!+2.27472\!\!$ & $\!\!-99.4500\!\!$ & $\!\!-11.2441\!\!$ & $\!\!+3.95594\!\!$ & $\!\!+0.138949\!\!$ & $\!\!+0.00447390\!\!$& $\!\!-0.000140794\!\!$  \\
\spp $\!\!(0,10')\!\!$ & $\!\!+104.982\!\!$ & $\!\!-4549.99\!\!$ & $\!\!-73.8320\!\!$ & $\!\!+2.77725\!\!$ & $\!\!+0.370546\!\!$ & $\!\!-0.0131493\!\!$ & $\!\!+0.00113558\!\!$ \\
\spp $\!\!(7,7)\!\!$ & $\!\!-3487.35\!\!$ & $\!\!-591.758\!\!$ & $\!\!+157.560\!\!$ & $\!\!-57.9418\!\!$ &  $\!\!+11.0662\!\!$ & $\!\!-1.12697\!\!$ & $\!\!+0.0470920\!\!$  \\
\spp $\!\!(7,7')\!\!$ & $\!\!+6381006\!\!$  & $\!\!-7188264\!\!$ & $\!\!+4269996\!\!$ & $\!\!-1437676\!\!$ & $\!\!+275442\!\!$ & $\!\!-27951.3\!\!$ & $\!\!+1164.94\!\!$  \\
\spp $\!\!(7,9)\!\!$ & $\!\!+504.942\!\!$ & $\!\!+22.2744\!\!$ & $\!\!-52.3132\!\!$ & $\!\!+7.14332\!\!$ & $\!\!-1.60871\!\!$ & $\!\!+0.161286\!\!$ & $\!\!-0.00668765\!\!$   \\
\spp $\!\!(7,9')\!\!$ & $\!\!+46001.8\!\!$ & $\!\!+4619.49\!\!$ & $\!\!-2289.01\!\!$ & $\!\!+755.115\!\!$ & $\!\!-145.752\!\!$ & $\!\!+14.8194\!\!$ & $\!\!-0.618975\!\!$  \\
\spp $\!\!(9,9)\!\!$ & $\!\!-0.263978\!\!$ & $\!\!+11.5410\!\!$ & $\!\!+1.30486\!\!$ & $\!\!-0.459081\!\!$ & $\!\!-0.0161249\!\!$ &  $\!\!-0.000519191\!\!$ & $\!\!+0.0000163389\!\!$  \\
\spp $\!\!(9,9')\!\!$ & $\!\!-24.3660\!\!$ & $\!\!+1056.04\!\!$ & $\!\!+17.1362\!\!$ & $\!\!-0.644594\!\!$ & $\!\!-0.0860028\!\!$ & $\!\! +0.00305192\!\!$ & $\!\!-0.000263564\!\!$ \\
\hline
\end{tabular}
}
\caption{Coefficients of the polynomial functions $F_{(i,j)}$ entering $A_{\mathrm{T}}^{(2)}$.}
\label{tab:coeffAT2-F}
}

\TABLE{
\centering
\footnotesize{
\begin{tabular}{|c|c|c|c|c|c|c|c|}
\hline
\spp $\!\!(i,j)\!\!$  & $1$ & $s$ & $s^2$ & $s^3$ & $s^4$ & $s^5$ & $s^6$  \\
\hline
\spp dim & 1 & $\rm GeV^{-2}$ & $\rm GeV^{-4}$ & $\rm GeV^{-6}$ & $\rm GeV^{-8}$ & $\rm GeV^{-10}$ & $\rm GeV^{-12}$ \\
\hline
\spp $\!\!(7,7)\!\!$ & $\!\!+3190503\!\! $ & $\!\!-3594132\!\! $ & $\!\!+2134998\!\!$ & $\!\!-718838\!\!$ &  $\!\!+137721\!\!$ & $\!\!-13975.7\!\! $ & $\!\!+582.471\!\!$  \\
\spp $\!\!(7,7')\!\!$ & $\!\!-6974.70\!\!$  & $\!\!-1183.52\!\!$ & $\!\!+315.119\!\!$ & $\!\!-115.884\!\!$ & $\!\!+22.1324\!\!$ & $\!\!-2.25393\!\!$ & $\!\!+0.0941840\!\!$  \\
\spp $\!\!(9,9)\!\!$ & $\!\!-12.1830\!\!$ & $\!\!+528.020\!\!$ & $\!\!+8.56811\!\!$ & $\!\!-0.322297\!\!$ & $\!\!-0.0430014\!\!$ &  $\!\!+0.00152596\!\!$ & $\!\!-0.000131782\!\!$   \\
\spp $\!\!(9,9')\!\!$ & $\!\!-0.527956\!\!$ & $\!\!+23.0821\!\!$ & $\!\!+2.60973\!\!$ & $\!\!-0.918163\!\!$ & $\!\!-0.0322497\!\!$ & $\!\!-0.00103838\!\!$ & $\!\!+0.0000326779\!\!$  \\
\hline
\end{tabular}
}
\caption{Coefficients of the polynomial functions $G_{(i,j)}$  entering $A_{\mathrm{T}}^{(2)}$.}
\label{tab:coeffAT2-G}
}

After computing this asymmetry with our inputs, we have fitted the results to a simple parametrisation of the following form
\begin{equation}
A_{\mathrm{T}}^{(2)}(q^2)=A_{\mathrm{T}}^{(2),\, CV}(q^2)^{+ \delta_u(q^2)}_{-\delta_d(q^2)}
\end{equation} 
with the central value
\begin{equation} \label{at2cv}
A_{\mathrm{T}}^{(2),\,CV}(q^2)=\frac{ \sum_{i=0,7,7',9,9',10,10'} \sum_{j=i,..10'} F_{(i,j)}(q^2) \delta C_i \delta C_j}{ \sum_{i=0,7,7',9,9',10,10'} \sum_{j=i,..10'} G_{(i,j)}(q^2) \delta C_i \delta C_j}
\end{equation}
where we have introduced the definition $\delta C_0\equiv1$ to write down the constant and linear terms in the same way as the quadratic ones. The errors on the asymmetry are given with respect to the SM central value $F_{(0,0)}/G_{(0,0)}$:
\begin{eqnarray}
\delta_u(q^2) \equiv P_1(q^2) - \frac{F_{(0,0)}(q^2)}{G_{(0,0)}(q^2)}, \\
\delta_d(q^2) \equiv  \frac{F_{(0,0)}(q^2)}{G_{(0,0)}(q^2)} - P_2(q^2).
\end{eqnarray}
All the above functions of $q^2=s$ have been fitted to polynomials in this variable. 
The coefficients  corresponding to the functions $F_{(0,0)}$, $G_{(0,0)}$, $P_1$ and $P_2$  are given in table \ref{tab:coeffAT2-FG} and that of $F_{(i,j)}$ and $G_{(i,j)}$ in tables \ref{tab:coeffAT2-F} and \ref{tab:coeffAT2-G} respectively. All these coefficients are dimensionful (but can be easily turned into dimensionless quantities once $F,G$ are expressed as a function of $\tilde s \equiv s/m_B^2$) with the dimension indicated in the second row of table 6\footnote{As an example of how to read those tables, we provide here the function $F_{(0,0)}(s)$:
\begin{eqnarray} F_{(0,0)}(s)=&+&12904.2-17256.7 \, {\rm GeV}^{-2} \times s+10543.8 {\rm GeV}^{-4}\, \times s^2 -3519.19 {\rm GeV}^{-6} \, \times s^3 
\nonumber \cr
&+& 667.247\, {\rm GeV}^{-8} \times s^4 -67.3536\, {\rm GeV}^{-10} \times s^5 +2.78209\, {\rm GeV}^{-12} 
\times s^6\,.\nonumber\end{eqnarray}

}.

All entries of  the matrices $F$ and $G$ should be taken to be zero, except for those provided in tables 
\ref{tab:coeffAT2-FG},
 \ref{tab:coeffAT2-F} and \ref{tab:coeffAT2-G} and those related to them through the following equations
\begin{eqnarray} 
F_{(7',7')}=F_{(7,7)} , \quad F_{(7',9')}=F_{(7,9)}, &&  F_{(10,10')}=F_{(9,9')}, \cr
F_{(7',9)}=F_{(7,9')} \quad {\rm and}   \quad F_{(9',9')}&=& F_{(10',10')}=F_{(10,10)}=F_{(9,9)}.
\label{AT2Frel}
 \end{eqnarray}
Most of these symmetries, and the following ones between different $F_{(i,j)}$ ($G_{(i,j)}$) elements, are easily understood once the large recoil limit of the spin amplitudes is inserted into the definition of the observable~\cite{Kruger:2005ep} (see Appendix~\ref{sec:symmetries} for details).      
Similarly for the $G_{(i,j)}$ functions we have
\begin{eqnarray} G_{(7',9)}=G_{(7,9')}, && G_{(10,10')}=G_{(9,9')}, \cr
G_{(7',7')}=G_{(7,7)} , \!\!\quad G_{(7',9')}=G_{(7,9)} \, &{\rm and}&  \, G_{(9',9')}= G_{(10',10')}=G_{(10,10)}=G_{(9,9)} \quad \, \, 
\label{AT2Grel}
\end{eqnarray}
together with the relations between the $G_{(i,j)}$ and $F_{(i,j)}$ functions:
\begin{eqnarray}  
G_{(0,7')}=F_{(0,7)}, \quad  G_{(7,9')}=F_{(7,9)}, &\quad& G_{(0,9')}=F_{(0,9)}, \quad G_{(0,10')}=F_{(0,10)}, \cr
G_{(0,7)}=F_{(0,7')}, \quad G_{(7,9)}=F_{(7,9')},  &\quad& G_{(0,9)}=F_{(0,9')}, \quad G_{(0,10)}=F_{(0,10')} .
 \label{AT2FGrel}
\end{eqnarray}
In conclusion, the total number of non-zero entries of the matrices $F_{(i,j)}$ and $G_{(i,j)}$ entering eq.~(\ref{at2cv}) is 20 for each matrix\footnote{For instance, the 20 non-zero elements for the matrix $F_{(i,j)}$ correspond to the values of \begin{eqnarray} (i,j)=&\{&\!\!(0,0),(0,7),(0,7'),(0,9),(0,9'),(0,10),(0,10'),(7,7),(7,7'),(7,9),(7,9'),(9,9),(9,9'),
\nonumber \cr &\,&
\!\!(7',7'),(7',9'),(7',9),(9',9'),(10,10),(10,10'),(10',10') \, \}\nonumber\end{eqnarray}}.

As stated earlier, there is no current measurement of this asymmetry, but we will present in Sec.~\ref{sec:results}
the predicted $A_{\mathrm{T}}^{(2)}$ value for each of the allowed regions in our different scenarios. 

\subsection{Class III}\label{sec:classIII}

Here we consider observables affected by ${\cal O}_{7}$, ${\cal O}_{7^\prime}$, ${\cal O}_{9,10}$, ${\cal O}_{9^\prime,10^\prime}$, and in principle other kinds of NP operators such as scalars or tensors.
The most important observable in this category is ${\cal B}(B \to X_s \ell^+ \ell^-)$ due to its limited sensitivity to non-perturbative physics. In the same category fall also other observables defined through the angular distribution of $B \to K^*(\to K \pi) \ell^+ \ell^-$, in particular the forward-backward asymmetry $A_{\mathrm{FB}}$ and the longitudinal polarisation fraction $F_{\mathrm{L}}$.

$\bullet$ ${\cal B}(B \to X_s \mu^+ \mu^-)$ will be used only  in the low-$q^2$ region (from 1 GeV$^2$ to 6 GeV$^2$) as the theoretical 
prediction in the high-$q^2$ (above 14.4 GeV$^2$) region suffers from further theoretical uncertainties \cite{Ghinculov:2003bx,Neubert:2000ch,Bauer:2001rc,Huber:2007vv}. In the low-$q^2$ region, the branching ratio is measured to be \cite{Huber:2007vv}:
\begin{equation}
{\cal B} ( {\bar{B}} \to X_s \, \mu^+ \, \mu^-)_{{\text{low-}}q^2}=\left\{ \begin{array}{ll}
\left( 1.49 \pm 0.50^{+0.41}_{-0.32} \right) \times
10^{-6}~ & (\rm Belle)~, \\
\left( 1.8 \pm 0.7 \pm 0.5 \right) \times 10^{-6} ~ & (\rm
BaBar)~, \\
\left( 1.60 \pm 0.50 \right) \times 10^{-6}~ & (\rm Average)
~. \\
\end{array} \right. 
\end{equation}
The SM prediction for ${\cal B} ( {\bar {{B}}} \to {X}_{s} \, \mu^+ \, \mu^-)$ 
 is $(1.59\pm0.11) \times 10^{-6}$~\cite{Huber:2007vv}.
With our inputs and including $m_s$-suppressed terms (see App.~\ref{sec:btoxsmumu} for more details),
we obtain the corresponding expression for the integrated branching ratio at the 
scale $\mu_b=4.8$ GeV in the low-$q^2$ region (from 1 to 6 GeV$^2$):
\begin{equation}\label{btoxsmumu}
{\cal B}(B \to X_s \mu^+ \mu^-)= 
10^{-7}\times \left[
\sum_{i,j=0,7,7^\prime,9,9^\prime,10,10^\prime} b_{(i,j)} \delta C_i \delta C_j   \pm \delta_b \right]
\end{equation}
The values of the non-vanishing coefficients $b$ are listed in 
table~\ref{tab:coeffBtoXsll}.
\begin{table}
\begin{center}
\footnotesize{\begin{tabular}{|l|l|l|}
\hline
\multicolumn{3}{|c|}{$b_{(0,0)}= 15.86 \quad \delta_b=1.51$} \\
\hline
\spp $b_{(0,7)}    = -0.517  $ &  $ b_{(0,9)}   =  2.663  $ & $b_{(0,10)}   = -4.679  $ \\
\spp $b_{(0,7')}   = -0.680  $ & $b_{(0,9')}   = -0.049  $  & $ b_{(0,10')}   = 0.061  $ \\
\spp $b_{(7,7)}   =b_{(7',7')}   =27.776  $    & $b_{(9,9)}   = b_{(9',9')}   =  0.534  $ & $b_{(10,10)} =b_{(10',10')} =  0.543  $\\
\spp $ b_{(7,7')}   = -0.399  $ &   $ b_{(9,9')}   =-0.014 $  & $ b_{(10,10')} =-0.014   $ \\
\spp  $ b_{(7,9)}  = b_{(7',9')}  =  4.920   $  & $ b_{(7,9')}   =b_{(7',9)}   =-0.113  $ & \\
\hline
\end{tabular}}
\caption{Coefficients describing the dependence of ${\mathcal B}(B\to X_s \mu^+\mu^-)$ on $C_{7,9,10}$ and $C_{7',9',10'}$.
\label{tab:coeffBtoXsll}
}
\end{center}
\end{table}

$\bullet$ $A_{\mathrm{FB}}(q^2)$. The forward-backward asymmetry in $\bar{B}_d \to \bar{K}^{*0} \ell^+ \ell^-$ is defined by:
\begin{equation}
A_{\mathrm{FB}}(q^2) = -\frac{1}{d\Gamma/dq^2}\left( \int_0^1 d ({\rm cos} \theta_l) \frac{d^2 \Gamma}{d q^2 d {\rm cos} \theta_l}- \int_{-1}^0 d ({\rm cos} \theta_l) \frac{d^2 \Gamma}{d q^2 d {\rm cos} \theta_l}\right)\,, \label{defafb}
\end{equation}
with $\theta_l$ the angle between the positively charged lepton in dimuon rest frame and the direction of the dilepton in the $\bar{B}_d$ rest frame.
This asymmetry can also be written in terms of spin amplitudes~\cite{Egede:2008uy} inside our framework as\footnote{Notice that while eq.~(\ref{defafb}) is valid in general, eq.~(\ref{AFBqsq}) is valid only within our framework, which means that one should add extra amplitudes in eq.~(\ref{AFBqsq}) when scalar operators are included.}
\begin{equation}
A_{\mathrm{FB}}(q^2) =-\frac{3}{2} \beta_\mu \frac{1}{d\Gamma/dq^2 }
 \left[{\mathrm{Re}}(A_{\| L} A_{\perp L}^*) - {\mathrm{Re}}(A_{\| R} A_{\perp R}^*)\right].  \label{AFBqsq}
\end{equation}
(See Appendix~\ref{btokllap} for definitions). The overall minus sign with respect to eq.~(4.4) in ref.~\cite{Egede:2008uy} stems from the definition of $A_{\mathrm{FB}}(q^2)$ in eq.~(\ref{defafb}) chosen to match the plots in refs.~\cite{Belle:2009zv, Aaltonen:2011cn}. The expression of $d \Gamma/dq^2$ in terms of $K^*$ spin amplitudes (including the muon mass terms) can be found in eq.~(\ref{dGamma}).
The QCDF framework at NLO is well suited to compute $A_{\mathrm{FB}}$, just as we did previously with $A_{\mathrm{T}}^{(2)}$, including an estimate of $\Lambda/m_b$ corrections. However, unlike $A_{\mathrm{T}}^{(2)}$, $A_{\mathrm{FB}}$ can receive not only contributions from the operators ${\cal O}_i$, ${\cal O}_i^\prime$ with $i=7,9,10$ but also from scalar and tensor operators~\cite{Alok:2009tz,Alok:2010zd}. Another important difference between $A_{\rm T}^{(2)}$ and $A_{\rm {FB}}$ is that $A_{\rm {FB}}$ is not protected at LO from soft form factor uncertainties contrary to $A_{\rm T}^{(2)}$. 
Besides, $A_{\rm T}^{(2)}$ exhibits the same remarkable features as $A_{\rm {FB}}$ like, for instance, the presence
or absence of a zero (in the presence of right-handed currents) \cite{Egede:2010zc,talk,talk2}. $A_{\rm {FB}}$ has
been under scrutiny lately, as a consequence of the Belle measurement suggesting that, contrary to SM prediction, it might not display a zero in the low-$q^2$ region,
triggering many proposals to explain this behaviour~\cite{Alok:2009tz,Alok:2010zd}. 

We define the integrated forward-backward asymmetry in the low-$q^2$ region to agree with the experimental determination:
\begin{equation}
{\tilde A}_{FB}=\frac{\int_{1 {\mathrm{GeV}}^2}^{6 {\mathrm{GeV}}^2}  \frac{d \Gamma}{dq^2} A_{\mathrm{FB}} (q^2) dq^2}{\int_{1 {\mathrm{GeV}}^2}^{6 {\mathrm{GeV}}^2}\frac{d \Gamma}{dq^2}},
\label{eq:IntAFB}
\end{equation}
while the average of the measured values by Belle \cite{Belle:2009zv} and CDF collaborations \cite{Aaltonen:2011cn} is
\begin{equation}
{\tilde A}_{FB}^{\mathrm{exp}}= 0.33 ^{+0.22}_{-0.24}.
\end{equation}


\begin{table}
\begin{center}
\footnotesize{
\begin{tabular}{|c|c|c|c|c|c|c|c|}
\hline
\spp $$  & $1$ & $s$ & $s^2$ & $s^3$ & $s^4$ & $s^5$ & $s^6$ \\
\hline
\spp dim & 1 & $\rm GeV^{-2}$ & $\rm GeV^{-4}$ & $\rm GeV^{-6}$ & $\rm GeV^{-8}$ & $\rm GeV^{-10}$ & $\rm GeV^{-12}$ \\
\hline
\spp  $\!\!H_{(0,0)}\!\!$ &  $\!\!+35333.6\!\!$ & $\!\!-311396\!\!$ & $\!\!+119428\!\!$ & $\!\!-30281.3\!\!$ & $\!\!+8546.83\!\!$ & $\!\!-1169.16\!\!$ & $\!\!+65.2322\!\!$  \\
\spp  $\!\!I_{(0,0)}\!\!$ &  $\!\!+773134\!\!$ & $\!\!-72762.1\!\!$ & $\!\!+280788\!\!$ & $\!\!-88514.3\!\!$ & $\!\!+24423.2\!\!$ & $\!\!-3375.38\!\!$ & $\!\!+188.8567\!\!$ \\
\spp  $\!\!P_3\!\!$ &  $\!\!+.118304\!\!$ & $\!\!-.602706\!\!$ & $\!\!+.410711\!\!$ & $\!\!-.125244\!\!$ & $\!\!+.0214497\!\!$ & $\!\!-1.98680\cdot10^{-3}\!\!$ & $\!\!+7.74701\cdot10^{-5}\!\!$ \\
\spp  $\!\!P_4\!\!$ &  $\!\!+.302083\!\!$ & $\!\!-1.13742\!\!$ & $\!\!+.847601\!\!$ & $\!\!-.299722\!\!$ & $\!\!+.0580893\!\!$ & $\!\!-5.87352\cdot10^{-3}\!\!$ & $\!\!+2.41917\cdot10^{-4}\!\!$  \\
\hline
\end{tabular}
}
\caption{Coefficients of the polynomial functions 
$H_{(0,0)}$ and $I_{(0,0)}$ entering SM prediction of $A_{\mathrm{FB}}$ and those of the 
 polynomials $P_3$ and $P_4$ corresponding to the associated upper and lower error bands respectively. 
}
\label{tab:coeffAFB-HI}
\end{center}
\end{table}

\begin{table}
\begin{center}
\footnotesize{
\begin{tabular}{|c|c|c|c|c|c|c|c|}
\hline
\spp $\!(i,j)\!$  & $1$ & $s$ & $s^2$ & $s^3$ & $s^4$ & $s^5$ & $s^6$  \\
\hline
\spp dim & 1 & $\rm GeV^{-2}$ & $\rm GeV^{-4}$ & $\rm GeV^{-6}$ & $\rm GeV^{-8}$ & $\rm GeV^{-10}$ & $\rm GeV^{-12}$ \\
\hline
\spp  $\!(0,7)\!$ & $-28429.0$ & $+636004$ & $+11547.1$ & $-654.500$ & $-35.5189$ & $-0.448945$ & $-0.0797274$ \\
\spp $\!(0,7')\!$ &  $+309.261$ & $-6889.37$ & $-1839.42$ & $+195.417$ & $+6.25264$ & $+0.200244$ & $-0.0220619$   \\
\spp $\!(0,9)$ &  $-5.09654$ & $-595.133$ & $+13614.3$ & $+237.012$ & $-13.3497$ & $-0.0975163$ & $-0.0602829$ \\
\spp $\!(0,10)\!$ & $-8200.84$ & $+72274.3$ & $-27719.1$ & $+7028.21$ & $-1983.70$ & $+271.360$& $-15.1402$  \\
\spp $\!(0,10')\!$ & $-50.4373$ & $-146.677$ & $-31.1282$ & $+62.7360$ & $-22.6837$ & $+3.00773$ & $-0.162833$ \\
\spp $\!(7,10)\!$ & $+6598.30$ & $-147615$ & $-2680.06$ & $+151.908$ &  $+8.24386$ & $+0.104199$ & $+0.0185045$  \\
\spp $\!(7,10')\!$ & $+71.7787$ & $-1599.01$ & $-426.925$ & $+45.3559$ &  $+1.45122$ & $+0.0464761$ & $-0.00512052$  \\
\spp $\!(9,10)\!$ & $+1.18289$ & $+138.129$ & $-3159.85$ & $-55.0098$ &  $+3.09843$ & $+0.0226333$ & $+0.0139915$  \\
\hline
\end{tabular}
}
\caption{Coefficients of the polynomial functions $H_{(i,j)}$ entering $A_{\mathrm{FB}}$.}
\label{tab:coeffAFB-H}
\end{center}
\end{table}

\begin{table}
\begin{center}
\footnotesize{
\begin{tabular}{|c|c|c|c|c|c|c|c|}
\hline
\spp $(i,j)$  & $1$ & $s$ & $s^2$ & $s^3$ & $s^4$ & $s^5$ & $s^6$\\
\hline
\spp dim & 1 & $\rm GeV^{-2}$ & $\rm GeV^{-4}$ & $\rm GeV^{-6}$ & $\rm GeV^{-8}$ & $\rm GeV^{-10}$ & $\rm GeV^{-12}$ \\
\hline
\spp  $\!(0,7)\!$ & $-3468590$ & $+813560$ & $+227870$ & $-94496.6$ & $+25300.8$ & $-3459.65$ & $+192.642$ \\
\spp  $\!(0,7')\!$ & $-85589.1$ & $-122670$ & $-69994.6$ & $+28153.7$ & $-7862.34$ & $+1093.91$ & $-61.8971$ \\
\spp  $\!(0,9)\!$ & $+20442.1$ & $-22730.3$ & $+69374.6$ & $-22297.5$ & $+6185.70$ & $-856.470$ & $+48.1719$ \\
\spp  $\!(0,9')\!$ & $-12916.9$ & $-74730.4$ & $-32300.8$ & $+13192.5$ & $-3605.75$ & $+501.316$ & $-28.4231$ \\
\spp  $\!(0,10)\!$ & $+261.790$ & $-121102$ & $-25790.2$ & $-176.716$ & $+45.8313$ & $-0.759850$ & $+0.113787$ \\
\spp  $\!(0,10')\!$ & $-273.106$ & $+122339$ & $+7232.91$ & $-179.752$ & $-13.9840$ & $+1.84675$ & $-0.0526165$ \\
\spp $\!(7,7)\!$ & $+4577553$ & $+174071$ & $-20355.8$ & $+6184.18$ &  $-1315.63$ & $+135.290$ & $-5.76932$  \\
\spp $\!(7,7')\!$ & $+329.213$  & $-145167$ & $-9858.44$ & $-33.5940$ & $+28.1525$ & $-1.45144$ & $+0.0810919$  \\
\spp $\!(7,9)\!$ & $-567.508$ & $+254709$ & $+6889.36$ & $-155.447$ & $-43.8097$ & $+2.35024$ & $-0.130858$   \\
\spp $\!(7,9')\!$ & $+125.219$ & $-55064.8$ & $-3710.40$ & $-58.9805$ & $+9.25472$ & $-0.632663$ & $+0.0321131$   \\
\spp $\!(9,9)\!$ & $-34.1907$ & $+14218.8$ & $+2996.38$ & $+20.4260$ & $-5.33164$ &  $+0.0886881$ & $-0.0132486$   \\
\spp $\!(9,9')\!$ & $+64.3379$ & $-28435.7$ & $-1680.38$ & $+41.4648$ & $+3.23993$ & $-0.429103$ & $+0.0122379$  \\
\spp $\!(10,10)\!$ & $-30.3804$ & $+14053.7$ & $+2992.92$ & $+20.5077$ & $-5.31866$ & $+0.0881797$ & $-0.0132048$  \\  
\spp $\!(10,10')\!$ & $+63.3872$ & $-28394.6$ & $-1678.74$ & $+41.7200$ & $+3.24565$ & $-0.428626$ & $+0.0122122$  \\
\hline
\end{tabular}
}
\caption{Coefficients of the polynomials functions $I_{(i,j)}$ entering $A_{\mathrm{FB}}$.}
\label{tab:coeffAFB-I}
\end{center}
\end{table}

We can provide a semi-numerical expression for this observable in a similar way to $A_{\mathrm{T}}^{(2)}$. Starting from the unintegrated asymmetry
\begin{equation}
{A}_{FB}(q^2)={A}_{FB}^{\,CV}(q^2)^{+ \delta_u(q^2)}_{-\delta_d(q^2)},
\label{eq:AFBqsq}
\end{equation}
where the central value ($CV$) is 
\begin{equation} \label{afbcv}
A_{\mathrm{FB}}^{\,CV}(q^2)=\frac{ \sum_{i=0,7,7',9,9',10,10'} \sum_{j=i,\ldots,10'} H_{(i,j)}(q^2) \delta C_i \delta C_j}{ \sum_{i=0,7,7',9,9',10,10'} \sum_{j=i,\ldots,10'} I_{(i,j)}(q^2) \delta C_i \delta C_j}
\end{equation}
(using again $\delta C_0=1$) and the uncertainties are given with respect to the SM central value curve ($H_{(0,0)}/I_{(0,0)}$):
\begin{eqnarray}
\delta_u(q^2) \equiv P_3(q^2) - \frac{H_{(0,0)}(q^2)}{I_{(0,0)}(q^2)}, \\
\delta_d(q^2) \equiv  \frac{H_{(0,0)}(q^2)}{I_{(0,0)}(q^2)} - P_4(q^2).
\label{AFBqsqerrors}
\end{eqnarray}
After integrating over the low-$q^2$ experimental kinematic range ($1 \leq q^2 \leq 6 \, {\rm GeV^2}$), following eq. (\ref{eq:IntAFB}) we obtain
\begin{equation}
{\tilde A}_{FB}={{\tilde A}_{FB}^{\,CV}\phantom{.}}^{+ \tilde \delta_u}_{- \tilde \delta_d},
\end{equation}
where the central value can be split into SM and NP contributions:
\begin{equation}
{\tilde A}_{FB}^{\, CV}={\tilde A}_{FB}^{SM} + {\tilde A}_{FB}^{NP},
\end{equation}
with
\begin{eqnarray}
{\tilde A}_{FB}^{SM}&=& \frac{\int_{1 {\mathrm{GeV}}^2}^{6 {\mathrm{GeV}}^2}  H_{(0,0)} (q^2) k(q^2) dq^2}{\int_{1 {\mathrm{GeV}}^2}^{6 {\mathrm{GeV}}^2} I_{(0,0)}(q^2) k(q^2) dq^2}, \label{intAFBSMcentral}\\
{\tilde A}_{FB}^{NP}&=& \frac{\int_{1 {\mathrm{GeV}}^2}^{6 {\mathrm{GeV}}^2} \sum_{i=0,7,7',9,9',10,10'} \sum_{j=i,..10'} H_{(i,j)}(q^2) k(q^2) \delta C_i \delta C_j  dq^2}{\int_{1 {\mathrm{GeV}}^2}^{6 {\mathrm{GeV}}^2} \sum_{i=0,7,7',9,9',10,10'} \sum_{j=i,..10'} I_{(i,j)}(q^2) k(q^2) \delta C_i \delta C_j dq^2} - {\tilde A}_{FB}^{SM},\quad
\end{eqnarray}
and the uncertainties are defined, according to eq.~(\ref{AFBqsqerrors}), as
\begin{eqnarray}
{\tilde \delta_u}&=& \frac{\int_{1 {\mathrm{GeV}}^2}^{6 {\mathrm{GeV}}^2}  I_{(0,0)} (q^2) k(q^2) P_3 (q^2) - H_{(0,0)}(q^2) k(q^2) dq^2}{\int_{1 {\mathrm{GeV}}^2}^{6 {\mathrm{GeV}}^2} I_{(0,0)}(q^2) k(q^2) dq^2}, \label{intAFBSMupper} \\
{\tilde \delta_d}&=& \frac{\int_{1 {\mathrm{GeV}}^2}^{6 {\mathrm{GeV}}^2}  H_{(0,0)}(q^2)k(q^2) - I_{(0,0)} (q^2) k(q^2) P_4 (q^2) dq^2}{\int_{1 {\mathrm{GeV}}^2}^{6 {\mathrm{GeV}}^2} I_{(0,0)}(q^2) k(q^2) dq^2}.\label{intAFBSMlower} 
\end{eqnarray}
The coefficients of the polynomials $H_{(0,0)}$, $I_{(0,0)}$, $P_3$ and $P_4$ can be found in table \ref{tab:coeffAFB-HI} and those of $H_{(i,j)}$ and $I_{(i,j)}$ are in tables \ref{tab:coeffAFB-H} and \ref{tab:coeffAFB-I} respectively.\footnote{Notice that a normalization factor $k(q^2)$ has been introduced in Eqs.(\ref{intAFBSMcentral})-(\ref{intAFBSMlower}). This factor cancels exactly in the differential $A_{FB}(q^2)$ or $F_L(q^2)$ expressions, but should be considered for the integrated $\tilde A_{FB}$ and $\tilde F_L$ observables. Its value is: $k(s)=64.7723- 76.8789\, s + 45.3645\, s^2 - 11.9407\, s^3 - 0.970171\, s^4 + 
 1.79656\, s^5 - 0.631163\, s^6 + 0.120215\, s^7 - 0.0135664\, s^8 + 
 0.000856554\, s^9 - 0.0000234364\, s^{10}$ with $s=q^2$. This normalization function is obtained by requesting that the SM prediction for the differential decay rate is exactly  reproduced, shown in Fig.\ref{dGammaPlot}, i.e.,  $10^{21} \times d\Gamma/ dq^2=I_{(0,0)} (q^2) \times  k(q^2) \times 10^{-6}$.}
 
 All components of  the  matrices $H$ and $I$ are taken to be zero (as it was done for $A_{\mathrm{T}}^{(2)}$) except for those provided in these tables and those related to them via the equations
\begin{eqnarray}  \label{AFBHrel}
H_{(7',10')}&=&-H_{(7,10)}, \quad H_{(9',10')} = - H_{(9,10)} \quad {\rm and} \quad H_{(7',10)} = - H_{(7,10')} .
 \end{eqnarray}
and
\begin{eqnarray} 
I_{(7',9')}&=&I_{(7,9)}, \quad I_{(7',9)}=I_{(7,9')},  \nonumber \\
I_{(7',7')}&=& I_{(7,7)}, \quad I_{(9',9')}=I_{(9,9)} \quad {\rm and} \quad I_{(10',10')}=I_{(10,10)},
\label{AFBIrel}
\end{eqnarray}
which leaves finally $12$ $H_{(i,j)}$ and $20$ $I_{(i,j)}$ non-zero functions entering eq.~(\ref{afbcv})-eq.~(\ref{intAFBSMlower}).

Using eqs.~(\ref{intAFBSMcentral}), (\ref{intAFBSMupper}), (\ref{intAFBSMlower}) and table \ref{tab:coeffAFB-HI} we get the following prediction for the integrated forward-backward asymmetry  ($\tilde{A}_{\mathrm{FB}}$) in the SM:
\begin{equation}
\tilde{A}_{FB}^{SM}=-0.0316^{+0.0269}_{-0.0303}.
\end{equation}


$\bullet$ $F_{\mathrm{L}}$: The longitudinal polarization fraction of the $K^*$ in the exclusive $B \to K^* \ell^+ \ell^-$ decay is defined in terms of the spin amplitudes as
\begin{equation}
F_{\mathrm{L}}= \frac{|A_0|^2}{\frac{d \Gamma}{dq^2}}\,.
\label{FLamplitudes}
\end{equation}
in absence of scalar and tensor operators~\cite{Alok:2010zd}, with $d \Gamma/dq^2$ given by eq.~(\ref{dGamma}). $F_{\mathrm{L}}$ can also be computed in QCDF and, as before, an estimate of $\Lambda/m_b$ corrections has been added to the other sources of uncertainty of this observable.

The integrated version of this observable in the low-$q^2$ region can be defined as in eq.~(\ref{eq:IntAFB})
\begin{equation}
\tilde{F}_L=\frac{\int_{1 {\mathrm{GeV}}^2}^{6 {\mathrm{GeV}}^2} \frac{d\Gamma}{dq^2} F_{\mathrm{L}}(q^2) dq^2} {\int_{1 {\mathrm{GeV}}^2}^{6 {\mathrm{GeV}}^2} \frac{d\Gamma}{dq^2}},
\end{equation}
and the average of the data measured by Belle \cite{Belle:2009zv} and CDF collaborations \cite{Aaltonen:2011cn} from this observable yields
\begin{equation}
{\tilde F}_L^{\mathrm{exp}}=  0.60^{+0.18}_{-0.19} \,.
\end{equation}


\begin{table}
\begin{center}
\footnotesize{
\begin{tabular}{|c|c|c|c|c|c|c|c|}
\hline
\spp $ $  & $1$ & $s$ & $s^2$ & $s^3$ & $s^4$ & $s^5$ & $s^6$ \\
\hline
\spp dim & 1 & $\rm GeV^{-2}$ & $\rm GeV^{-4}$ & $\rm GeV^{-6}$ & $\rm GeV^{-8}$ & $\rm GeV^{-10}$ & $\rm GeV^{-12}$ \\
\hline
\spp  $\!\!J_{(0,0)}\!\!$ &  $\!\!+42950.7\!\!$ & $\!\!+326107\!\!$ & $\!\!+137315\!\!$ & $\!\!-54729.1\!\!$ & $\!\!+14915.6\!\!$ & $\!\!-2078.06\!\!$ & $\!\!+ 117.102\!\!$  \\
\spp  $\!\!I_{(0,0)}\!\!$ &  $\!\!+773134\!\!$ & $\!\!-72762.1\!\!$ & $\!\!+280788\!\!$ & $\!\!-88514.3\!\!$ & $\!\!+24423.2\!\!$ & $\!\!-3375.38\!\!$ & $\!\!+188.857\!\!$  \\
\spp  $\!\!P_5\!\!$ &  $\!\!-.0792139\!\!$ & $\!\!+.952685\!\!$ & $\!\!-.395205\!\!$ & $\!\!+.0821238\!\!$ & $\!\!-.00911051\!\!$ & $\!\!+4.67994\cdot10^{-4}\!\!$ & $\!\!-6.09404\cdot10^{-6}\!\!$  \\
\spp  $\!\!P_6\!\!$ &  $\!\!-.133068\!\!$ & $\!\!+.720264\!\!$ & $\!\!-.154064\!\!$ & $\!\!-.0186277\!\!$ & $\!\!+.0121348\!\!$ & $\!\!-1.77815\cdot10^{-3}\!\!$ & $\!\!+8.87194\cdot10^{-5}\!\!$ \\
\hline
\end{tabular}
}
\caption{Coefficients of the polynomial functions 
$J_{(0,0)}$ and $I_{(0,0)}$ entering SM prediction of $F_{\mathrm{L}}$ and those of the 
 polynomials $P_5$ and $P_6$ corresponding to the associated upper and lower error bands respectively. 
 }\label{tab:coeffFL-JI}
\end{center}
\end{table}

\begin{table}
\begin{center}
\footnotesize{
\begin{tabular}{|c|c|c|c|c|c|c|c|}
\hline
\spp $(i,j)$  & $1$ & $s$ & $s^2$ & $s^3$ & $s^4$ & $s^5$ & $s^6$ \\
\hline
\spp dim & 1 & $\rm GeV^{-2}$ & $\rm GeV^{-4}$ & $\rm GeV^{-6}$ & $\rm GeV^{-8}$ & $\rm GeV^{-10}$ & $\rm GeV^{-12}$ \\
\hline
\spp  $(0,7)$ & $+21257.2$ & $+146631$ & $+65353.0$ & $-26003.8$ & $+7140.37$ & $-997.823$ & $+56.4467$ \\
\spp $(0,9)$ &  $+10438.4$ & $+73176.5$ & $+32041.6$ & $-12776.8$ & $+3503.48$ & $-489.692$ & $+27.7109$ \\
\spp $(0,10)$ & $+2821.90$ & $-122131$ & $-7771.38$ & $+178.016$ & $+20.9504$ & $-0.329242$& $+0.0223758$  \\
\spp $(7,7)$ & $-1326.63$ & $+57405.1$ & $+3778.94$ & $-61.4705$ &  $-9.92262$ & $+0.105597$ & $-0.01443264$  \\
\spp $(7,9)$ & $-1318.24$ & $+57047.6$ & $+3692.75$ & $-72.1696$ & $-9.82836$ & $+0.128021$ & $-0.0122690$   \\
\spp $(9,9)$ & $-327.478$ & $+14173.1$ & $+901.859$ & $-20.6586$ & $-2.43126$ &  $+0.0382081$ & $-0.00259668$   \\
\hline
\end{tabular}
}
\caption{Coefficients of the polynomial functions $J_{(i,j)}$ entering $F_{\mathrm{L}}$.}
\label{tab:coeffFL-J}
\end{center}
\end{table}

The analysis of $A_{\mathrm{FB}}$  and $\tilde{A}_{\mathrm{FB}}$    performed in eqs.~(\ref{eq:AFBqsq})-(\ref{intAFBSMlower}) can be repeated, step by step, for $F_{\mathrm{L}}$ and $\tilde{F}_{\mathrm{L}}$ with the substitutions $H_{(0,0)} \rightarrow J_{(0,0)}$, $H_{(i,j)} \rightarrow J_{(i,j)}$, 
$P_3 \rightarrow P_5$, $P_4 \rightarrow P_6$ and, obviously, $A_{\mathrm{FB}} \rightarrow F_{\mathrm{L}}$, ${\tilde A}_{\mathrm{FB}} \rightarrow {\tilde F}_{\mathrm{L}}$ and exactly the same normalization factor $k(q^2)$.
Table \ref{tab:coeffFL-JI} contains the coefficients of $J_{(0,0)}$, $I_{(0,0)}$ (for completeness), $P_5$ and $P_6$, while the different non-zero $J_{(i,j)}$ are either shown in table \ref{tab:coeffFL-J} or given by
\begin{eqnarray}
J_{(7',7')}&=&J_{(7,7)} , \quad J_{(7',9')} = - J_{(7,9')} = - J_{(7',9)}= J_{(7,9)}, \cr
J_{(9',9')}&=& J_{(10',10')}=J_{(10,10)}=J_{(9,9)}, \quad J_{(10,10')}=J_{(9,9')}, \cr
J_{(0,7')}&=& - J_{(0,7)}, \quad J_{(0,9')}= - J_{(0,9)}, \quad J_{(0,10')} = - J_{(0,10)}, \cr
J_{(7,7')}&=& -2 J_{(7,7)}, \quad {\rm and} \quad J_{(9,9')}=-2 J_{(9,9)},  \label{FLJrel}
 \end{eqnarray}
rendering $20$ entries $J_{(i,j)}$ different from zero entering $F_{\rm L}$.

Therefore, the value of the integrated polarization fraction ($\tilde{F}_L$) in the SM can be computed theoretically using our inputs to get
\begin{equation}
\tilde{F}_{L}^{SM} = 0.725^{+0.024}_{-0.037}.
\end{equation}

\subsection{${\cal{B}}(B_s \to \mu^+ \mu^-)$}

\begin{figure}
\begin{center}
\includegraphics[width=0.50\textwidth]{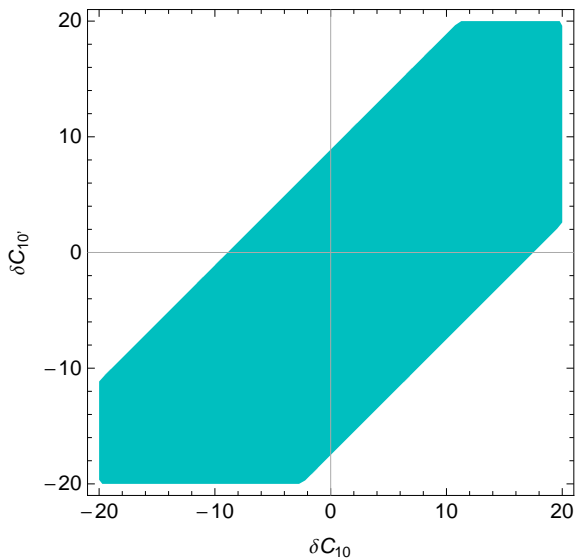}
\caption{Constraint imposed by ${\cal{B}}(\bar{B}_s \to \mu^+ \mu^-)$ to the values of the Wilson coefficients in the $(\delta C_{10},\delta C_{10^{\prime}})$ plane.\label{fig:BrBsmumu}}
\end{center}
\end{figure}

The branching ratio of $\bar{B}_s^0 \to \mu^+ \mu^-$ in presence of only NP axial operators (relevant to this analysis) is given, at leading order, by~\cite{Alok:2009tz, Altmannshofer:2008dz, Hurth:2008jc}
\begin{equation}
{\cal{B}}(\bar{B}_s \to \mu^+ \mu^-)|_{\rm axial}=\frac{G_F^2 \alpha^2}{16\pi^3} f_{B_s}^2 m_{B_s} \tau_{B_s} |V_{tb} V_{ts}^*|^2 m_\mu^2 \sqrt{1-\frac{4m_\mu^2}{m_{B_s}^2}} |C_{10}-C_{10^{\prime}}|^2
\label{eq:BrBsmumu}
\end{equation}
Using the inputs in table~\ref{wilson} and~\ref{tab:inputs} we get the SM prediction
\begin{equation}
{\cal{B}}(\bar{B}_s \to \mu^+ \mu^-)^{\mathrm{SM}} = (3.44 \pm 0.32) \cdot 10^{-9},
\end{equation}
which is one order of magnitude smaller than the most recent experimental averaged upper bound, obtained at the $90\%$ confidence level in ref.~\cite{Asner:2010qj}\footnote{The LHCb Collaboration has just released a paper \cite{Aaij:2011rj} where the upper limit on the branching ratio is set to ${\cal{B}}(\bar{B}_s \to \mu^+ \mu^-) < 5.6 \cdot 10^{-8}$ at $95\%$ confidence level for an integrated luminosity of $37\,{\rm pb}^{-1}$. Since this upper bound is larger than the one obtained by the CDF collaboration \cite{Asner:2010qj} we are not using it in this work.}:
\begin{equation}
{\cal{B}}(\bar{B}_s \to \mu^+ \mu^-)^{\mathrm{exp}} < 3.2 \cdot 10^{-8}.
\end{equation}
Eq.~(\ref{eq:BrBsmumu}) can be used to compute a semi-numerical expression for this observable that will impose constrains in the $(\delta C_{10},\delta C_{10^{\prime}})$ plane (see Fig.~\ref{fig:BrBsmumu}),
\begin{equation}
{\cal{B}}(\bar{B}_s \to \mu^+ \mu^-) = 1.8525 \cdot 10^{-10}\left[ |-4.3085 + \delta C_{10} - \delta C_{10^\prime} |^2  \pm 1.7274 \right].
\label{eq:BrBsmumu_sn}
\end{equation}
We have employed eq.~(\ref{eq:BrBsmumu_sn}) to check that the values of $\delta C_{10}$ and $\delta C_{10^{\prime}}$ used in Scenarios B and C (see below) were compatible with the constraints coming from ${\cal{B}}(\bar{B}_s \to \mu^+ \mu^-)$. Since the experimental upper bound is still much larger than the SM prediction, no further cuts in the parameter space of Wilson coefficients have been found.

\section{Results}\label{sec:results}

In this section we obtain the allowed regions systematically from all the 
observables discussed previously. Since we aim first at illustrating how much our conclusions
vary depending on the precise framework adopted to analyse the data, we will not
adopt a sophisticated statistical approach (see refs.~\cite{Hurth:2008jc,Deschamps:2009rh} for examples of such approaches in similar contexts), and we will stick
to a scanning approach, combining the $1 \,\sigma$ theoretical and experimental ranges for each observable
linearly to draw the corresponding constraint. For instance, if an observable 
$\hat{X}_i$ has been measured experimentally $X_i \pm \delta X_i$ and has the theoretical prediction $Y_i(\delta C_j) \pm \delta Y_i$, we draw the projection of the
region corresponding to the constraint $|X_i - Y_i| \leq  (\delta X_i + \delta Y_i)$.

\subsection{$(C_7,C_{7'})$ plane}

\begin{figure*}
\begin{center}
\includegraphics[width=0.7\textwidth]{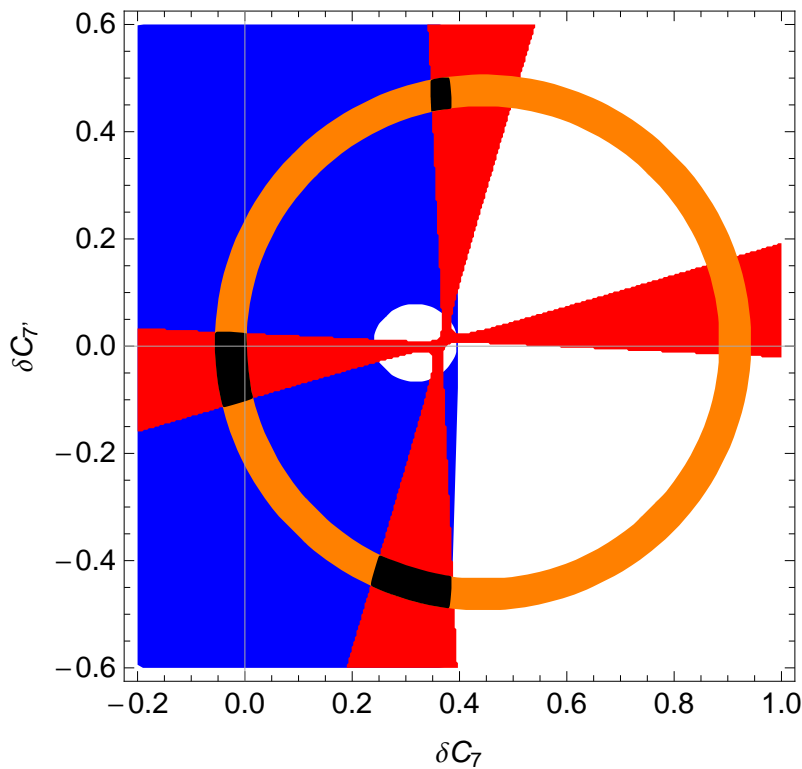}
\end{center}
\caption{Class I observables at $1\, \sigma$: $A_I$ (solid blue region with a white disk), {\cal{B}}($B \to X_s \gamma$) (orange ring) and $S_{K^* \gamma}$ (red cross). The three disconnected regions allowed by the intersection of these three observables are depicted in black. The SM value is given by the crossing of light gray lines at $(\delta C_7,\delta C_{7^{\prime}})=(0,0)$ point.
All plots of Wilson coefficients are taken at $\mu_b=4.8$ GeV.
}
\label{plot1}
\end{figure*}

As discussed in the introduction, we focus first on the $C_7, C_{7^\prime}$ plane, which will be the starting point of our discussion.
Therefore, we consider the three Class-I observables which only depend on the electromagnetic operators $C_7, C_{7^\prime}$,
leading to Fig.~\ref{plot1}. If one considers only ${\cal B}(B \to X_s \gamma)$ (ring in Fig.~\ref{plot1}) 
and $S_{K^* \gamma}$ (cross in Fig.~\ref{plot1}), four regions 
remain allowed: the SM one sitting around the origin, the ``flipped-sign'' solution~\cite{Gambino:2004mv} discussed in the introduction  around $(\delta C_7,\delta C_{7'})=(0.9,0)$, and two non SM-like solutions with $\delta C_7\simeq 0.35$ and $\delta C_{7'}$ around $\pm 0.5$. The flipped-signed solution does not correspond exactly to $C_7^{\rm eff} \to -C_7^{\rm eff}$ (and $C_{7'} \simeq 0$), due to interference terms between the electromagnetic operator and the four-quark operators in the observables considered here. The discriminating power
of the isospin asymmetry in $B \to K^* \gamma$ is quite obvious at this stage, as it discards this flipped-sign solution at $1 \, \sigma$ without
requiring further assumptions concerning NP for other operators. To recover this solution one needs to enlarge  both theoretical and experimental uncertainties up to $1.59 \, \sigma$. In our language, we disfavour this solution, working at $1 \, \sigma$, on the sole basis of Class-I operators,
contrary to ref.~\cite{Gambino:2004mv, Bobeth:2008ij} which needed Class-III quantities [${\cal B}(B\to X_s\ell^+\ell^-)$] and thus obtained conclusions with more restrictive 
assumptions concerning the manifestations of NP.

We will use the three identified black regions in Fig.~\ref{plot1} as the reference or primary regions:
\begin{itemize}
\item the region around $(\delta C_7,\delta C_{7^\prime})=(0,0)$, referred  to as the ``Central" or SM-like solution;
\item the upper region  around  $(\delta C_7,\delta C_{7^\prime})=(0.35,0.45)$, referred to as the ``Upper" region;
\item the lower region  around  $(\delta C_7,\delta C_{7^\prime})=(0.30,-0.45)$, referred to as the ``Lower" region.
\end{itemize}
The last two regions will be commonly called non SM-like solutions in the following.
These regions constitute the starting point to study the impact of Class-II and Class-III observables under the three different scenarios (A, B and C) presented in the introduction, each more general than the previous one.

It is important to remark that the two non SM-like primary regions  of Fig.~\ref{plot1} contain an interesting subset of solutions for $C_7$ with a flipped sign with respect to the SM. These solutions are characterised by a small modulus of $C_7$ and the addition of a larger contribution from $C_{7^\prime}$ to get agreement with data.

\subsection{Scenario A}

\begin{figure}
\centering
\subfloat[]{\label{plot2ax}
\includegraphics[width=0.33\textwidth]{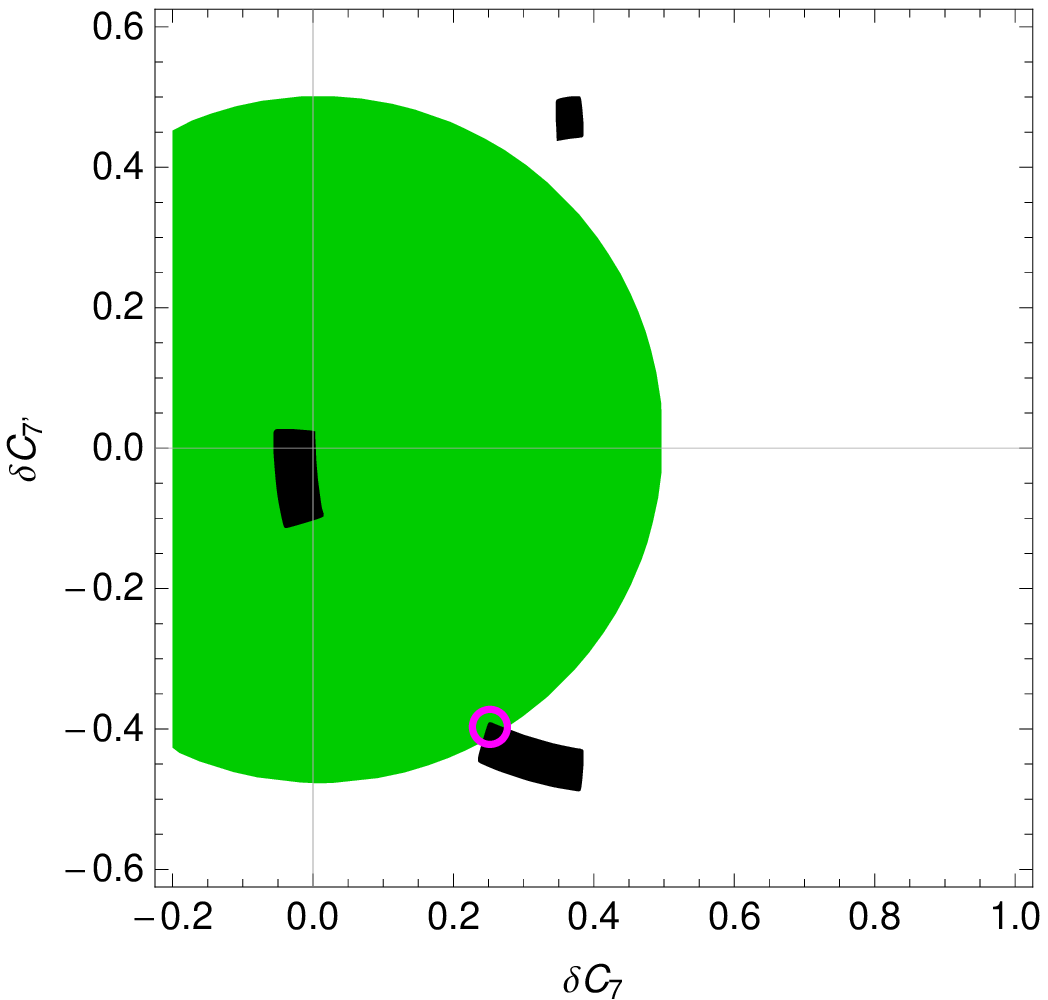}}                
\subfloat[]{\label{plot3ax}
\includegraphics[width=0.33\textwidth]{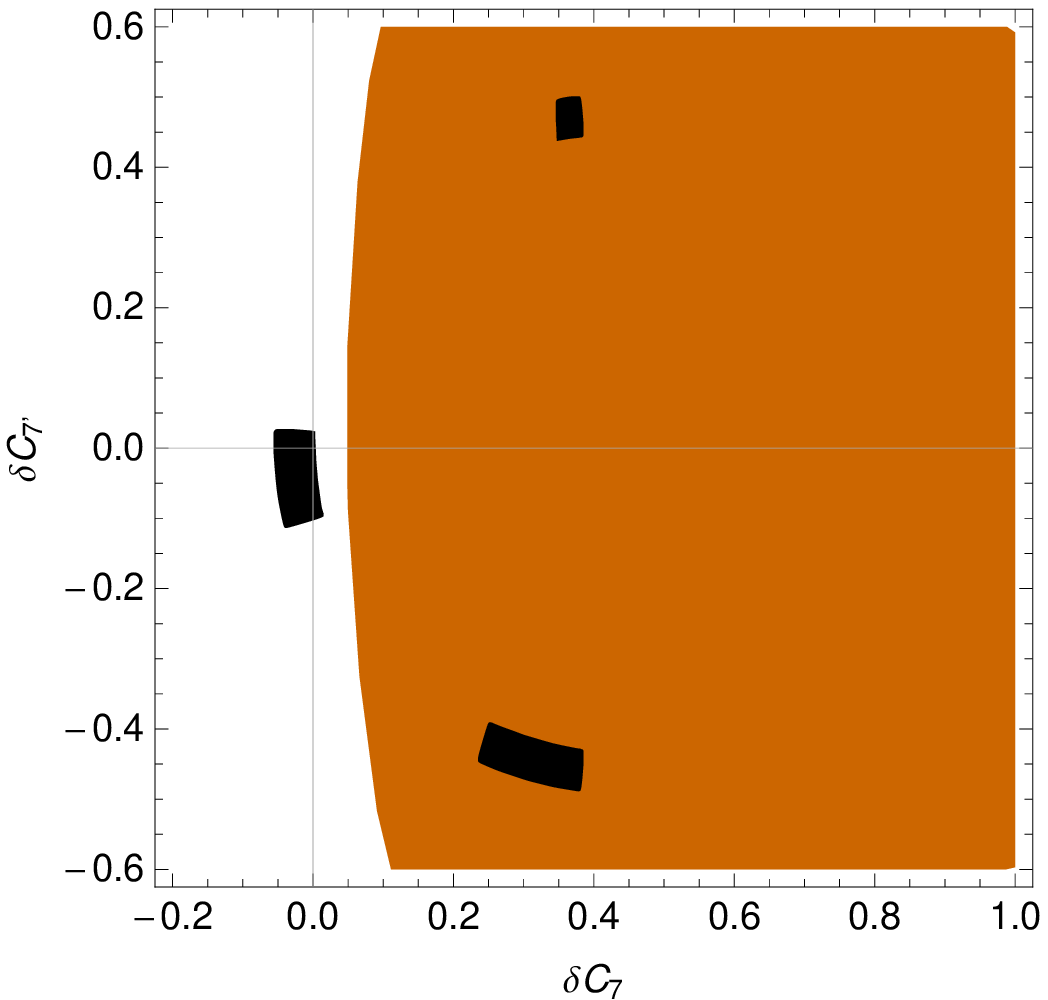}}
\subfloat[]{\label{plot4ax}
\includegraphics[width=0.33\textwidth]{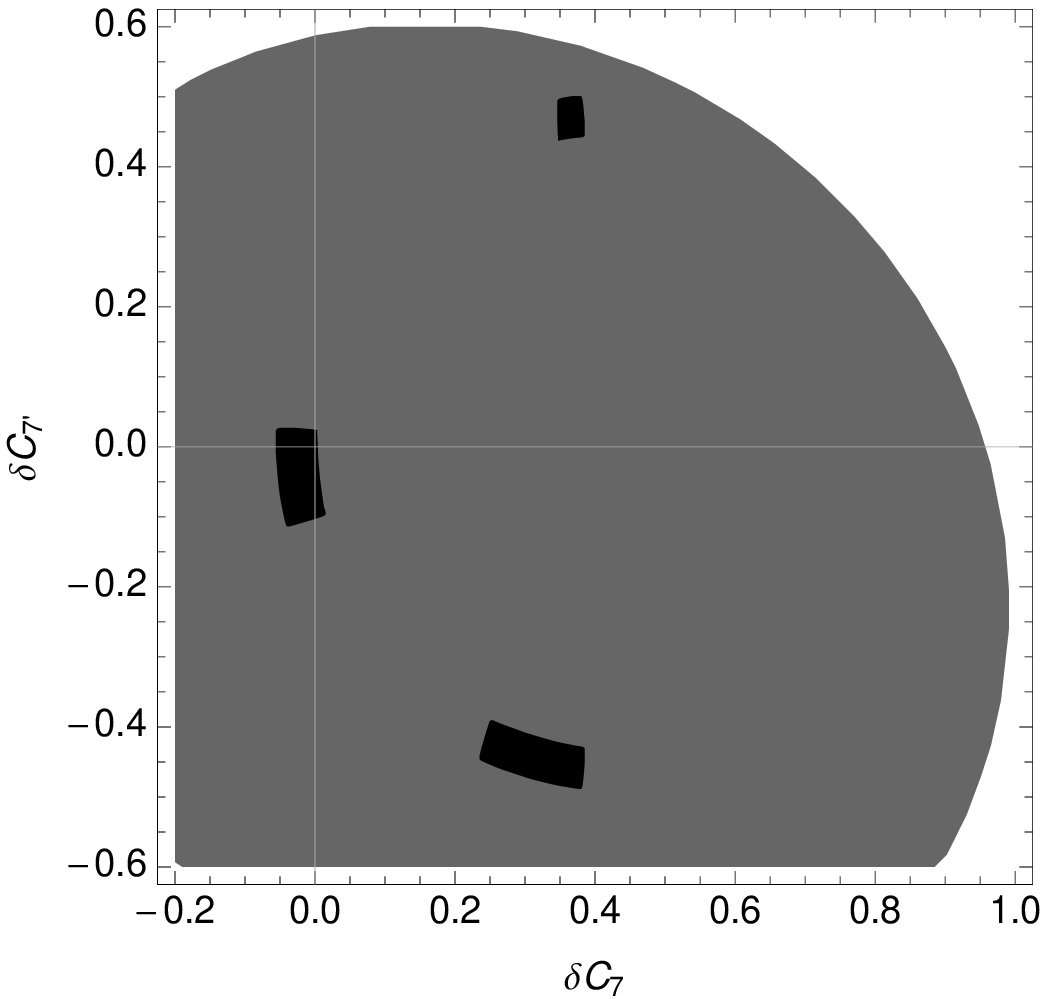}}
\caption{Constraint from Class-III observables {\cal{B}}($B \to X_s \mu^+ \mu^-$) (left), $\tilde{A}_{FB}$ (middle) and $\tilde{F}_L$
at $1\, \sigma$ in the $(\delta C_7,\delta C_{7^\prime})$ plane in Scenario A together with the three (black) regions allowed by Class-I observables. The magenta circle centered at $(0.25, -0.40)$ on the first plot indicates the tiny allowed region in this Scenario A.
}
\end{figure}

Let us start with Scenario A. If we consider the Class-III observables ${\cal B}(B  \to X_s \mu^+ \mu^-)$,
 $\tilde{A}_{FB}$ and $\tilde{F}_L$ for $B \to K^* \mu^+ \mu^-$ in the low-q$^2$ region,
we obtain the constraints shown in Figs.~\ref{plot2ax}, \ref{plot3ax} and \ref{plot4ax} respectively. One observes that the three observables favour different regions
of the $(C_7,C_{7'})$ plane: the inclusive decay favours the SM region and a very small subregion inside one of the non-SM like solutions, whereas (as expected) the forward-backward asymmetry would favour the
flipped sign-solution (had it not disappeared due to the isospin asymmetry) but also the two non-SM like solutions. The longitudinal polarisation would agree with all the regions (cutting only a very small part of the flipped-sign solution region).

\begin{figure*}
\begin{center}
\includegraphics[width=0.7\textwidth]{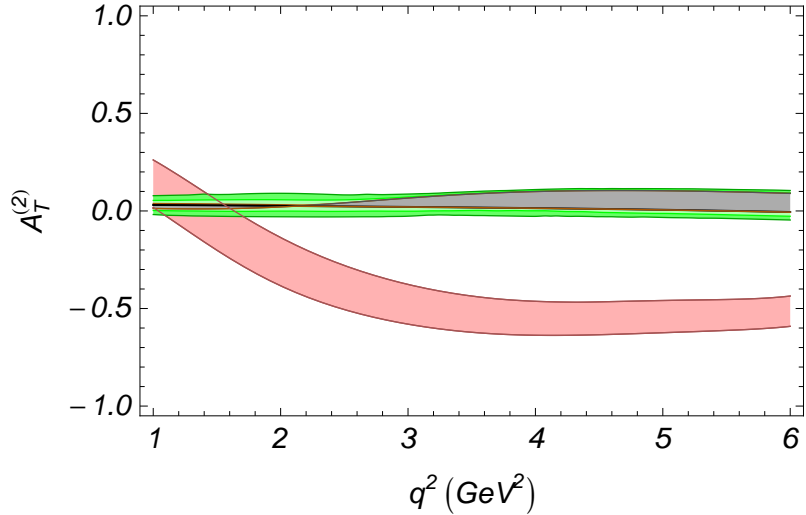}
\end{center}
\caption{Prediction for $A_{\rm T}^{(2)}$ under Scenario A (lower pink band), including error bars for all the allowed New Physics curves. The band around zero corresponds to the SM prediction.}
\label{at2A}
\end{figure*}

We see that Scenario A yields somewhat contradictory information from the various observables concerning which region in the $(C_7,C_{7^\prime})$ plane should be preferred. There is actually only a very small region in perfect agreement with all the observables measured (Class I and Class III), around $\delta C_7\simeq 0.25, \delta C_{7'} \simeq -0.40$ highlighted with a magenta circle in Fig.~\ref{plot2ax}, corresponding to the intersection of the lower black region with the $B\to X_s\mu^+\mu^-$ constraint.
It makes therefore sense to extend the set of operators potentially affected by NP and to consider Scenario B, including also New Physics in $C_9$ and 
$C_{10}$. Before leaving Scenario A, it is very interesting to compute the values for the (Class-II) observable $A_{\mathrm{T}}^{(2)}$ that is not yet measured, and turn it into a prediction.  Figure \ref{at2A} illustrates the prediction for this observable as a function of $q^2$ for the small set of points allowed by Scenario A. 
This leads to a very  precise prediction for the variation of $A_{\rm T}^{(2)}$ (including error bars) with $q^2$.  Notice that all the curves included in this region exhibit a zero in a range between 1 to 1.6 $\rm GeV^2$ which is controlled, at LO, by the same equation  that fixes the position of the zero of $A_{\rm {FB}}$ \cite{talk,talk2}. Given the small value of $C_7^{\rm eff}(\mu_b)\simeq-0.29+0.25=-0.04$, the position of this zero is shifted to the left with respect to the SM. Finally, another important prediction of this scenario is that $A_{\rm T}^{(2)}(q^2)$ would clearly prefer negative values, due to the negative value of $C_{7^\prime} \simeq -0.4$. Therefore, a measured value for $A_{\rm T}^{(2)}$ different from the narrow prediction given here would be enough to rule out this scenario. On the contrary, a measurement consistent with this prediction would make Scenario A the most plausible one (compared to the other scenarios), and furthermore, would signal clearly the presence of right-handed currents in radiative decays.

\subsection{Scenario B}

\begin{figure}
\centering
\subfloat[]{\label{plot2dB}
\includegraphics[width=0.33\textwidth]{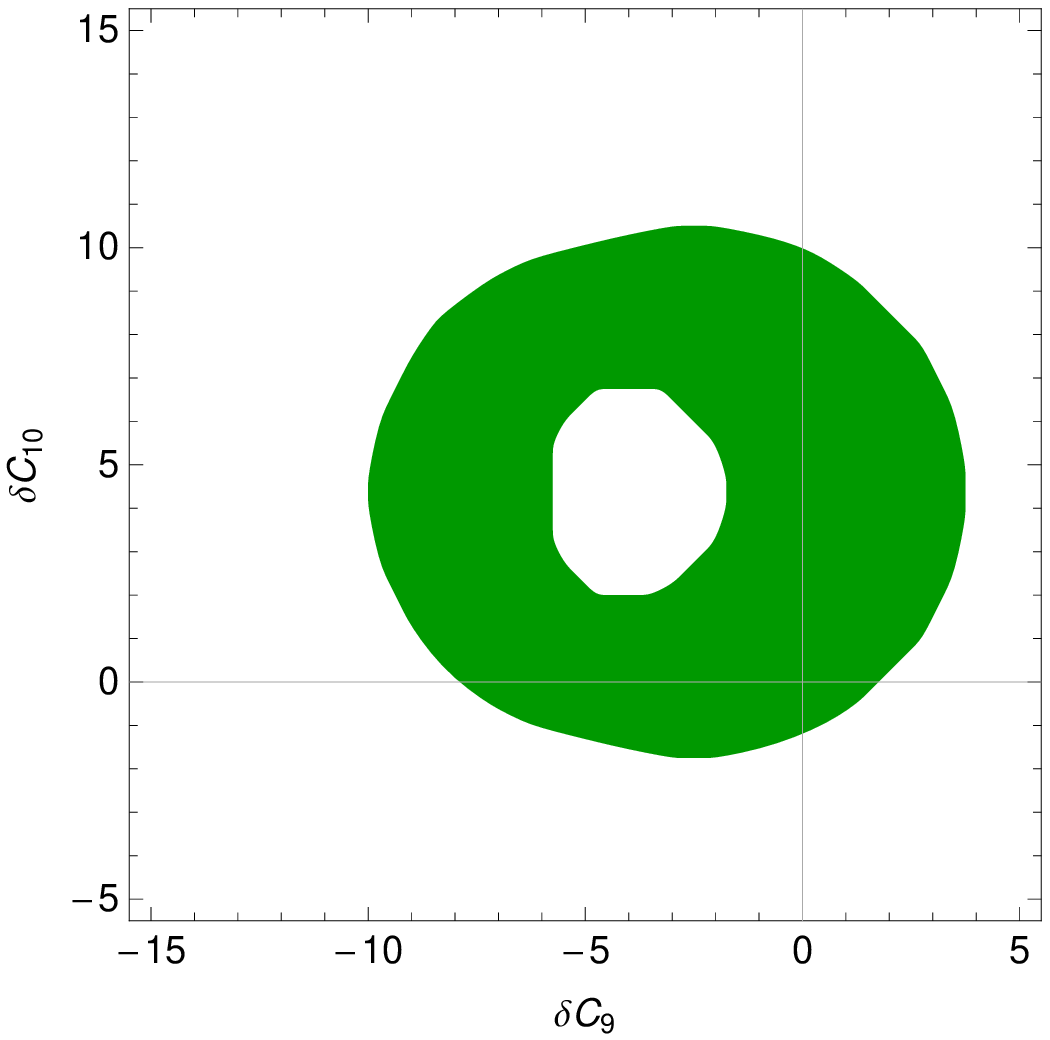}}                
\subfloat[]{\label{plot3d}
\includegraphics[width=0.33\textwidth]{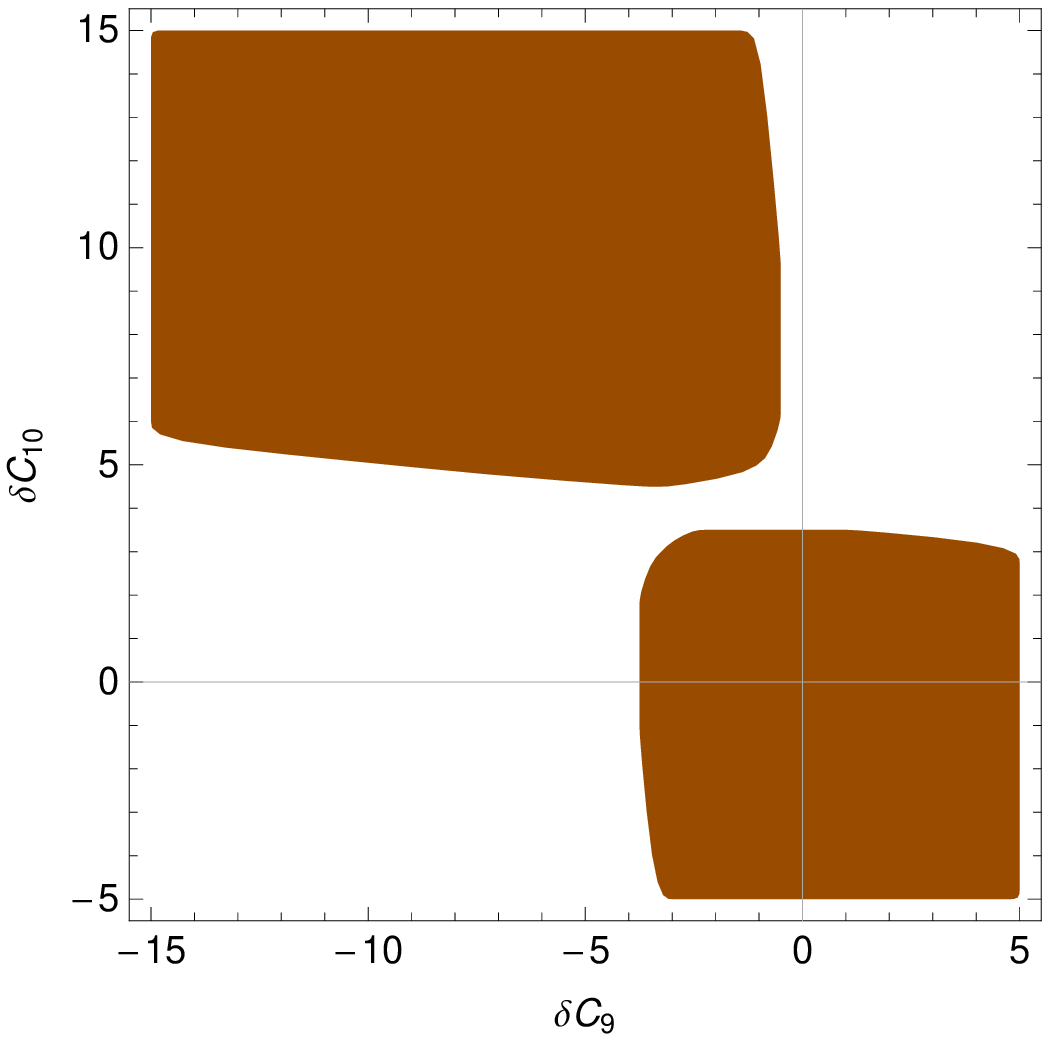}}
\subfloat[]{\label{plot4d}
\includegraphics[width=0.33\textwidth]{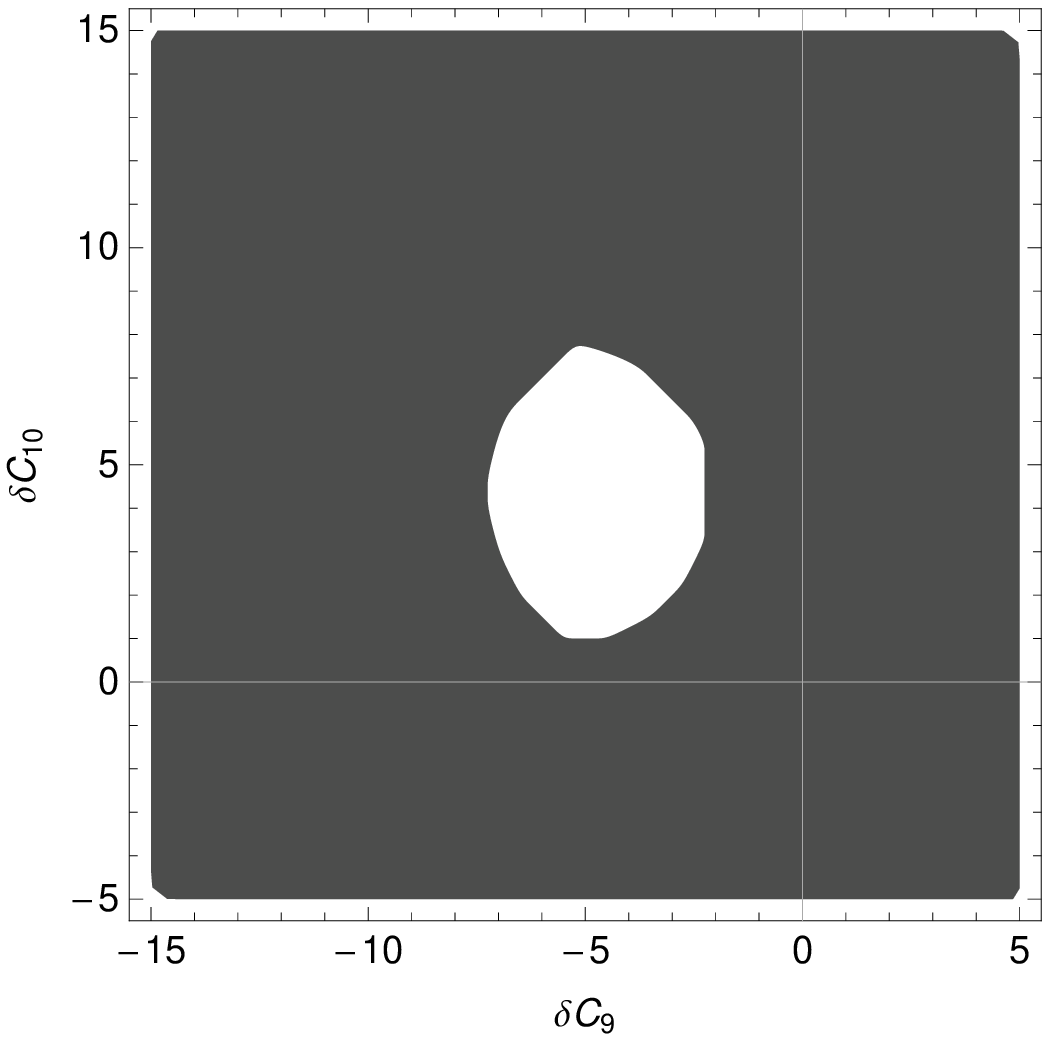}}
\caption{Constraint from Class-III observables {\cal{B}}($B \to X_s \mu^+ \mu^-$) (left), $\tilde{A}_{FB}$ (middle) and $\tilde{F}_{L}$ (right) at $1\, \sigma$ in the $(\delta C_9,\delta C_{10})$ plane in Scenario B. The region shown is compatible with the constraints on $\delta C_7$ and $\delta C_{7^\prime}$ imposed by Class-I observables.}
\end{figure}

In case of Scenario B,
the regions permitted by the Class-III observables ${\cal B}(B  \to X_s \mu^+ \mu^-)$, $\tilde{A}_{FB}$ and $\tilde{F}_L$ in Figs.~\ref{plot2ax}, \ref{plot3ax}, \ref{plot4ax} become extended to the whole plane,
and thus are not constraining anymore either $C_7$ or $C_{7'}$. 
In this scenario, the three primary (black) regions in Fig.~\ref{plot1} allowed by the Class-I observables are  compatible with all the Class-III observables considered and become the allowed region for $C_7$ and $C_{7'}$ in this scenario.
This obviously does not mean that the observables of class III mentioned above do not provide any constraint on NP, just that these constraints are not visible in this particular subspace of NP parameters.
As emphasized in the introduction, the $(C_7,C_{7'})$ plane is a summary that does not provide the full information on NP. It is thus interesting to turn to the 
$(C_9,C_{10})$ plane.
Figs.~\ref{plot2dB}, \ref{plot3d} and \ref{plot4d} are obtained taking the values of the (now) permitted  three primary (black) regions in Fig.~\ref{plot1} for $(C_7,C_{7^\prime})$ and determining the values of $C_9$ and $C_{10}$ that are then allowed for ${\cal B}(B  \to X_s \mu^+ \mu^-)$, $\tilde{A}_{FB}$ and $\tilde{F}_L$, respectively.
 
\begin{figure}
\centering
\includegraphics[width=0.7\textwidth]{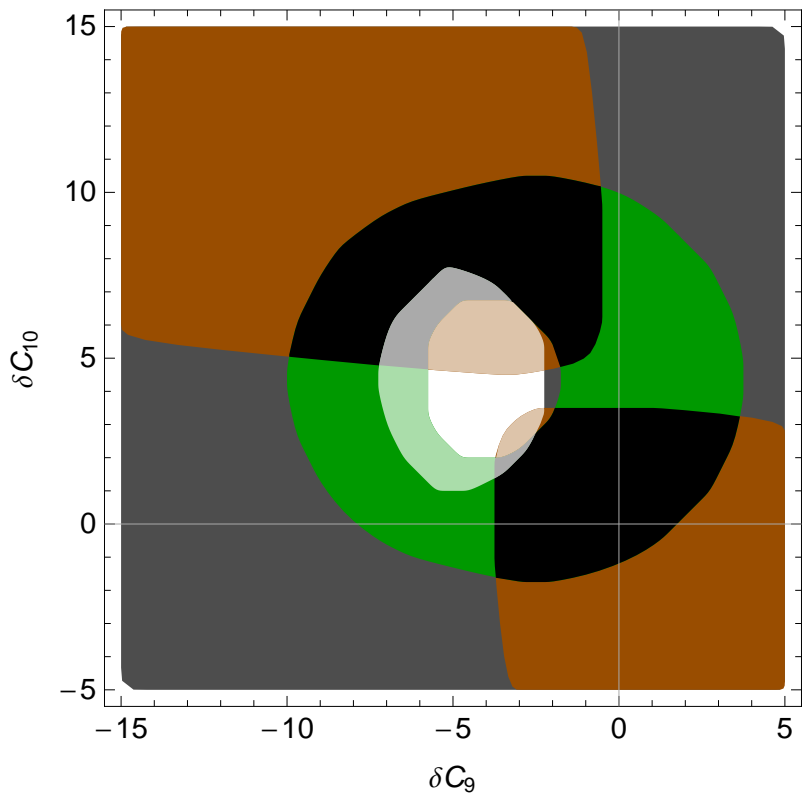}
\caption{Overlap of the constraints from Class-III observables {\cal{B}}($B \to X_s \mu^+ \mu^-$) (green ring), $A_{\mathrm{FB}}$ (upper and lower ``hyperbolic-like" brown regions; see Fig. 5b) and $F_{\mathrm{L}}$ (dark gray area with a central inlet) at $1\, \sigma$ in the $(\delta C_9,\delta C_{10})$ plane in Scenario B. The constraints imposed by their intersection are shown as two black regions.}
\label{plot2d3d4d}
\end{figure}

It is quite interesting to notice that the region excluded by $\tilde{F}_L$ is very close to the central region excluded by ${\cal B}(B  \to X_s \mu^+ \mu^-)$. This is more striking once all constraints from the three observables are overlapped in one single  Fig.~\ref{plot2d3d4d}, where only two regions (shown in black) are allowed by all constraints. The nature of these two areas can be understood by in the following way:
\begin{itemize}
\item SM region: the region centered at the origin corresponds to deviations from SM values  for $(C_9,C_{10})$  keeping the same sign for these coefficients as in SM;
\item flipped-values region or non-SM region: this solution contains  a subregion with opposite sign values for  $C_9$ and $C_{10}$ with respect to the SM ones.
\end{itemize}

The existence of these two regions can be understood from the fact that most of the observables have an approximate symmetry consisting in 
changing the sign of $C_9,C_{10}$ altogether, as long as $C_7$ or $C_{7^\prime}$ remain small (see, for instance, the large recoil expression for $A_{\rm {FB}}$ in eq. (\ref{AFBnumLR}) of Appendix \ref{sec:symmetries}
with $C_{9'} = C_{10'} = 0$). We checked that each of the three
primary (black) regions in the $(C_7,C_{7'})$ plane yield Class-III constraints 
in the $(C_9,C_{10})$ plane that cover the two regions in Fig. 6 almost
entirely. It implies that the two regions in $(C_9,C_{10})$ plane exist
independently of the precise values for $C_7$ and $C_{7'}$, as long as any of the
latter remain small and have a limited impact on the leptonic observables. In our framework, this smallness is indeed ensured by the constraints in ($C_7,C_{7'}$) plane coming from ${\cal B}( {\bar B} \to X_s \gamma)$. 

\begin{figure}

\centering
\subfloat[]{
\!\!\!\includegraphics[width=0.33\textwidth]{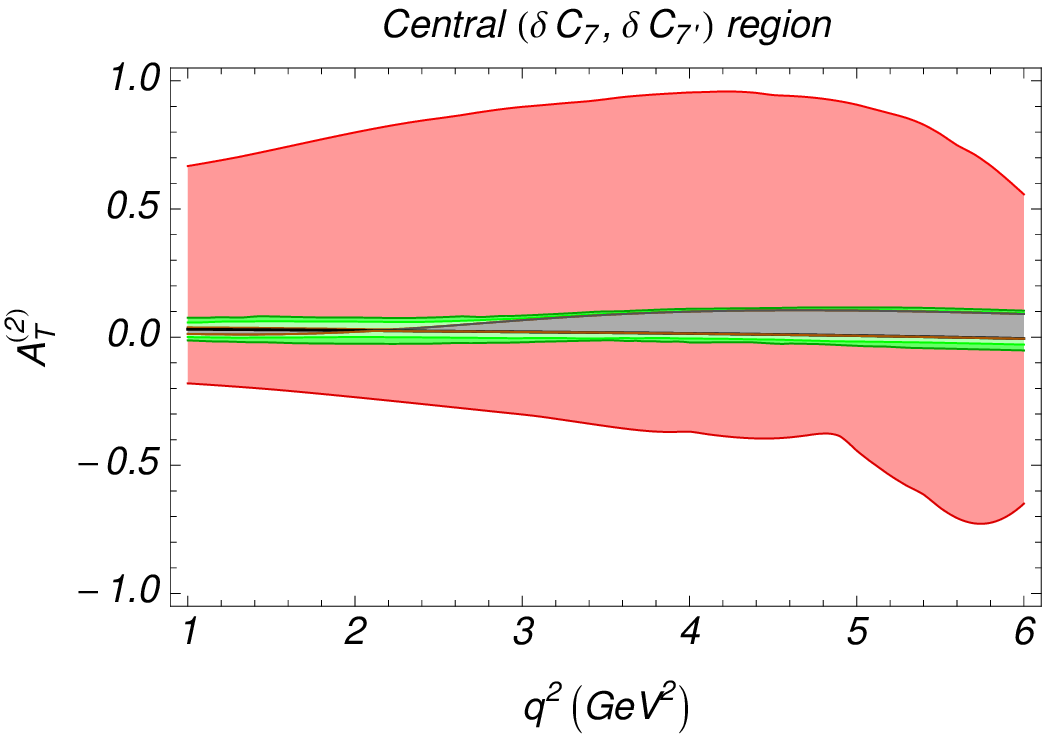}}                
\subfloat[]{
\includegraphics[width=0.33\textwidth]{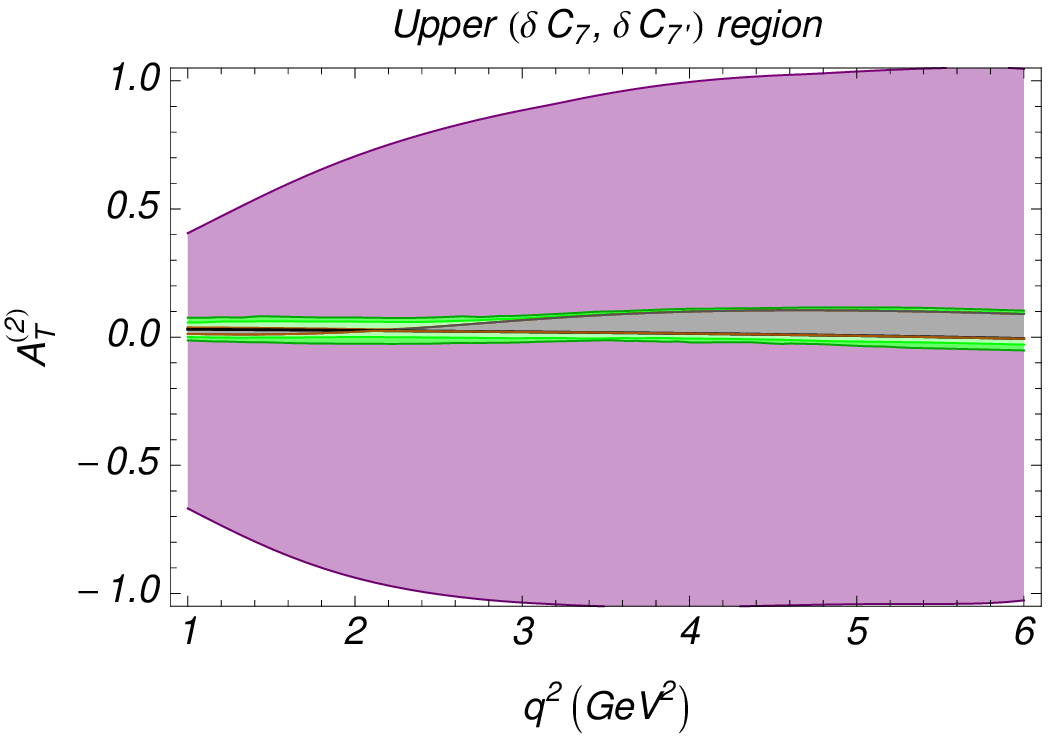}}
\subfloat[]{
\includegraphics[width=0.33\textwidth]{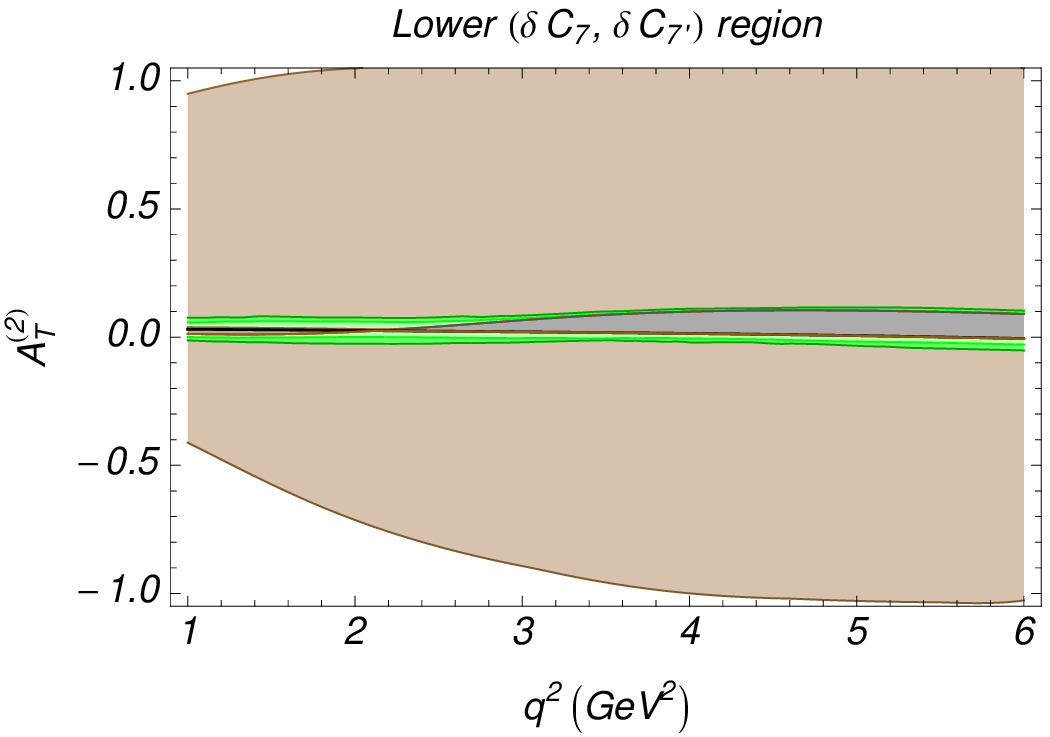}} \\
\subfloat[]{
\!\!\!\includegraphics[width=0.33\textwidth]{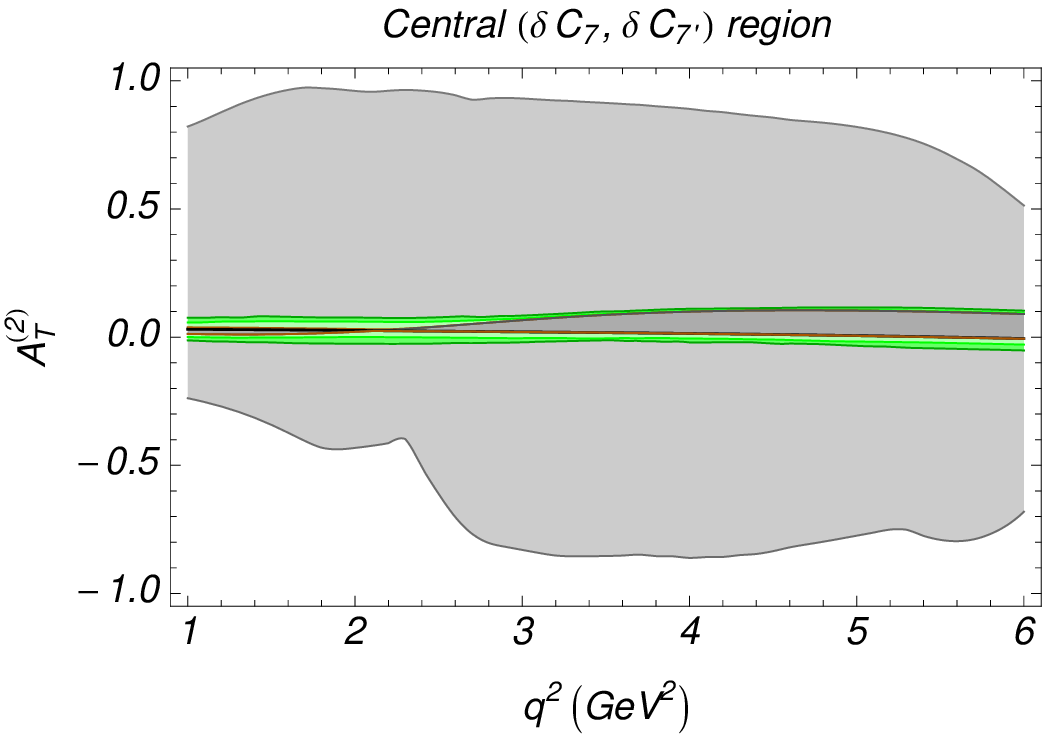}}                
\subfloat[]{
\includegraphics[width=0.33\textwidth]{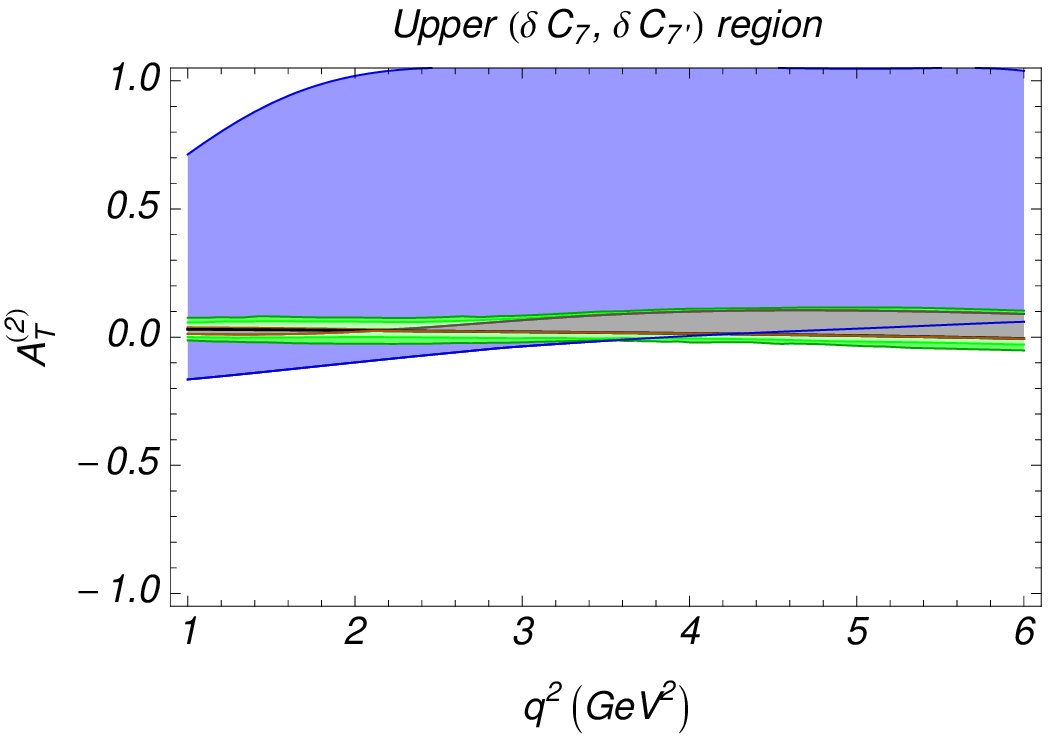}}
\subfloat[]{
\includegraphics[width=0.33\textwidth]{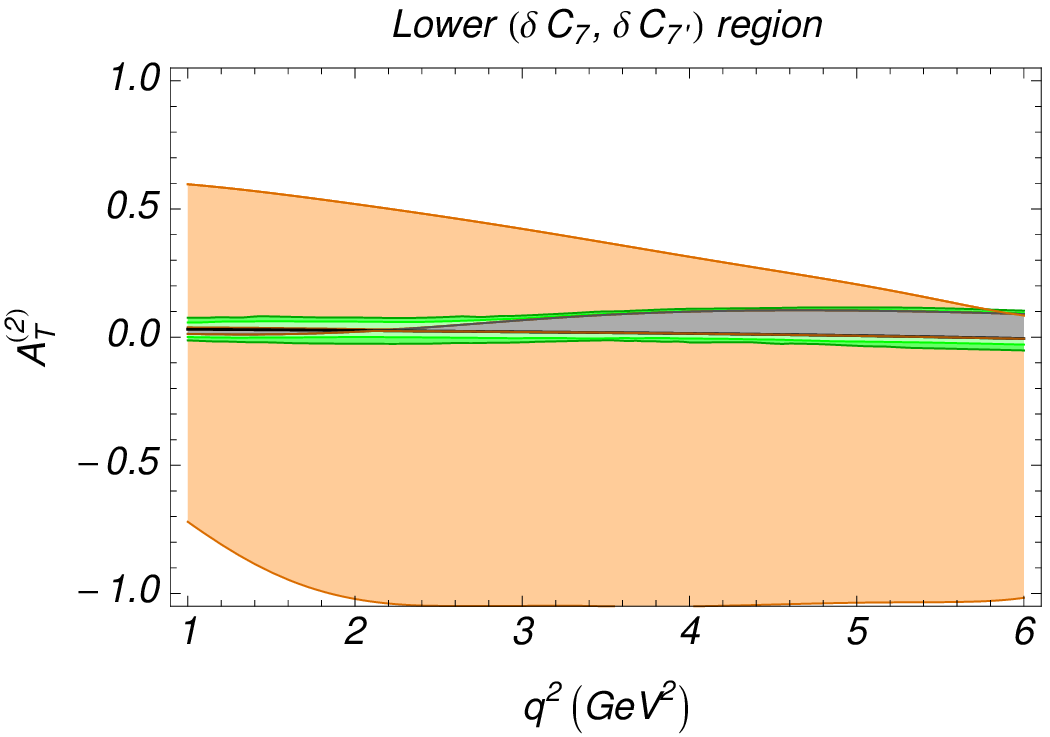}}
\caption{Prediction for $A_{\rm T}^{(2)}(q^2)$ corresponding to the ``flipped-values" region in $(C_9,C_{10})$ plane (first row of plots) and the SM-like region in $(C_9,C_{10})$ plane (second row) in Fig.~6. Each column corresponds to  SM-like Central region (left), non SM-like Upper region (center), non SM-like Lower region (right) for the $(C_7,
C_{7^\prime})$ plane allowed regions in Fig.~2.
}
\label{at2b}
\end{figure}

It is interesting to provide predictions for the (still not measured) asymmetry $A_{\mathrm{T}}^{(2)}$, using as inputs the WCs associated to the three black regions allowed in $(C_7,C_{7'})$ plane, together with the corresponding set of values  in the $(C_9,C_{10})$ plane (two black regions). This is shown in Fig.~\ref{at2b}. We can see there that the large allowed areas for $(C_9,C_{10})$ lead to wide bands in $A_{\rm T}^{(2)}(q^2)$. The Upper non-SM like $(C_7,C_{7'})$ region
associated to the SM-like $(C_9,C_{10})$ area gives a clear prediction for the sign of $A_{\rm T}^{(2)}$, which is just opposite to the one preferred by Scenario A. Also the Central (SM-like) $(C_7,C_{7'})$ region
associated to the non SM-like $(C_9,C_{10})$ area (Fig.~7a) and the Lower $(C_7,C_{7'})$ region associated to the SM-like $(C_9,C_{10})$ area (Fig.~7f) yield constraints on $A_{\rm T}^{(2)}$, though less stringent than those in Fig.~7e.

In conclusion, in this scenario  the upper region of $(C_7,C_{7^\prime})$ with the corresponding SM-like region for $(C_9,C_{10})$ could be discriminated clearly only if the sign of $A_{\rm T}^{(2)}$ would turn out to be negative, as predicted by Scenario A. Besides, high-$q^2$ measurements, not included in the present analysis, could shrink the 
allowed $(C_9,C_{10})$ region and thus reduce the range of possibilities 
 for $A_{\rm T}^{(2)}$ in this scenario.

\subsection{Scenario C}

\begin{figure}
\centering
\subfloat[]{\label{plot2dC}
\includegraphics[width=0.5\textwidth]{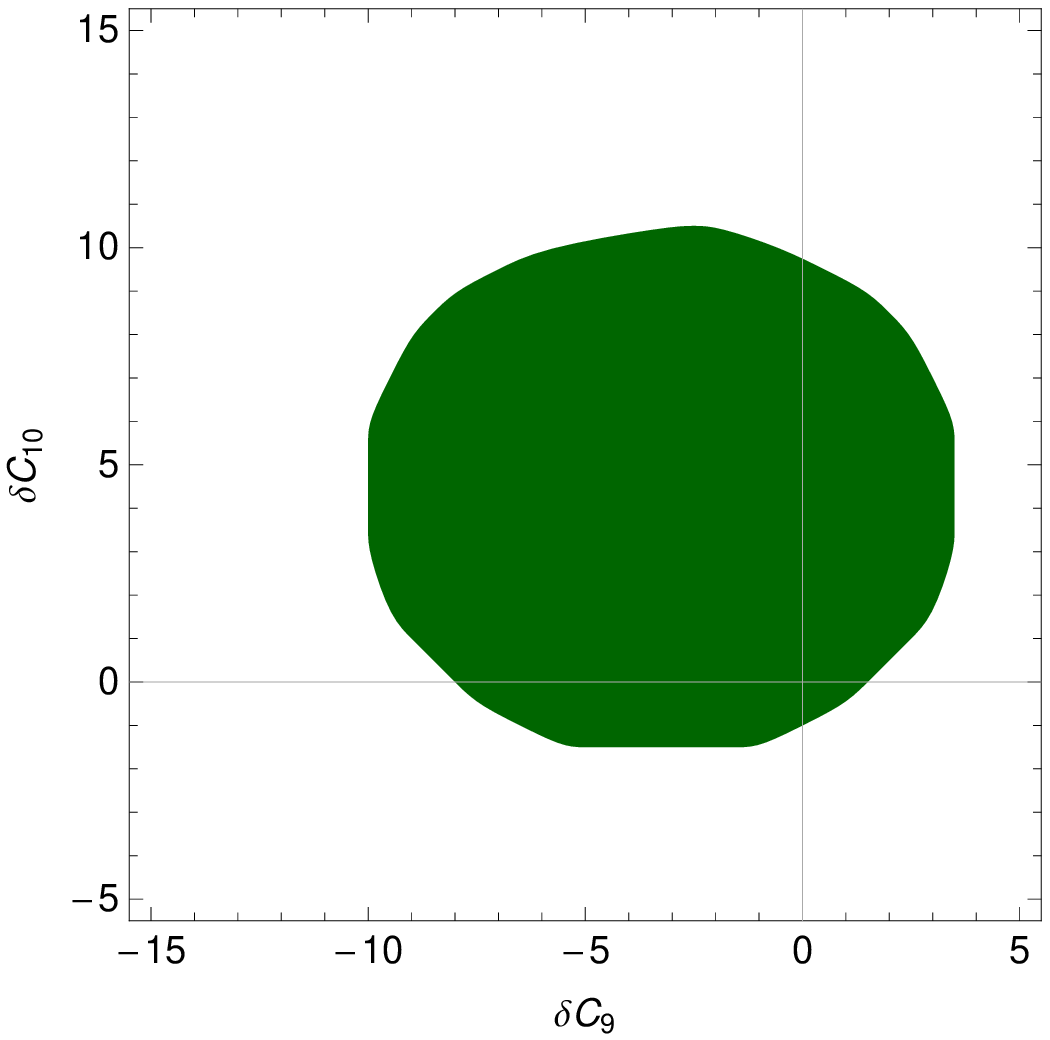}}                
\subfloat[]{\label{plot2e}
\includegraphics[width=0.5\textwidth]{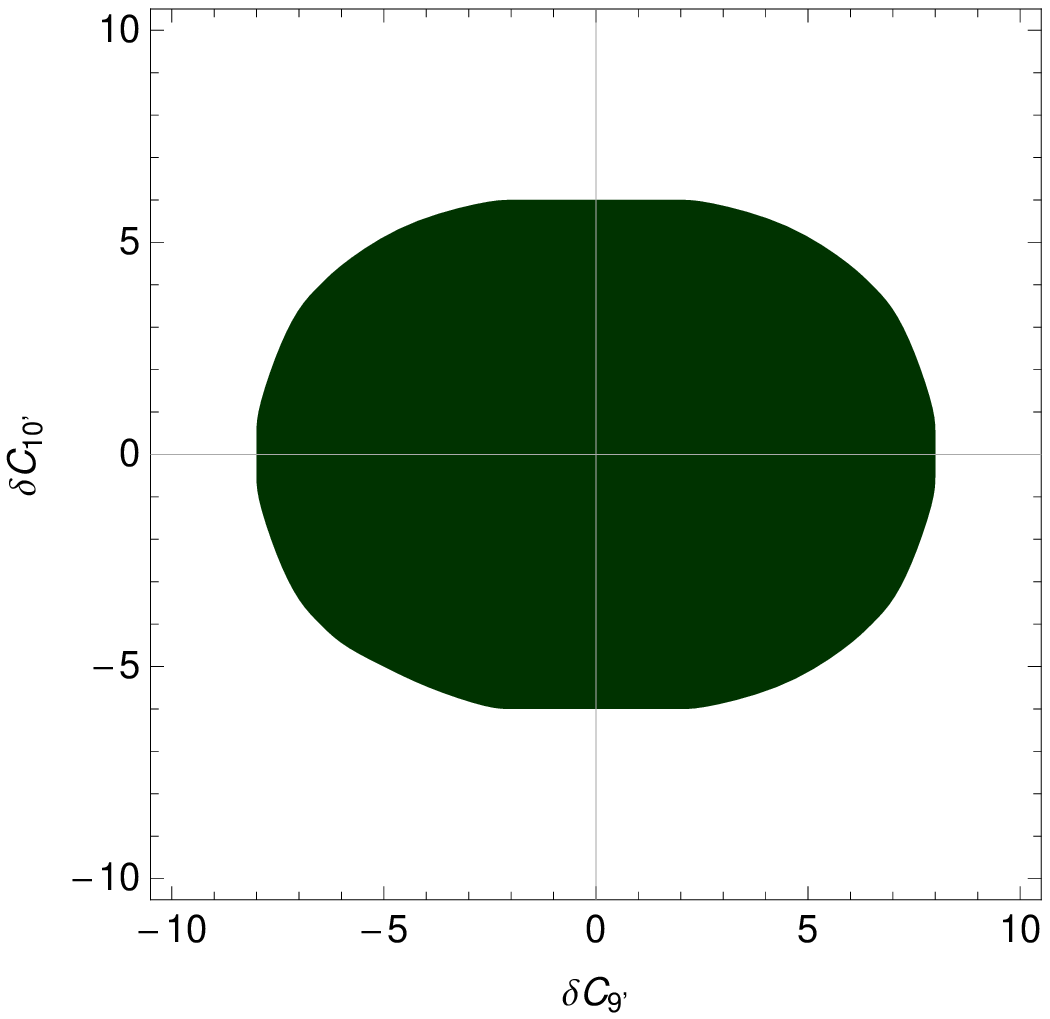}}
\caption{
Constraints from Class-III observable {\cal{B}}($B \to X_s \mu^+ \mu^-$) at $1\, \sigma$ in the $(\delta C_9,\delta C_{10})$ and $(\delta C_{9^{\prime}},\delta C_{{10}^{\prime}})$ planes in Scenario C. The regions shown are compatible with the constraints on $\delta C_7$ and $\delta C_{7^{\prime}}$ imposed by Class-I observables.
}
\end{figure}

Finally, we could imagine that the previous constraints did not overlap as nicely as in Fig.~\ref{plot2d3d4d}. 
We would then turn to Scenario C, allowing for chirally-flipped semileptonic operators.
For $(\delta C_7, \delta C_{7^\prime})$, we take all the 
model-independent allowed values from the three regions of Fig.~\ref{plot1}.  Among all the constraints considered 
previously from Class-III observables, only ${\cal B}(B\to X_s \mu^+\mu^-)$ still provides a constraint on the semileptonic (primed and unprimed) Wilson coefficients.
Indeed, when NP contributions in $C_{9^\prime}$ and $C_{10^\prime}$ are also considered, the 
empty region in the middle of Fig.~\ref{plot2dB} gets filled up but the 
minimum and maximum values of $\delta C_9 $ and $\delta C_{10}$ allowed do not change perceptibly, as can be seen in Fig.~\ref{plot2dC}. 
In Fig.~\ref{plot2e} we show the allowed region in the 
$(\delta C_{9^\prime}, \delta C_{10^\prime})$ plane in the same scenario. It is not very surprising to obtain such oval shapes in the various planes of interest, since it corresponds to the projections of the quadratic (elliptic) constraint given by eq.~(\ref{btoxsmumu}). In conclusion in Scenario C, the allowed region for $(\delta C_7,\delta C_{7^\prime})$ is given by the three black regions in Fig.~\ref{plot1}, and the corresponding ones for the planes $(\delta C_9,\delta C_{10})$ and $(\delta C_{9^\prime},\delta C_{10^\prime})$ are given by Figs.~\ref{plot2dC} and \ref{plot2e} respectively.

We have not given the predictions for $A_{\rm T}^{(2)}$ under this scenario, as the extra freedom provided by $C_{9'}$ and $C_{10'}$ is likely to fill the whole parameter space available.

\subsection{$2 \, \sigma$ constraints}

\begin{figure}[ht]
\centering
\subfloat[]{\label{plot12sigma}
\includegraphics[width=0.5\textwidth]{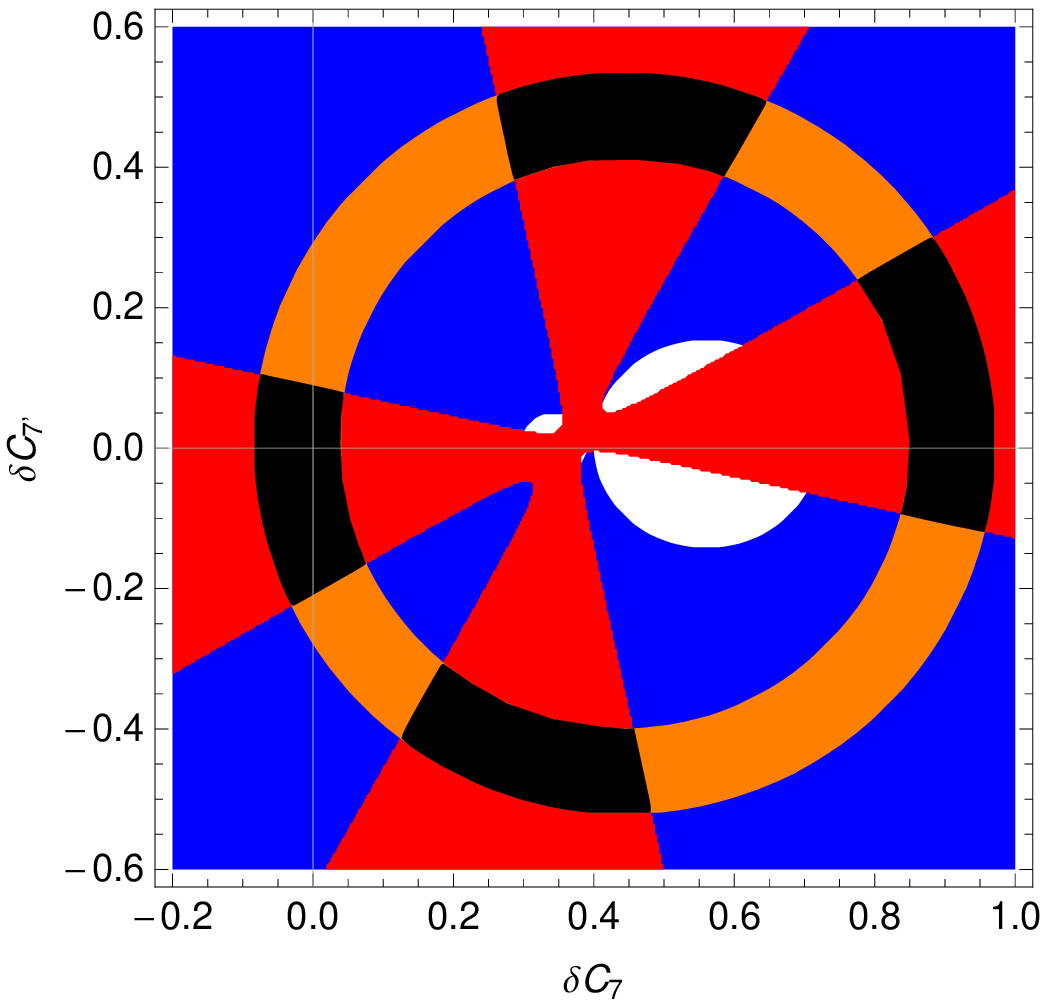}}                
\subfloat[]{\label{plot2a2sigma}
\includegraphics[width=0.5\textwidth]{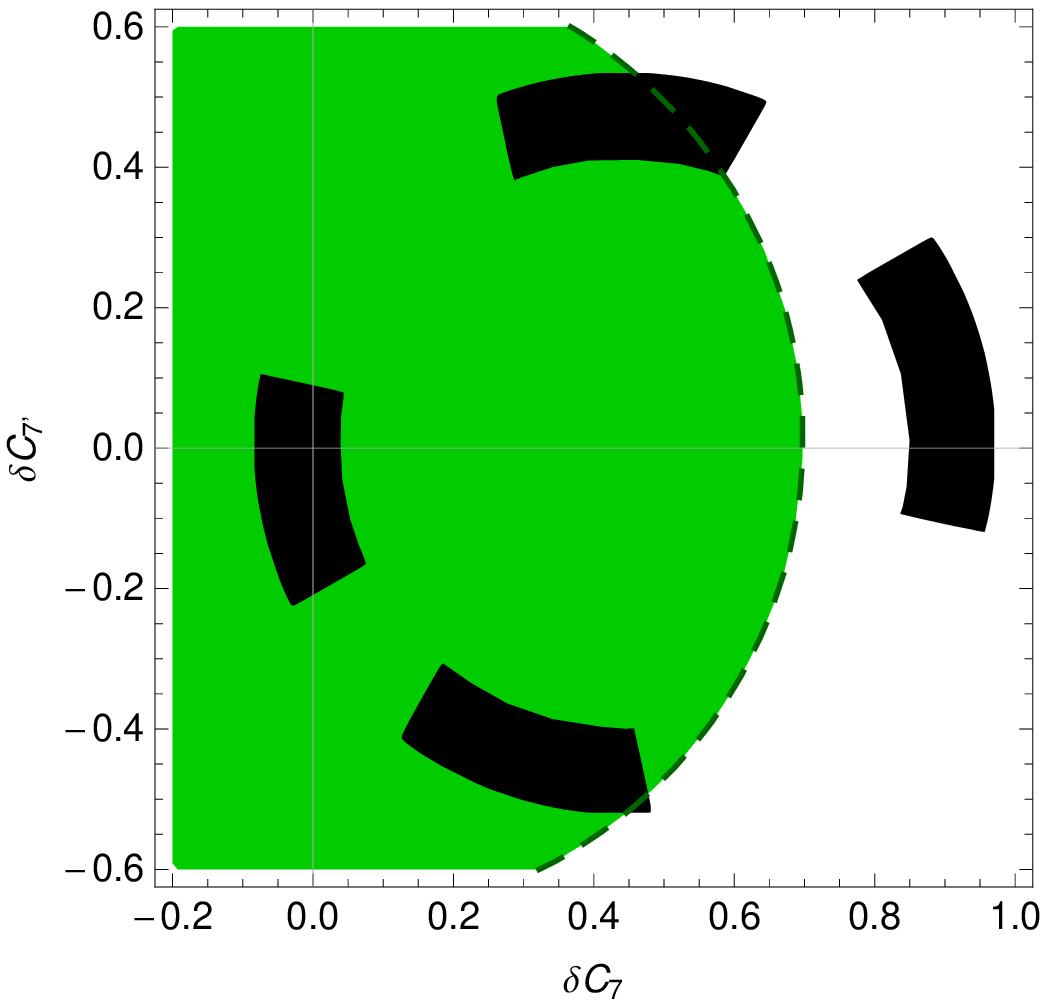}}
\caption{On the left, $2\, \sigma$ constraint from Class-I observables: $A_I$ (background solid blue region with two white disks -partially hidden-), {\cal{B}}($B \to X_s \gamma$) (orange ring) and $S_{K^* \gamma}$ (red cross). The three disconnected regions allowed by the intersection of these three observables are depicted in black. On the right, $2\, \sigma$ constraint from Class-III observable $B\to X_s\mu^+\mu^-$. The SM value is given by the crossing of light gray lines at $(\delta C_7,\delta C_{7^{\prime}})=(0,0)$ point.}
\end{figure}

When the uncertainty in both theoretical and experimental results is increased to $2 \, \sigma$, the regions allowed in the $(\delta C_7,\delta C_{7^\prime})$ plane are enlarged, as ${\cal{B}}(B \to X_s \gamma)$, $S_{K^* \gamma}$ and $A_I$ yield larger overlapping regions. More importantly, the whole region corresponding to the ``flipped-sign" solution is no longer excluded by Class I observables (see Fig.~\ref{plot12sigma}). We have followed the procedure explained before and used the resulting four disconnected regions to explore the behaviour of Class-II and Class-III observables under scenarios A, B and C.

In Scenario A, ${\cal{B}}(B \to X_s \mu^+ \mu^-)$ excludes the whole ``flipped-sign" solution region, a sizeable portion of the upper region and small part of the lower one, as shown in Fig. \ref{plot2a2sigma}, whereas neither ${\tilde A}_{\mathrm{FB}}$ nor ${\tilde F}_{\mathrm{L}}$ provide further constraints, since they fill the whole of the $(\delta C_7, \delta C_{7^{\prime}})$ area explored.
\begin{figure}[ht]
\centering
\subfloat[]{\label{plot2d2sigma}
\includegraphics[width=0.5\textwidth]{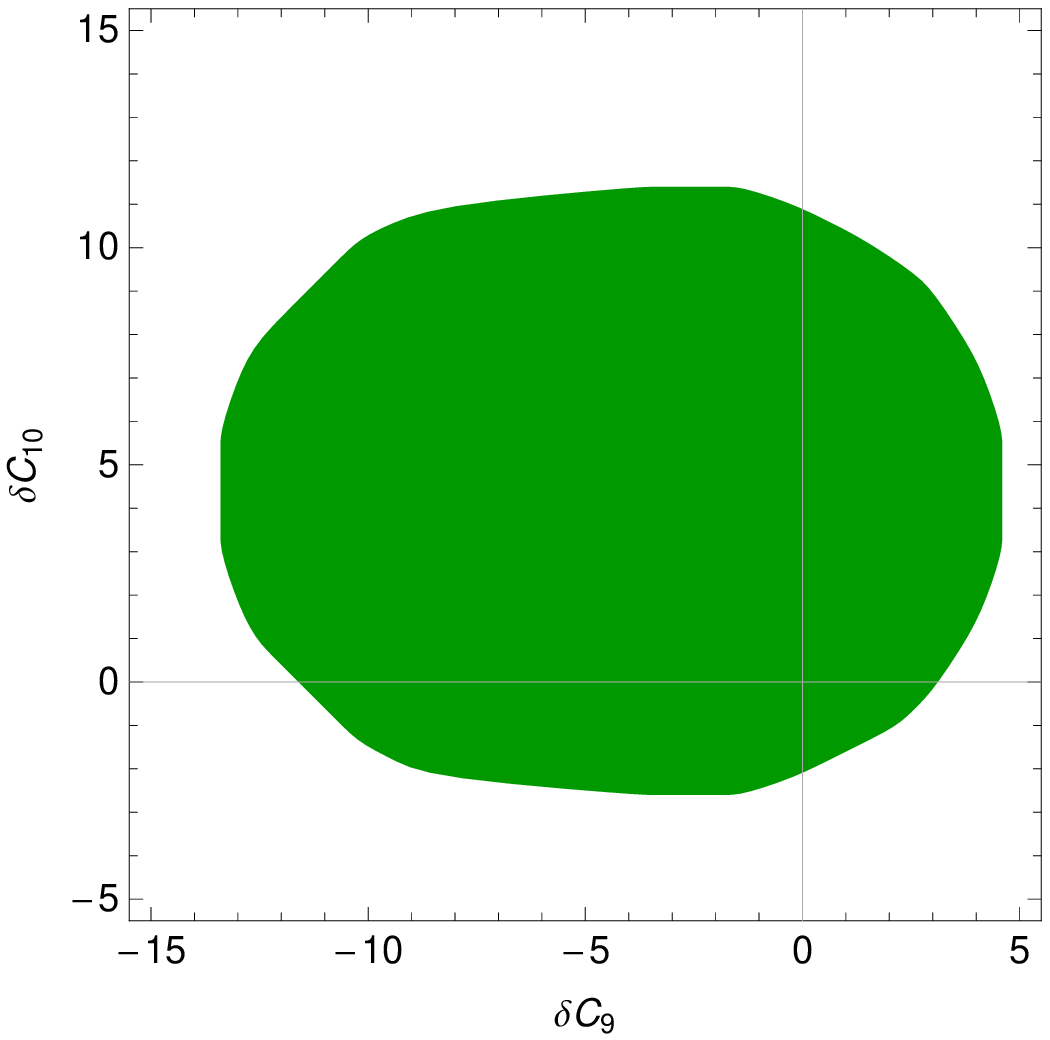}}                
\subfloat[]{\label{plot4d2sigma}
\includegraphics[width=0.5\textwidth]{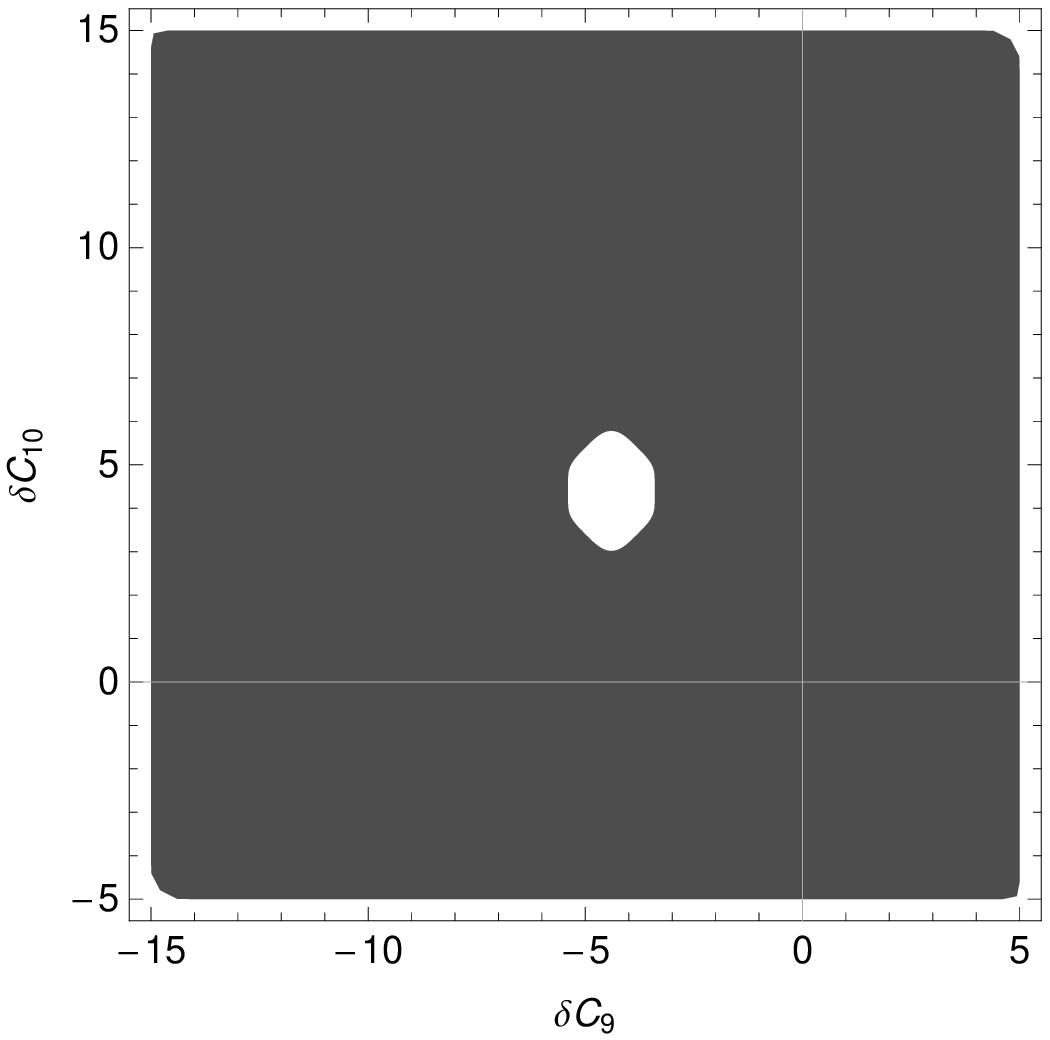}}
\caption{Constraints from Class-III observables {\cal{B}}($B \to X_s \mu^+ \mu^-$) (left) and $\tilde{F}_{L}$ (right) at $2\, \sigma$ in the $(\delta C_9,\delta C_{10})$ plane in Scenario B.}
\end{figure}

\begin{figure}[ht]
\centering
\includegraphics[width=0.7\textwidth]{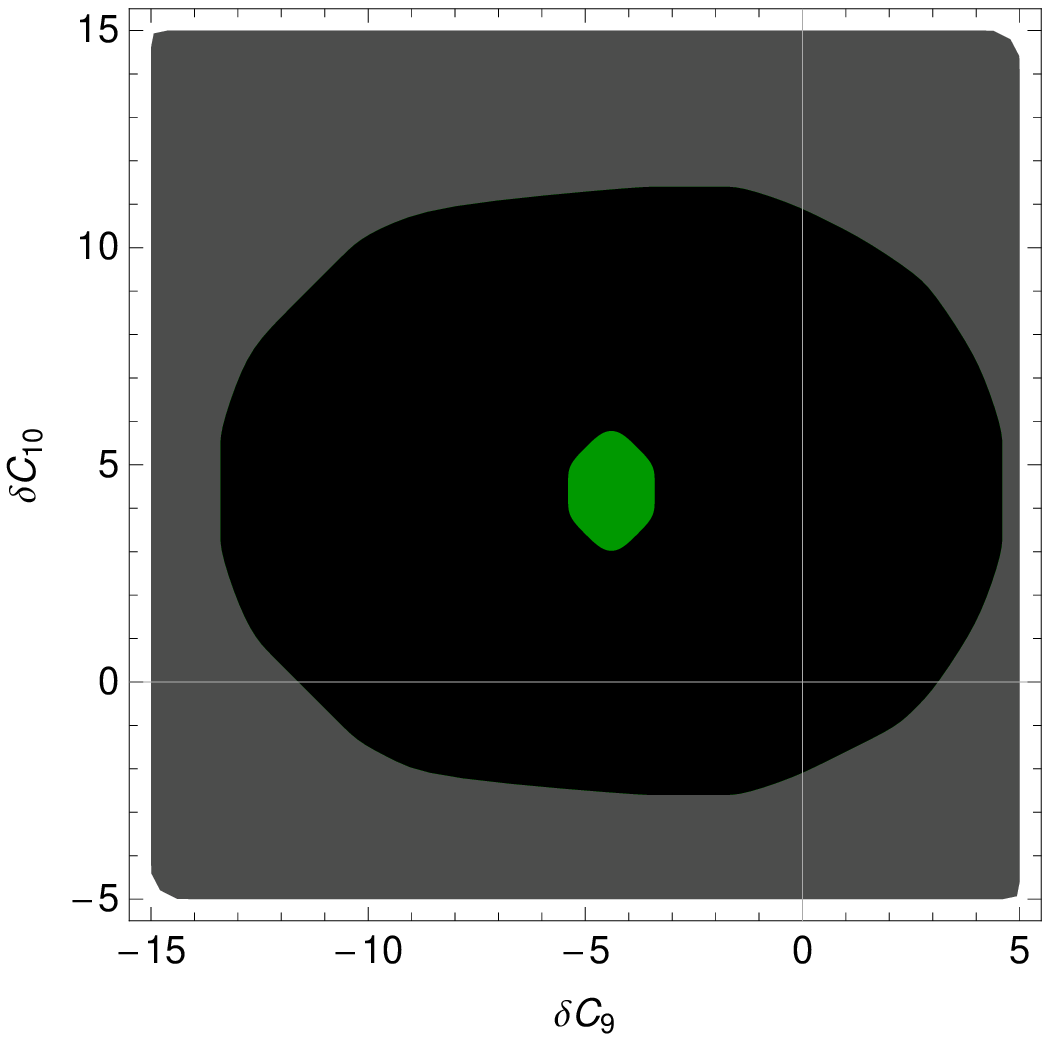}
\caption{
Overlap of the constraints from Class-III observables {\cal{B}}($B \to X_s \mu^+ \mu^-$) and ${\tilde F}_{\mathrm{L}}$ at $2\, \sigma$ in the $(\delta C_9,\delta C_{10})$ plane in Scenario B. The constraints imposed by their intersection are shown as a black region.
}
\label{plot2d4d2sigma}
\end{figure}

Next we move to Scenario B and include possible NP contributions to $(\delta C_9, \delta C_{10})$, as depicted in Figs. \ref{plot2d2sigma} and \ref{plot4d2sigma}. The region allowed by ${\cal{B}}({ B} \to X_s \mu^+ \mu^-)$ becomes enlarged by about a $40\%$ with respect to the $1 \, \sigma$ plot and the central region, previously forbidden, becomes filled altogether. In this scenario, ${\tilde A}_{\rm FB}$ does not provide extra constraints but $\tilde{F}_{\mathrm{L}}$ maintains an excluded central zone, although much reduced in area. Fig.~\ref{plot2d4d2sigma} shows (in black) the regions allowed by the overlapping of these two observables. Moreover, the``flipped sign solution" for the $(\delta C_7,\delta C_{7'})$ plane is now allowed under this scenario and the following one. 

\begin{figure}[ht]
\centering
\subfloat[]{\label{plot2e2sigma}
\includegraphics[width=0.5\textwidth]{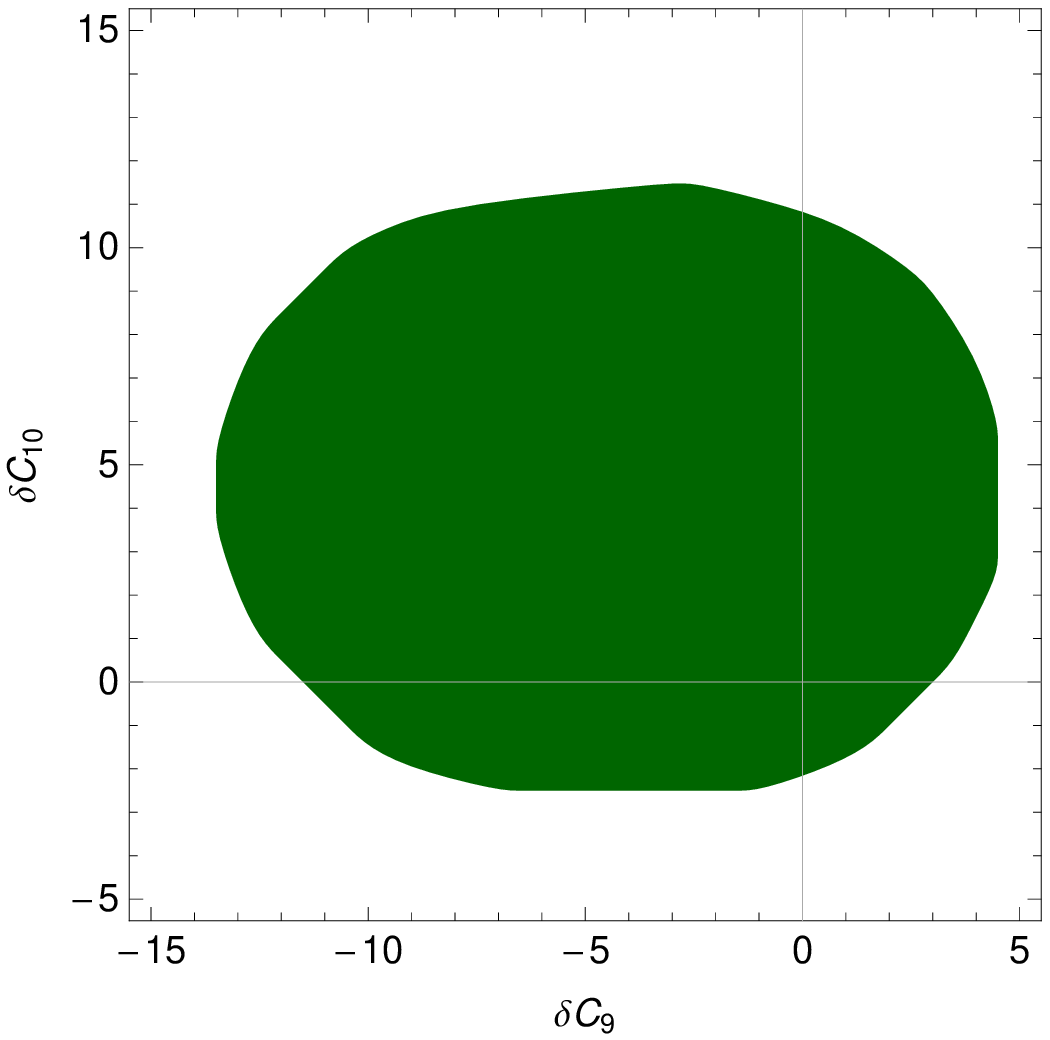}}                
\subfloat[]{\label{plot2f2sigma}
\includegraphics[width=0.5\textwidth]{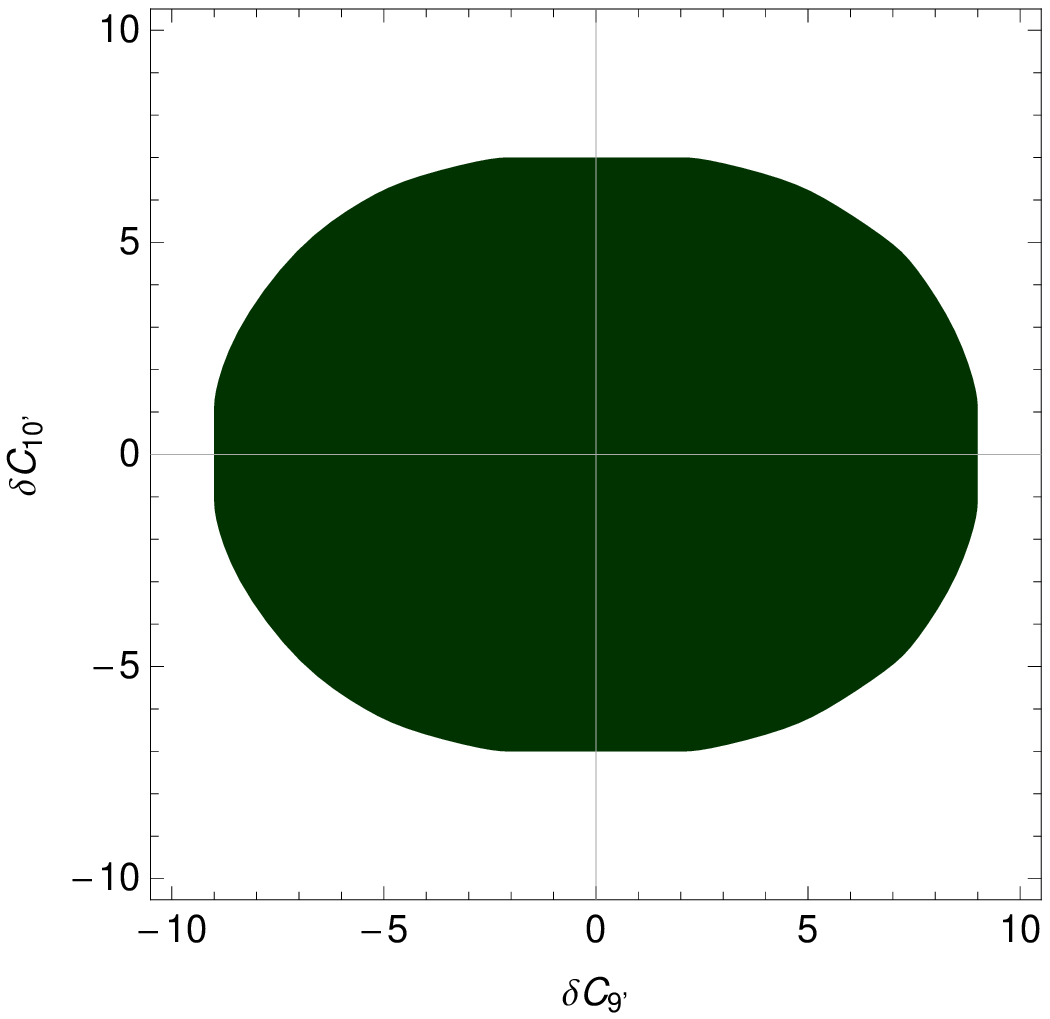}}
\caption{ Constraints from Class-III observable {\cal{B}}($B \to X_s \mu^+ \mu^-$) at $2\, \sigma$ in the $(\delta C_9,\delta C_{10})$ and $(\delta C_{9^\prime},\delta C_{10^\prime})$ planes in Scenario C. The regions shown are compatible with the constraints on $\delta C_7$ and $\delta C_{7^\prime}$ imposed by Class-I observables.
}
\end{figure}

We come finally to Scenario C. Besides $(\delta C_9, \delta C_{10})$, we must also allow for  NP in the Wilson coefficients $C_{9^\prime}$ and $C_{10^\prime}$, while $(\delta C_7, \delta C_{7^\prime})$ remain confined to the four black regions of Fig.~\ref{plot12sigma}. ${\cal{B}}({ B} \to X_s \mu^+ \mu^-)$ is again the only observable that imposes constraints in the Wilson coefficients related to ${\cal{O}}_i$ and ${\cal{O}}_{i^\prime}$ (with $i=9,10$) as shown in Figs.~\ref{plot2e2sigma} and \ref{plot2f2sigma}.
\subsection{Generalization to extended frameworks} \label{generalframe}

Let us
assume, for instance, that we want also to include  contributions from scalar operators (like those defined in \cite{Alok:2009tz}). Consequently the scenarios will also be enlarged: Scenario A (${\cal{O}}_{7},{\cal{O}}_{7^\prime}$), B (${\cal{O}}_{7},{\cal{O}}_{7^\prime}, {\cal{O}}_9, {\cal{O}}_{10}$), C (${\cal{O}}_{7},{\cal{O}}_{7^\prime}$, scalars), D (${\cal{O}}_{7},{\cal{O}}_{7^\prime},{\cal{O}}_9,{\cal{O}}_{10},{\cal{O}}_{9^\prime}$, ${\cal{O}}_{10^\prime}$), E (${\cal{O}}_{7},{\cal{O}}_{7^\prime},{\cal{O}}_9,{\cal{O}}_{10}$, scalars), F (all operators). 
 We would then proceed again along the same steps as before, up to certain changes:
\begin{enumerate}
\item We classify again the observables according to this new framework. This may move some
observable from Class-I to higher classes, because they have sensitivity to
scalars, like the $K^*\gamma$ observables $A_I$ or $S_{K*\gamma}$. Only ${\cal B}({\bar B} \to X_s \gamma)$ will remain.
\item We determine the new reference region for $C_7$ and $C_{7'}$ defined by the (now reduced) set of Class-I observables. The new
primary regions will be larger than in the previous framework because some observables are not included in the new Class-I.
\item At this stage, and working in Scenario A, it is interesting to define two types of Class-II observables, Class-IIa,
only sensitive to dipole, semileptonic and chirally flipped (our observables in  Class II of the previous framework will be here) and Class-IIb, only sensitive to dipole operator (and its chirally flipped counterpart) and scalars. These observables may shrink the new reference regions, leading to allowed regions of different shapes for Class-IIa and Class-IIb. If we add now Class-III observables with sensitivity to the whole list of operators in the framework, this will generate a further cut on the primary region. If the same set of observables as in the previous framework has been included it is clear that, even if re-classified, the allowed region under Scenario A will be exactly the same as in the previous framework, even if the primary regions are different.

\item The main differences arise when dealing with the rest of scenarios. We should repeat the same analysis under Scenario B till F. It is clear that Scenario B and Scenario C, for instance, may select different subregions inside the primary regions, and that Scenario D will enlarge the region for Scenario B, and the same will happen between scenarios E and C. 
Finally Scenario F will cover all previous ones, defining the largest allowed subregion inside the primary regions. 
This  region maybe larger that in the previous framework (since more freedom in the value of the  WC already studied  is provided by the introduction of scalar contributions).
 
\item This systematic procedure that subdivides the primary regions in different subregions may help to disentangle the importance of each set of operators: dipole, semileptonic, chirally flipped, scalar, when confronting theory with data.
In particular, certain observables like $A_{\rm T}^{(2)}$ and its generalization, may discriminate between the different subregions.

\end{enumerate}

This procedure can be generalized to other frameworks following the same steps. 
Defining intermediate steps between the dipole-only case and the full-fledged 
scenario for the introduction of New Physics helps in understanding the importance of the NP contribution for each observable. In the present paper, we have restricted ourselves to the framework where NP arises in dipole, semileptonic operators and their chirally-flipped partners.

\section{Discussion and outlook}

We have exploited the $(\delta C_7, \delta C_{7^\prime})$ plane as a starting point to investigate the pattern of
NP in the Wilson coefficients for radiative $\Delta B=1$ transitions. We have defined several classes of observables
to help us in this task, selecting only observables with a good theoretical control over hadronic uncertainties (or a significant discriminating power for our NP scenarios) and providing numerical expressions for these quantities as functions of $\delta C_{7,7',9,9',10,10'}$. We defined reference regions for $(\delta C_7, \delta C_{7^\prime})$ from Class-I observables, then studied several scenarios of NP involving chirality-flipped operators with the help of Class-II and Class-III observables. 

As far as the theory and experimental errors of the measured observables remain inside
the $1\,\sigma$ range we can draw the following conclusions.
Scenario A, where only $(C_7, C_{7'})$ receive large NP contributions, is a  predictive
scenario. Class-I observables provide three different regions (one corresponding to the SM case, two other ones with almost vanishing $C_7$ values and large $C_{7'}$ value).  Once Class-III observables are included only a very small subregion (inside one of the two non-SM like regions) is allowed if we keep all the constraints at $1\,\sigma$. Consequently, only those theories that can provide values for $(C_7,C_{7'})\simeq (C_7^{SM}+0.25,-0.4)$ are compatible (within $1\,\sigma$) with current (Class-I and Class-III) data, due to the
interplay between the inclusive decay $B\to X_s\mu^+\mu^-$ and the forward-backward asymmetry $\tilde{A}_{FB}$. 
Notice that the SM is not one of such theories. 
This motivated us to enlarge the set of operators
where NP contributions can be sizeable, leading to constraints on the semileptonic operators. 
Scenario B constitutes the first extension, allowing for NP in $C_{7,7',9,10}$. In this case, the previous constraints from Class-III observables are transferred from the $(\delta C_7, \delta C_{7'})$ plane to the $(\delta C_9, \delta C_{10})$ one. There are two distinctive regions allowed, corresponding to the SM solution, but also to a flipped-value configuration, where $C_9$ and $C_{10}$ have some values opposite to the SM. It is interesting to notice that $B\to X_s\mu^+\mu^-$ and $\tilde{F}_L$
exclude almost the same central area in the $(\delta C_9,\delta C_{10})$ plane.
Scenario C (with NP in $C_{7,7',9,9',10,10'}$) would be an interesting extension if the previous experimental constraints shift in the future, or if the measurement of the (Class-II) asymmetry $A_{\mathrm{T}}^{(2)}$ shows a discrepancy with the pattern of Wilson coefficients exhibited in Scenario B, once more data and constraints have been added. The (Class-I) constraints on $(C_7,C_{7'})$ remain unchanged with respect to Scenario B,
whereas Class-III observables provide only limited constraints on the largest set of Wilson coefficients considered.
Currently, only $B\to X_s\mu^+\mu^-$ provides constraints on $C_{9,9',10,10'}$.

We have also indicated how the (Class-II) asymmetry $A_{\mathrm{T}}^{(2)}$ gives a very precise prediction for Scenario A, that can be used either to confirm it or to rule it out. It  may also help, depending on its sign, to discriminate 
among the allowed regions in Scenario B. $A_{\rm T}^{(2)}$  exhibits a strong sensitivity to the allowed regions for $(C_9,C_{10})$; further cuts in these regions using high-$q^2$ measurements, will improve the predictive power of $A_{\rm T}^{(2)}$ in this scenario. Under Scenario C, there is too much freedom with all WC switched on to be able to cut on precise regions as it happens for most of the other observables.

We also have shown that Class-I observables alone allow us to dismiss the flipped-sign solution at $1.59 \,\sigma$, even in a NP scenario much more general than in ref.~\cite{Gambino:2004mv}, allowing for NP in dipole and semileptonic operators, but also in their chirally-flipped counterparts. We achieved this by trading the Class-III observable $B\to X_s\mu^+\mu^-$ (considered in ref.~\cite{Gambino:2004mv}, and sensitive to many NP contributions apart from those in the dipole ones) for the Class-I isospin asymmetry in $B\to K^*\gamma$ (even though the theoretical control on hadronic uncertainties is less satisfying for this observable).

A summary of the maximal and minimal values of the WC analyzed in the different scenarios is provided in table~\ref{summarytable}\footnote{For the internal 4-d and 6-d correlations involving 4 WCs (Scenario B) and 6 WCs (Scenario C) we can provide a datafile with the correlated points upon request.}.


\begin{table}
\renewcommand{\arraystretch}{1.4}
\addtolength{\arraycolsep}{1pt}
\footnotesize{
\begin{tabular}{| c || c | c || c | c || c | c |}
\hline
& $\delta C_7(\mu_b)$ &   $\delta C_{7^\prime}(\mu_b)$ &  $\delta C_9(\mu_b)$ &  
$\delta C_{10}(\mu_b)$& $\delta C_{9^\prime}(\mu_b)$ &  $\delta C_{{10}^\prime}(\mu_b)$ \\
\hline
\multicolumn{7}{|c|}{Overlap of the 1 $\sigma$ constraints} \\
\hline
\textnormal{Sc. A} & $[0.244,0.274]$  & $[-0.417,-0.39]$&$0$&$0$&$0$&$0$\\ 
\hline
\multirow{3}{*}{\textnormal{Sc. B}} & $[0.346,0.385]$ & [0.435,0.501] & \multirow{2}{*}{$[-9.75,-0.5]$} & \multirow{2}{*}{$[4.75,10.5]$}  & \multirow{3}{*}{$0$} & \multirow{3}{*}{$0$} \\
 & $ [-0.056,0.016]$ & $[-0.114,0.027]$ & \multirow{2}{*}{$[-3.75,3.5]$} & \multirow{2}{*}{$[-1.75,3.5]$}  & &\\
 & $[0.235,0.385]$ & $[-0.489,-0.39]$ & &  & &  \\
 \hline
\multirow{3}{*}{\textnormal{Sc. C}} & $[0.346,0.385]$ & [0.435,0.501] & \multirow{3}{*}{$[-10,3.5]$} & \multirow{3}{*}{$[-1.5,10.5]$}  & \multirow{3}{*}{$[-8, 8]$} & \multirow{3}{*}{$[-6,6]$} \\
 & $ [-0.056,0.016]$ & $[-0.114,0.027]$ & &  & &\\
  & $[0.235,0.385]$ & $[-0.489,-0.39]$ & &  & &  \\
  \hline
\multicolumn{7}{|c|}{Overlap of the 2 $\sigma$ constraints} \\
\hline
\multirow{3}{*}{\textnormal{Sc. A}} & $[0.262,0.586]$ & [0.381,0.531] & \multirow{3}{*}{$0$} & \multirow{3}{*}{$0$}  & \multirow{3}{*}{$0$} & \multirow{3}{*}{$0$} \\
 & $ [-0.083,0.076]$ & $[-0.225,0.105]$ &  &   & &\\
 & $[0.124,0.475]$ & $[-0.519,-0.306]$ & &  & &  \\
 \hline
\multirow{4}{*}{\textnormal{Sc. B}} & $[0.262,0.646]$ & [0.381,0.534] & \multirow{4}{*}{$[-13.4,4.5]$} & \multirow{4}{*}{$[-2.5,11.4]$}  & \multirow{4}{*}{$0$} & \multirow{4}{*}{$0$} \\
 & $ [-0.083,0.076]$ & $[-0.225,0.105]$ &  &   & &\\
 & $[0.775,0.97]$ & $[-0.12,0.3]$ & &  & &  \\
  & $[0.124,0.481]$ & $[-0.519,-0.306]$ & &  & &  \\
 \hline
\multirow{4}{*}{\textnormal{Sc. C}} & $[0.262,0.646]$ & [0.381,0.534] & \multirow{4}{*}{$[-13.5,4.6]$} & \multirow{4}{*}{$[-2.6,11.5]$}  & \multirow{4}{*}{$[-9,9]$} & \multirow{4}{*}{$[-7,7]$} \\
 & $ [-0.083,0.076]$ & $[-0.225,0.105]$ &  &   & &\\
 & $[0.775,0.97]$ & $[-0.12,0.3]$ & &  & &  \\
  & $[0.124,0.481]$ & $[-0.519,-0.306]$ & &  & &  \\
 \hline
\end{tabular}
\renewcommand{\arraystretch}{1}
\addtolength{\arraycolsep}{-1pt}
\caption[]{Summary table of the maximum and minimum Wilson coefficients values allowed by the three different scenarios within our framework. The table is organized in three independent blocks corresponding to the pairs $(\delta C_7,\delta C_{7\prime})$, $(\delta C_9,\delta C_{9^\prime})$ and $(\delta C_{10},\delta C_{10^\prime})$ respectively.  Notice that the correlations between different WCs are more complex than those summarised in this table. In order to recover the exact 2d-correlations, one should look at Fig.~3a (Scenario A), Figs.~1, 6 (Scenario B) and Figs.~1, 6, 8a, 8b (Scenario C) at $1\,\sigma$, and at Fig.~9b (Scenario A), Figs.~9a, 11 (Scenario B) and Figs.~9a, 11, 12a, 12b (Scenario C) at $2\,\sigma$.} \label{summarytable}}
\end{table}

In ref.~\cite{Bobeth:2008ij}, an analysis of various NP contributions was considered, allowing either for New Physics in $(C_7,C_{7'})$ (both of them being real), or $C_{10}$ (considered as potentially complex). In particular, our findings concerning Scenario A (NP only in $C_7$ and $C_{7'}$) are in agreement with Fig.~2 in ref.~\cite{Bobeth:2008ij} concerning $S_{K^*\gamma}$, as well as the fact that the flipped-sign solution is excluded (even though the conclusion is based on different observables). However, the other scenarios discussed in \cite{Bobeth:2008ij}  considered NP entering in one Wilson coefficient at a time, and thus provide only a particular section of the parameter space of Wilson coefficients. Another related study was performed in ref.~\cite{Bobeth:2010wg}, where $B\to K^*\ell^+\ell^-$ at large and low recoil (which was not considered here) was combined with $B\to X_s \ell^+\ell^-$ to study the $(C_9,C_{10})$ plane, considering $C_7=\pm C_7^{SM}$. This led to two regions in $(C_9, C_{10})$ similar to the ones obtained in our case, however smaller partly due to the additional constraints put on $C_7$ (and $C_{7'}$) in this reference.  

In ref.~\cite{Hurth:2008jc}, a global analysis of $\Delta B=1$ observables was performed in a minimal flavour violating framework that included the possibility of sizable scalar contributions (but no chirally flipped operators). The combination of the various observables was performed using a Bayesian statistical approach. Even though the inputs and the underlying assumptions concerning the structure of NP are different (scalar versus chirality-flipped operators), we observe some common features. Two different regions for $(C_7,C_9,C_{10})$ are allowed, corresponding approximately to a change of sign for the Wilson coefficients  (Fig.~1 in ref.~\cite{Hurth:2008jc}). Once NP is allowed for  $(C_9,C_{10})$ (Scenario B), there is a ring-like constraint from ${\cal B}(B\to X_s\ell^+\ell^-)$ in the $(C_9,C_{10})$ plane, with only two regions surviving once the  forward-backward asymmetry $\tilde{A}_{FB}$ is included (Fig.~4 in ref.~\cite{Hurth:2008jc}). This is in basic agreement with our own plots, even though we should highlight that the non-SM region in the $(C_9,C_{10})$ plane corresponds to different allowed values for the electromagnetic operators: in ref.~\cite{Hurth:2008jc}, this region corresponds to the SM and the ``flipped-sign'' solution ($C_7\simeq -C_7^{SM}$, $C_{7'}\simeq 0$) disfavoured  by $B\to X_s \mu^+\mu^-$ in their framework, whereas 
our region corresponds to the SM solution and to the flipped-value regions where $C_7\simeq 0$ and $|C_{7'}|\simeq |C_{7}^{SM}|$.

Our approach could be extended to other, more involved, scenarios of New Physics, including contributions to the chromomagnetic, scalar and/or tensors operators as explained in detail in Sec \ref{generalframe}, allowing us to assess the impact of each observable in a controlled way. Such a task is left for future work.

\section*{Acknowledgements}

The authors would like to thank A. Dighe, T. Feldmann, U. Haisch, J. Kamenik, E. Lunghi and M. Misiak  for fruitful exchanges. SDG would like to thank UAB where part of this work was completed under project 2009PIV00066. DG would like to thank specially A. Dighe for encouragement. JM thanks the Tata Institute for Fundamental Research for their hospitality. JM acknowledges financial support from FPA2008-01430, SGR2009-00894. MR also thanks A. Khodjamirian, R. Miquel, Ll. Galbany and P. Mart\'i for enlightening discussions. MR work has been supported by Universitat Aut\`onoma de Barcelona.

\appendix

\section{Inputs}\label{app:inputs}

We have followed the discussion in refs.~\cite{Huber:2005ig, Gambino:2003zm, Gorbahn:2004my, Bobeth:2003at, Misiak:2006ab} concerning the matching and the running of the Wilson coefficients from the high scale $\mu_0=2M_W$ down to the low scale $\mu_b=4.8$ GeV. We were able to reproduce at the 1\% level the tables 3, 4 and 5 in ref.~\cite{Huber:2005ig} (apart from $C_7^{(11)},C_9^{(22)}, C_{10}^{(22)}$)  and the table 5 in ref.~\cite{Misiak:2006ab} for the Wilson coefficients, providing a check that we control the scale dependence of the Wilson coefficients accurately. Contrary to other analyses in the literature, we have expressed the deviations from the SM Wilson coefficients at the low scale $\mu_b$ around 4.8 GeV. However, the evolution from $\mu_0$ to $\mu_b$ can be determined as the linear combinations:
\begin{eqnarray}
\delta C_7(\mu_b)&=&0.575 \times \delta C_7(\mu_0)\,,\nonumber \\
\delta C_9(\mu_b)&=&1.021 \times \delta C_9(\mu_0) + 0.008 \times \delta C_{10}(\mu_0)\,,\nonumber  \\
\delta C_{10}(\mu_b)&=&0.008 \times \delta C_9(\mu_0) +1.038 \times \delta C_{10}(\mu_0)\,.
\end{eqnarray}
Several schemes have been used to define the quark masses:
\begin{itemize}
\item For $m_t$ and $m_c$, we used the $\overline{\rm MS}$ scheme scheme at the required scale (respectively $\mu_0$ and $m_c$). We convert $m_t^{pole}$ into $m_t^{\overline{\rm MS}}$ using the conversion formulae in refs.~\cite{Chetyrkin:1996vx,Chetyrkin:1997dh}.
\item For $m_b$, two different masses are needed: the mass in the 1S scheme (or an equivalent scheme with infrared subtraction) is required whenever the $b$-quark is close to the mass shell, whereas the pole mass is used for normalisation purposes as well as for loop computations where the $b$-quark is off-shell. Follwing ref.~\cite{Huber:2005ig,Misiak:2006ab}, we take the value of $m_b^{1S}$ obtained from fits to hadronic and leptonic moments of the differential branching ratio for the inclusive decay $B\to X_c\ell\nu$~\cite{Bauer:2004ve}, and we determine the pole mass using the conversion formulae in ref.~\cite{Hoang:2000fm}.
\item For $m_s$, we use the strange quark mass in the $\overline{\rm MS}$ scheme, taken at the scale $\mu_b$. We are aware that there is an ambiguity in the scheme and scale chosen for this mass (this ambiguity would be resolved by going to higher orders in perturbation theory, which are not included in the present analysis). We used $m_s/m_b$ both in the $\overline{\rm MS}$ scheme to evaluate the SM value of $C_{7'}$ (however we kept the $m_b^{\rm pole}$ normalisation to determine $\hat{m}_s=m_s/m_b$ needed for $B\to X_s \ell^+\ell^-$).
\end{itemize}

The running of the quark masses in the $\overline{\rm MS}$ is performed following ref.~\cite{Huber:2005ig}. The strong and electromagnetic coupling constants are determined by their value at $M_Z$, and their running is given by the equations in ref.~\cite{Huber:2005ig}.

\section{Extension to chirally-flipped operators}\label{app:numerics}

\subsection{$B\to X_s\gamma$}

The branching ratio for $B\to X_s \gamma$ for a photon energy larger than $E_0=$1.6 GeV can be written as~\cite{Misiak:2006ab}:
\begin{equation}
{\mathcal B}(B\to X_s\gamma)_{E_\gamma>E_0,SM}
 ={\mathcal B}(B\to X_c e \bar\nu)\left|\frac{V_{ts}^*V_{tb}}{V_{cb}}\right|^2
 \frac{6\alpha_{\rm em}}{C \pi} [P(E_0)+N(E_0)]\,,\\ \label{eq:pe0}
 \end{equation}
 where
\begin{eqnarray}
  C&=&\left|\frac{V_{ub}}{V_{cb}}\right|^2 \frac{\Gamma(\bar{B}\to X_c e\bar\nu)}{\Gamma(\bar{B}\to X_u e\bar\nu)}, \label{eq:C}\\
  P(E_0)&=& \sum_{i,j=1\ldots 8} C_i^{\rm eff}(\mu) C_j^{{\rm eff}*}(\mu) K_{ij}(E_0,\mu) \,.
\end{eqnarray}
Concerning $B\to X_s\gamma$, we were able to reproduce, not only the central value and uncertainty for the branching ratio, but also the results from the three different interpolation procedures and the scale dependence on $\mu_0$ and $\mu_b$ described in ref.~\cite{Misiak:2006ab} as well as the dependence on $C_{7,8}$ at the scale $\mu_0$ in eq.~(29) of ref.~\cite{Freitas:2008vh}. 
The contribution from the chirally-flipped operator ${\cal O}_{7'}$ should have the same structure as the SM operator ${\cal O}_7$ and there are no interferences between the two contributions, leading to an additional contribution to eq.~(\ref{eq:pe0}) of the form:
\begin{equation}
P(E_0) \to P(E_0)+ (C_{7'})^2 [1 + \tilde\alpha_s(\mu) K_{77}^{(1)} 
  + \tilde\alpha_s(\mu)^2 K_{77}^{(2)}]\,,
\end{equation}
where $K_{77}^{(i)}$ are the coefficients of the perturbative expansion of the kernel $K_{77}(E_0,\mu)$.

\subsection{$B\to K^* \gamma$ isospin asymmetry}
Concerning the isospin asymmetry, we reproduced the central value of the isospin asymmetry quoted in ref.~\cite{Kagan:2001zk}, following the formalism discussed in ref.~\cite{Feldmann:2002iw}:
\begin{eqnarray}
A_I[B\to K^*\gamma]_{SM}&=&{\rm Re}[b_d^\perp(0)-b_u^\perp(0)]_{SM}\,,\\
b^\perp_{q,SM}(0)&=&\frac{12\pi^2 f_B e_q}{m_b {\mathcal{C}_7} \xi_\perp(0)}\left[
\frac{f_{K^*}^\perp}{m_B}K_1^\perp(0)+\frac{f_{K^*} m_{K^*}}{6\lambda_B m_B} K_2^\perp(0)\right]\,,
\end{eqnarray}
where ${\mathcal{C}_7}=C_7^{\rm eff}+O(\alpha_s)$ includes NLO corrections to the amplitude for $B\to K^*\gamma$, computed in ref.~\cite{Beneke:2001at}. In $K_{1,2}^\perp(0)$, we have included the Cabibbo-suppressed power corrections discussed in App. A.3 in ref.~\cite{Beneke:2004dp} and neglected in  ref.~\cite{Feldmann:2002iw}, performing the replacements
\begin{eqnarray}
K_{1,2}^{\perp(c)}&\to& K_{1,2}^{\perp(c)}
  +\frac{\lambda_u}{\lambda_t} K_{1,2}^{\perp(c)}[F_V \to F_V^{(u)}]\,,\\
F_V^{(u)}(s=\bar{u}m_B^2)&=&\frac{3}{4}\left(C_2-\frac{C_1}{6}\right)[h(s,m_c)-h(s,0)]\,,
\end{eqnarray}
following the notation in ref.~\cite{Feldmann:2002iw}.

\begin{figure}[t]
\begin{center}
\scalebox{0.4}{\includegraphics{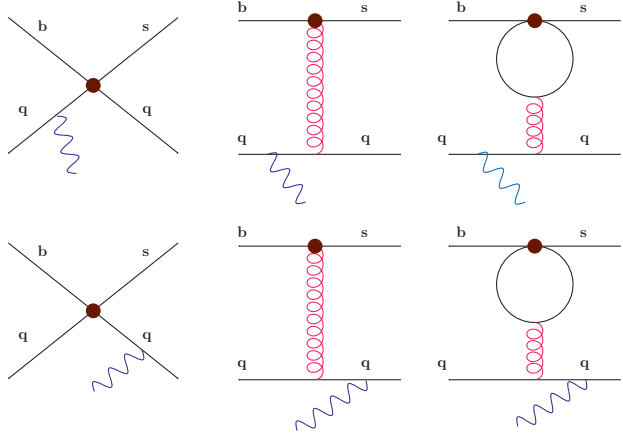}}
\end{center}
\caption{Annihilation topologies involving operators ${\cal O}_{1-6}$(left).
Hard spectator interaction involving operator ${\cal O}_8$(center) and 
${\cal O}_{1-6}$(right).\label{Fig1}}
\end{figure}

Unfortunately, the hard-spectator scattering involving the chromomagnetic operator ${\cal O}_8$ exhibits an endpoint divergence indicating a breakdown of QCD factorisation. We follow refs.~\cite{Kagan:2001zk,Feldmann:2002iw} to regularise the divergent integral
\begin{equation}
\int_0^1 du \to (1+\rho e^{i\phi})\int_0^{1-\Lambda_h/m_B} du\,,
\label{regHSS}
\end{equation} 
where $\rho$ is assumed to be smaller than 1 for our numerical estimations, and the phase $\phi$ is arbitrary.

Once we add chirality-flipped operators, ${\cal O}_{7'}$ will contribute to the branching ratio of $B\to K^*\gamma$. It is not difficult to check that its contribution is the same as the one from ${\cal O}_7$, and that there are no interferences between the two contributions. We will neglect the contributions from SM operators to the amplitude for a photon of right-handed helicity. On the other hand, the flipped operators considered in the present paper do not contribute to the spectator interactions
responsible for the isospin asymmetry (which are induced by the four-quark operators and the chromomagnetic operators). Therefore, the only change induced by chirality-flipped operators corresponds to modifying the normalisation, i.e., the denominator in the expression of the isospin asymmetry (at first order in isospin breaking)
\begin{equation}
A_I[B\to K^*\gamma]=\frac{{\rm Re}[b_d^\perp(0)-b_u^\perp(0)]}
 {1+|C_{7'}/\mathcal{C}_7|^2}\,.
\end{equation}

\subsection{$S_{K^*\gamma}$}\label{sec:skstargamma}

We define the decay amplitudes of $B_d$ mesons into $K^*$ and $\gamma_{L(R)}$ as in \cite{Ball:2006cva}:
\begin{equation}
\bar{\cal{A}}_{L(R)}=\bar{\cal{A}}(\bar{B}_d^0 \to \bar{K}^{*0}\gamma_{L(R)}), \qquad {\cal{A}}_{L(R)}={\cal{A}}(B_d^0 \to K^{*0}\gamma_{L(R)}).
\label{DecayAmplitudes}
\end{equation}
With the assumptions explained under eq.~(\ref{ACP}) and using eqs.~(\ref{DecayAmplitudes}), the mixing induced CP-asymmetry ($S$) and the direct CP asymmetry ($C$) can be written as
\begin{equation}
S=\frac{2\,{\mathrm{Im}}\left[r_d \left({\cal{A}}_L^* \bar{\cal{A}}_L + {\cal{A}}_R^* \bar{\cal{A}}_R \right)\right]}{|{\cal{A}}_L|^2 + |{\cal{A}}_R|^2 + |\bar{\cal{A}}_L|^2 + |\bar{\cal{A}}_R|^2}, \qquad
C=\frac{|{\cal{A}}_L|^2 + |{\cal{A}}_R|^2 - |\bar{\cal{A}}_L|^2 - |\bar{\cal{A}}_R|^2}{|{\cal{A}}_L|^2 + |{\cal{A}}_R|^2 + |\bar{\cal{A}}_L|^2 + |\bar{\cal{A}}_R|^2}.
\label{S&C}
\end{equation}
where $r_d = e^{-i \phi_d}$ and $\phi_d$ is the $\bar{B}^0_d-B^0_d$ mixing angle.

In ``na\"ive" factorisation, the decay amplitudes of eqs. (\ref{DecayAmplitudes}) are given by
\begin{subequations}
\begin{eqnarray}
 \bar{\cal{A}}_{L} \! &=& \! - \frac{4 G_F}{\sqrt{2}}\! \left[\lambda_u^{(s)} {\cal{C}}_7^{(u)} + \lambda_t^{(s)} {\cal{C}}_7^{(t)} \right]\! {\langle \bar{K}^* \gamma_L \vert {\cal O}_7^L \vert \bar{B} \rangle}, \\
 \bar{\cal{A}}_{R} \! &=& \! - \frac{4 G_F}{\sqrt{2}}\! \left[\lambda_u^{(s)} {\cal{C}}_{7^\prime,\,{\rm{SM}}}^{(u)} + \lambda_t^{(s)}\! \left({\cal{C}}_{7^\prime,\,{\rm{SM}}}^{(t)}+C_{7^\prime}^{(t)}\right)\right]\! {\langle \bar{K}^* \gamma_R \vert {\cal O}_7^R \vert \bar{B} \rangle},
 \label{AbarLR}
\end{eqnarray}
 \end{subequations}
and
\begin{subequations}
\begin{eqnarray}
{\cal{A}}_{L} \! &=& \! - \frac{4 G_F}{\sqrt{2}}\! \left[(\lambda_u^{(s)})^* {\cal{C}}_{7^\prime,\,{\rm{SM}}}^{(u)} + (\lambda_t^{(s)})^*\! \left({\cal{C}}_{7^\prime,\,{\rm{SM}}}^{(t)}+C_{7^\prime}^{(t)}\right)\right]\! {\langle K^* \gamma_L \vert \left({\cal O}_7^R\right)^\dagger \! \vert B \rangle}\!, \\
{\cal{A}}_{R} \! &=&  \! - \frac{4 G_F}{\sqrt{2}}\! \left[(\lambda_u^{(s)})^* {\cal{C}}_7^{(u)} + (\lambda_t^{(s)})^* {\cal{C}}_7^{(t)} \right]\! {\langle K^* \gamma_R \vert \left({\cal O}_7^L\right)^\dagger \! \vert B \rangle},
\label{ALR}
\end{eqnarray}
 \end{subequations}
where, we have used the short-hand notation introduced in eq.~(\ref{C7pSM})
\begin{equation}
{\cal{C}}_{7^\prime,\,{\rm{SM}}}^{(q)} = \frac{m_s}{m_b} {\cal{C}}_{7,\,{\rm{SM}}}^{(q)}\,
\end{equation}
with $q=u,t$. We have taken the notation and definitions from ref.~\cite{Beneke:2004dp}: $ {\cal{C}}_7^{(q)}$ are coefficients, defined as a ratio of full form factors and soft form factors, that can be computed in QCDF (${\cal{C}}_7^{(t)}$ is equivalent to $C_7^{\rm{eff}}$ at LO in $\alpha_s$ whereas ${\cal{C}}_7^{(u)}$ vanishes).
Setting ${\cal{C}}_{7^\prime,\,{\rm{SM}}}^{(q)} = 0$ and taking real Wilson coefficients $C_7^{\rm{eff}}$ and $C_{7^\prime}$, the mixing-induced CP-asymmetry yields the simple tree-level expression in ref.~\cite{Bobeth:2008ij, Grinstein:2004uu, Grinstein:2005nu}:
\begin{equation}
S_{K^* \gamma}^{\mathrm{(LO)}}= \frac{-2 \, \big\vert C_{7^\prime}/C_7^{{\rm eff}\,(0)}\big\vert }{1+\big\vert C_{7^\prime}/C_7^{{\rm eff}\,(0)}\big\vert^2} \sin \left(2 \beta -\arg\big(C_7^{{\rm eff}\,(0)} C_{7^\prime}\big)\right).
\label{SKsgammaLO}
\end{equation}
Eq.~(\ref{SKsgammaLO}) determines the cross-shaped plot of $S_{K^* \gamma}$ in the $(\delta C_7,\delta C_{7^\prime})$ plane (see Figure~\ref{plot1}) to a very good degree of approximation. We checked that $S_{K^* \gamma}^{\mathrm{(LO)}}$ allows us to recover, at $2 \, \sigma$, the shape of Figure 2 (left) in \cite{Bobeth:2008ij} using their input parameters. Notice, however, that our actual computation, used for the plots in the present article, is performed including NLO QCDF corrections.

Some comments are in order here. On the one hand, the operators ${\cal O}_7^{L(R)}$ are given by
\begin{equation}
{\cal O}_7^{L(R)}=\frac{e}{16 \pi^2} m_b \bar{s} \sigma_{\mu \nu} \frac{1 \pm \gamma_5}{2} b F^{\mu \nu},
\end{equation}
and generate the left- (right-) handed photons in the $b \to s \gamma$ decay. Following refs.~\cite{Ball:2006cva, Ball:2006eu} we express the matrix elements in eqs.~(\ref{AbarLR}) and (\ref{ALR}) in terms of the form factor $T_1^{B \to K^*}(q^2)$ as
\begin{eqnarray}
\lefteqn{{\langle \bar{K}^*(p,\eta) \gamma_{L(R)}(q,e) \vert {\cal O}_7^{L(R)} \vert \bar{B} \rangle} =}\hspace*{1cm}\nonumber\\
&=& -\frac{e}{8 \pi^2}\, m_b T_1^{B \to K^*}(0) \left\{\epsilon^{\mu\nu\rho\sigma} e_\mu^* \eta_\nu^* p_\rho q_\sigma \pm i \left[ (e^* \eta^*) (pq) - (e^* p)(\eta^* q)\right]\right\}\nonumber\\
&\equiv& -\frac{e}{8 \pi^2}\, m_b T_1^{B \to K^*}(0) S_{L(R)}\,, \\
\lefteqn{{\langle K^* (p,\eta) \gamma_{L(R)}(q,e) \vert ({\cal O}_7^{R(L)})^\dagger \vert B \rangle}  = -\frac{e}{8 \pi^2}\, m_b T_1^{B \to K^*}(0) S_{L(R)}\,,}\hspace*{1cm}
\label{matrixelements}
\end{eqnarray}
where $S_{L,R}$ are the helicity amplitudes corresponding, respectively, to left- and right-handed photons and $e_\mu(\eta_\mu)$ is the polarisation four-vector of the photon $(K^*)$. 

\begin{figure}
\begin{center} 
\includegraphics[width=0.50\textwidth]{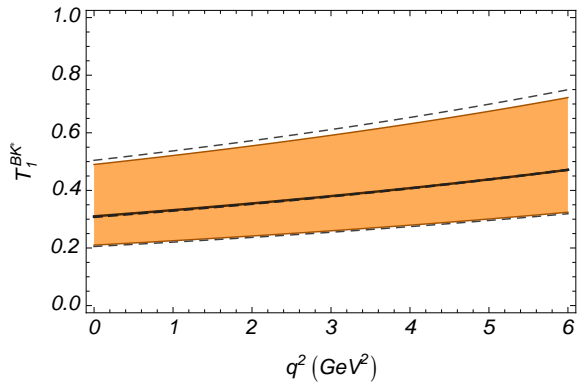}
\end{center}
\caption{Form factor $T_1^{B \to K^*}(q^2)$ in the $0-6\,{\mathrm{GeV}}^2$ energy range. The orange bands represent the full form factor with its associated errors given by the parametrisation in Appendix B.4 of \cite{Khodjamirian:2010vf} and the gray dashed lines depict $T_1^{B \to K^*}(q^2)$ computed from $\xi_\perp(q^2)$ using the large-recoil expressions in refs.~\cite{Kruger:2005ep, Beneke:2000wa}.} \label{fig: T1vsXiperp}
\end{figure}

On the other hand, since the photon emitted in the decay $b \to s \gamma$ is real, only the operators ${\cal O}_{1,\ldots,8}$ of the weak effective Hamiltonian contribute to this process. In particular, those that build up $C_7^{\mathrm{eff}}$ (see eq.~(\ref{C7effC8eff})) appear at $O(\alpha_s^0)$, while the rest of the operators enter the NLO QCDF corrections. Even though there is just one form factor contributing to this process, we have used the corresponding soft form factor computed by means of eq.~(\ref{SFF}) to be consistent with the QCDF formalism applied to both $A_I$ and $\bar{B}_d^0 \to \bar{K}^{*0} \ell^+ \ell^-$ observables. This amounts to replacing $T_1^{B \to K^*}(0) \rightarrow \xi_\perp(0)$ in eqs. (\ref{matrixelements}), which is indeed a very good approximation, as we can see in Fig.~\ref{fig: T1vsXiperp}.

Contrary to ref.~\cite{Beneke:2004dp} we have chosen to keep the CKM-suppressed terms proportional to $\lambda_u^{(s)}$.  In ``na\"ive" factorisation, both ${\cal{C}}_7^{(u)}$ and ${\cal{C}}_{7^\prime,\,{\rm{SM}}}^{(u)}$ vanish at LO in $\alpha_s$. If NP is absent, $C_{7^\prime}^{(t)}$ vanishes, as we have split the ${\cal{C}}_{7^\prime,\,{\rm{SM}}}^{(q)}$ helicity-suppressed $\gamma_R$ terms already present in the SM (see eq.~(\ref{C7pSM})) from the ${\cal O}_{7^\prime}$ NP contribution. Therefore, including NP in the decay amplitudes can be obtained upon the following replacements in eqs.~(\ref{AbarLR}, \ref{ALR}):
\begin{eqnarray}
C_{7^\prime}^{(t)} \rightarrow \delta C_{7^\prime}, \quad 
{\cal{C}}_7^{(t)} \rightarrow {\cal{C}}_{7\,,{\rm{SM}}}^{(t)} + \delta C_7,
\end{eqnarray}
where, as said in the previous section, ${\cal{C}}_7^{(q)} = C_7^{{\rm{eff}}\,(q)} + O(\alpha_s)$ includes the NLO corrections to the decay amplitude $B \to K^* \gamma$ \cite{Beneke:2001at}. Therefore, the replacement $\xi_\perp(0) \, {\cal{C}}_7^{(q)} \rightarrow {{\cal{T}}^{(q)}_\perp}$ \cite{Kruger:2005ep,Beneke:2004dp,Beneke:2001at} in the expressions above will be enough to account for these corrections in QCDF. Using this framework, we have computed the $O(\alpha_s)$ factorisable and non-factorisable corrections to hard-spectator scattering diagrams, as well as to those diagrams that involve a $B \to K^*$ form factor \cite{Beneke:2001at}. We have also included the power-suppressed weak annihilation and hard-spectator scattering contributions following \cite{Beneke:2004dp, Seidel:2004jh}; the latter suffer from the same kind of endpoint divergence that we find in $A_I$, and they have been regularised by means of eq.~(\ref{regHSS}).

\subsection{$B\to X_s \ell^+\ell^-$}\label{sec:btoxsmumu}

The branching ratio for $B\to X_s \ell^+\ell^-$, normalised by $B\to X_c\ell \nu$ and integrated between 1 and 6 GeV$^2$ can be written in the following manner:
\begin{equation}
\frac{d{\mathcal B}(\bar{B}\to X_s\ell^+\ell^-)_{SM}}{d\hat{s}}
 ={\mathcal B}(B\to X_c e\bar{\nu}) \left|\frac{V_{ts}^* V_{tb}}{V_{cb}}\right|^2
      \frac{4}{C} \frac{\Phi_{\ell\ell}(\hat{s})}{\Phi_u}, \qquad \hat{s}=\frac{s}{m_{b,pole}^2}
\end{equation}
where
\begin{equation}
 \frac{\Phi_{\ell\ell}(\hat{s})}{\Phi_u}= \sum_{i\leq j}
 {\rm Re} \left[C_i^{\rm eff}(\mu) C_j^{{\rm eff}*}(\mu)
 \left(\sum_{A,B=7,9,10} M^A_i M^{B*}_j \tilde{S}_{AB} + \Delta H_{ij}\right)
        \right]
\end{equation}
and $C$ has already been defined in eq.~(\ref{eq:C}).

We were able to reproduce the central value and uncertainty 
of $B\to X_s\ell^+\ell^-$, but also the dependence on $C_{7,8,9,10}$ at the scale
$\mu_0$ in eq.~(12) of ref.~\cite{Huber:2005ig} (apart from the linear term in $C_7(\mu_0)$ which is very sensitive to small changes in the input parameters).

We have modified the building blocks $S$ following ref.~\cite{Guetta:1997fw} to include $m_s$ corrections and contributions from chirality-flipped operators in the following way\footnote{We checked and agreed with the expressions in ref.~\cite{Guetta:1997fw}, taking into account the fact that this reference uses a different definition of ${\cal O}_7$ and ${\cal O}_7'$ which mixes different chiralities, contrary to ours.}.
\begin{itemize}
\item 
For the functions involving only $A,B=7,9,10$, we modified the functions to include $m_s$-suppressed contributions to the phase space and to $O(\alpha_s^0)$ part. 
\item For the functions  involving only $A,B=7',9',10'$, we took the same expression as their unprimed counterparts, profitting from the fact that the expressions are symmetric with respect to the change $\gamma_5\to -\gamma_5$. 
\item For the functions involving both a SM operator and a chirally-flipped one, we took the expressions from ref.~\cite{Guetta:1997fw}, which include only $O(\alpha_s^0)$ contributions (contrary to the other functions that include also 
$O(\alpha_s)$ and $O(1/m_b^2)$ corrections).
\end{itemize}
\begin{eqnarray}
S_{77}=S_{7'7'}&=&N\left(1+\frac{2\hat{m}_\ell^2}{\hat{s}}\right)
 \left[
  -4\hat{s}-4(1+\hat{m}_s^2)+\frac{8(1-\hat{m}_s^2)^2}{\hat{s}} +O(\alpha_s,1/m_b^2)
  \right]\qquad
  \\
S_{79}=S_{7'9'}&=&N\left(1+\frac{2\hat{m}_\ell^2}{\hat{s}}\right) \cdot 12[1-\hat{m}_s^2-\hat{s}+O(\alpha_s,1/m_b^2)] \\
S_{99}=S_{9'9'}&=&N
 \Bigg[1+2\hat{m}_\ell^2-2\hat{m}_s^2+2\hat{m}_\ell^2\hat{m}_s^2+\hat{m}_s^4\\
 &&\qquad \qquad
    +\frac{2\hat{m}_\ell^2(1-\hat{m}_s^2)^2}{\hat{s}} + (1-4\hat{m}_\ell^2+\hat{m}_s^2)\hat{s}-2\hat{s}^2+O(\alpha_s,1/m_b^2)\Bigg]
\nonumber 
\end{eqnarray}
\begin{eqnarray}
S_{1010}=S_{10'10'}&=&N
 \Bigg[1-10\hat{m}_\ell^2-2\hat{m}_s^2-10\hat{m}_\ell^2\hat{m}_s^2+\hat{m}_s^4\\
 &&\qquad \qquad
    +\frac{2\hat{m}_\ell^2(1-\hat{m}_s^2)^2}{\hat{s}} + (1+8\hat{m}_\ell^2+\hat{m}_s^2)\hat{s}-2\hat{s}^2+O(\alpha_s,1/m_b^2)\Bigg]
\nonumber\\
S_{77'}&=& N\left(1+\frac{2\hat{m}_\ell^2}{\hat{s}}\right) (-48\hat{m}_s)\\
S_{79'}=S_{7'9}&=&
 N\left(1+\frac{2\hat{m}_\ell^2}{\hat{s}}\right)(-12\hat{m}_s)(1-\hat{m}_s^2+\hat{s})\\
 S_{99'}&=& N(-12 \hat{m}_s)(\hat{s}+2\hat{m}_\ell^2)\\
 S_{1010'}&=& N(-12 \hat{m}_s)(\hat{s}-6\hat{m}_\ell^2)
\end{eqnarray}
with the phase space factor
\begin{equation}
N=\sqrt{1+\hat{s}^2+\hat{m}_s^4-2\hat{s}-2\hat{m}_s^2-2\hat{s}\hat{m}_s^2}\sqrt{1-\frac{4\hat{m}_\ell^2}{\hat{s}}}
\end{equation}
For the quantities related to matrix elements $M^A_i$, we have taken the expressions of ref.~\cite{Huber:2005ig} for the unprimed operators. The situation is much simpler for chirally-flipped operators since only three of them are to be considered:
\begin{equation}
M^{7'}_i = \tilde{\alpha}_s \kappa \delta_{i,7'}, \quad 
M^{9'}_i = (1+\tilde{\alpha}_s \kappa f_9^{\rm pen}(\hat{s})) \delta_{i,9'}, \quad 
M^{10'}_i= \delta_{i,10'}\,.
\end{equation}
The uncertainty attached to the central value in table~\ref{tab:coeffBtoXsll} includes not only the uncertainties from the variation of the difference input parameters, but also a 5\% error estimated in
ref.~\cite{Huber:2005ig} as the uncertainty from non-perturbative $1/m_b$-suppressed contributions.

\subsection{$\bar{B} \to \bar{K}^{*0}\ell^+\ell^-$ observables} \label{btokllap}

\subsubsection{General considerations}

The differential decay amplitude of the exclusive process $\bar{B}_d \to \bar{K}^{*0} \ell^+ \ell^-$, with $\bar{K}^{*0} \to K^- \pi^+$ on the mass shell, can be characterised completely in terms of the dilepton pair invariant mass $q^2$, which is embedded  in the so-called \textit{angular coefficients}, and the three independent angles $\theta_l$, $\theta_K$ and $\phi$ (see Section 2.1 of \cite{Egede:2010zc}). These angular coefficients $J_i$ are observable quantities that depend on kinematical parameters, real combinations of the six complex $\bar{K}^{*0}$ spin amplitudes and the seventh transverse amplitude $A_t$ (in the presence of scalars an extra amplitude is required \cite{Altmannshofer:2008dz}).

Within our framework, the spin amplitudes can be expressed in terms of the seven $B \to K^*$ form factors and the Wilson coefficients $C_i$ of the weak effective Hamiltonian, that account for the short-distance interactions. Neglecting $O(\alpha_s)$ corrections and using the effective Wilson coefficient associated to ${\cal O}_7$ (which includes the contributions from the four-quark operators ${\cal O}_{1\ldots 8}$), as well as the numerically relevant coefficients $C_9$ and $C_{10}$ associated to ${\cal O}_9$ and ${\cal O}_{10}$ respectively, we find~\cite{Egede:2010zc}:
\begin{eqnarray}
A_{\perp}^{L,R} &=&  N \sqrt{2} \lambda^{1/2} 
 \bigg[ 
 \left\{ 
   (C_9 + C_{9^\prime}) \mp (C_{10} +  C_{10^\prime})
 \right\} \frac{ V(q^2) }{ m_B + m_{K^*}} +  \nonumber \\
 &&\quad +\frac{2m_b}{q^2} (C_7^{\mathrm{eff}} + C_{7^\prime}^{\mathrm{eff}}) T_1(q^2)
 \bigg] \,,\\
A_{\|}^{L,R}  &=& - N \sqrt{2}(m_B^2 - m_{K^*}^2) 
            \bigg[ \left\{ (C_9 - C_{9^\prime}) \mp (C_{10} -  C_{10^\prime})
            \right\} \frac{A_1(q^2)}{m_B-m_{K^*}}  +    \nonumber  \\
&& \quad +\frac{2 m_b}{q^2} (C_7^{\mathrm{eff}} - C_{7^\prime}^{\mathrm{eff}}) T_2(q^2) \bigg]  \,, 
 \end{eqnarray}
\begin{eqnarray}
A_{0}^{L,R} &=& - \frac{N}{2 m_{K^*} \sqrt{q^2}}  \,\,  \bigg[ \left\{ (C_9 - C_{9^\prime}) \mp (C_{10} -  C_{10^\prime}) \right\} \cdot  \nonumber \\   
&& \quad \cdot \left\{ (m_B^2 - m_{K^*}^2 - q^2) ( m_B + m_{K^*}) A_1(q^2) 
 -\frac{\lambda A_2(q^2)}{m_B + m_{K^*}}
\right\}  + \\
&& \quad  + {2 m_b}(C_7^{\mathrm{eff}} - C_{7^\prime}^{\mathrm{(eff)}}) \left\{
 (m_B^2 + 3 m_{K^*}^2 - q^2) T_2(q^2)
-\frac{\lambda}{m_B^2 - m_{K^*}^2} T_3(q^2) \right\}
\bigg] \,,\nonumber \\
A_t &=&    \frac{N \lambda^{1/2}}{\sqrt{q^2}}   \bigg[ 2 (C_{10} -  C_{10^\prime}) \bigg] A_0(q^2) \,,
\label{Kspin}
\end{eqnarray}
where
\begin{equation}
  \label{eq:Lambdadef}
  \lambda= m_B^4  + m_{K^*}^4 + q^4 - 2 (m_B^2 m_{K^*}^2+ m_{K^*}^2 q^2  + m_B^2 q^2),
\end{equation}
\begin{equation}
N=\sqrt{\frac{G_F^2 \alpha^2}{3\cdot 2^{10}\pi^5 m_B^3}
|V_{tb}V_{ts}^{*}|^2 q^2 \lambda^{1/2}
\beta_\mu},
\end{equation}
with 
\begin{equation}
\beta_\mu=\sqrt{1-\frac{4 m_\mu^2}{q^2}}.
\end{equation}
We have introduced the Wilson coefficients corresponding to the chirally flipped operators ${\cal O}_{7^\prime}$, ${\cal O}_{9^\prime}$ and ${\cal O}_{10^\prime}$, so we consider only  NP contributions stemming from the SM-like operators and their chirally-flipped partners (i.e. we assume there are neither scalar/pseudoscalar nor tensor/pseudotensor operators at work).

\subsubsection{Soft form factors}
\label{sffsection}

Concerning the $B \to K^*$ form factors, there are seven a priori independent hadronic form factors, encoding the non-perturbative long-distance interactions, that enter the $B \to K^*$ matrix elements, namely the vector current form factor $V(q^2)$, the three axial current form factors $A_0(q^2)$, $A_1(q^2)$, $A_2(q^2)$, the tensor form factor $T_1(q^2)$ and the pseudo-tensor form factors $T_2(q^2)$ and $T_3(q^2)$ \cite{Beneke:2000wa}. Although there are several computations of these form factors in the literature (see for instance ref.~\cite{Ball:2004rg}), we have chosen the parametrisation in Appendix B.4 of ref. \cite{Khodjamirian:2010vf} to remain more conservative in the estimation of the uncertainties associated to the fitting coefficients that describe them.
In the limit where the decaying hadron is heavy (as in $B_d$) and the recoiling meson acquires a large energy ($E_{K^*}$), the form factors can be expanded in the small ratios $\Lambda_{\mathrm{QCD}}/m_b$ and $\Lambda_{\mathrm{QCD}}/E_{K^*}$. Neglecting corrections of order $\Lambda_{\mathrm{QCD}}/m_b$ and $\alpha_s$, the seven $B \to K*$ form factors reduce to just two universal ``soft" form factors $\xi_\perp$ and $\xi_\|$ \cite{Beneke:2000wa,Charles:1998dr}. 

In this limit the $K^*$ spin amplitudes and $A_t$
acquire very simple forms which prove to be most useful to explain the symmetries between the fitting coefficients ($F$, $G$, $H$, $I$, $J$ and $K$) of $\bar{B}_d \to \bar{K}^{*0} \ell^+ \ell^-$ observables given in Secs.~\ref{sec:classII} and \ref{sec:classIII} \cite{Kruger:2005ep}
\begin{subequations}
  \begin{align}
A_{\perp}^{L,R} &=\sqrt{2} N m_B(1- \hat{s})\bigg[  
(C_9 + C_{9^\prime}) \mp (C_{10} +  C_{10^\prime})
+\frac{2\hat{m}_b}{\hat{s}} (C_7^{\mathrm{eff}} + C_{7^\prime}^{\mathrm{eff}}) 
\bigg]\xi_{\perp}(E_{K^*}),   \\
A_{\|}^{L,R}  &= -\sqrt{2} N m_B (1-\hat{s})\bigg[
 (C_9 - C_{9^\prime}) \mp (C_{10} -  C_{10^\prime}) 
+\frac{2 \hat{m}_b}{\hat{s}}(C_7^{\mathrm{eff}} - C_{7^\prime}^{\mathrm{eff}})  \bigg] \xi_{\perp}(E_{K^*})\, , \\
A_{0}^{L,R} &= -\frac{N m_B }{2 \hat{m}_{K^*} \sqrt{\hat{s}}} (1-\hat{s})^2\bigg[ (C_9 - C_{9^\prime})  \mp (C_{10} -  C_{10^\prime})
+ 2
\hat{m}_b (C_7^{\mathrm{eff}} - C_{7^\prime}^{\mathrm{eff}}) \bigg]\xi_{\|}(E_{K^*})\, ,\\
A_t  &= \frac{N m_B }{ \hat{m}_{K^*} \sqrt{\hat{s}}} (1-\hat{s})^2\bigg[ C_{10} -  C_{10^\prime} \bigg] \xi_{\|}(E_{K^*}) \, ,
  \end{align} \label{LEETeqs}
\end{subequations}
with $\hat{s} = q^2/m_B^2$ and $\hat{m}_i = m_i/m_B$.

The QCDF framework allows us to calculate the $\alpha_s$ corrections to form factors and decay amplitudes up to the NLO \cite{Beneke:2004dp,Beneke:2000wa,Beneke:2001at} in a systematic way but, since we have no means of computing the $1/m_b$-suppressed corrections, we decided to estimate them consistently using an ensemble method for the $K^{*0}$ spin amplitudes; an exhaustive discussion of all these issues can be found in sections 2.2 and 2.3 of ref.~\cite{Egede:2010zc}.

However, since QCDF uses only soft form factors and not the full form factors, we are restricted to the kinematic region in which $E_{K^*} \sim m_b$ (or equivalently, $q^2 \ll m_B$). Moreover, the longitudinal spin amplitude displays a logarithmic divergence as $q^2 \rightarrow 0$, which signals the breakdown of QCDF for energies below $1\, {\mathrm{GeV}}^2$. Further cuts are provided by the light (below $1\, {\mathrm{GeV}}^2$) and $J/\psi$ (over $6\, {\mathrm{GeV}}^2$) resonances. Thus, we have confined the analysis of $A_{\rm T}^{(2)}$, $A_{\rm {FB}}$ and $F_{\rm L}$ to the dilepton mass range, $1\,{\mathrm{GeV}}^2 \leqslant q^2\leqslant 6\,{\mathrm{GeV}}^2$.

We obtain the soft form factors demanded by the QCDF framework \cite{Beneke:2001at, Beneke:2004dp} from the full form factors $V(q^2), A_1(q^2)$ and $A_2(q^2)$~\cite{Khodjamirian:2010vf} using~\cite{Egede:2010zc, Bobeth:2008ij, Beneke:2004dp, Bobeth:2010wg}
\begin{eqnarray}
\label{SFF}
\xi_\perp(q^2) &=& \frac {m_B} {m_B + m_{K^*}} V(q^2), \cr
\xi_\|(q^2) &=& \frac{m_B + m_{K^*}} {2 E_{K^*}} A_1(q^2) - \frac{m_B - m_{K^*}} {m_B} A_2(q^2).
\end{eqnarray}
Our choice of ref.~\cite{Khodjamirian:2010vf} with sizeable error bars compared to other possible determinations is guided by our aim to be  conservative in our estimation of errors. Eq.~(\ref{SFF}), in particular, defines the value of the soft form factors at $q^2=0$ from the values of the full form factors taken from ref.~\cite{Khodjamirian:2010vf}.

\subsubsection{The differential decay distribution and uniangular projections} \label{ddd}

The angular dependence of the $\bar{B}_d \to \bar{K}^{*0} \ell^+ \ell^-$ differential decay distribution can be integrated out yielding, in terms of the $\bar{K}^{*0}$ spin amplitudes,
\begin{eqnarray}
\frac{d\Gamma}{dq^2}&=&\frac{1}{4} \left[(3 + \beta_\mu^2) (|A_\perp|^2 + |A_\||^2 + |A_0|^2) \right] + \nonumber\\
&+&\frac{3 m_\mu^2}{q^2} \left\{|A_t|^2 + 2 [{\mathrm{Re}}(A_{\perp L} A_{\perp R}^*) + {\mathrm{Re}}(A_{\| L} A_{\| R}^*) + {\mathrm{Re}}(A_{0 L} A_{0 R}^*)] \right\}
\label{dGamma}
\end{eqnarray}
where we have defined $A_i A_j^* \equiv A_{i L} A_{j L}^* + A_{i R} A_{j R}^*$, with $i,j = 0, \perp, \|$. 

\begin{figure}[ht]
\begin{center} 
\includegraphics[width=0.50\textwidth]{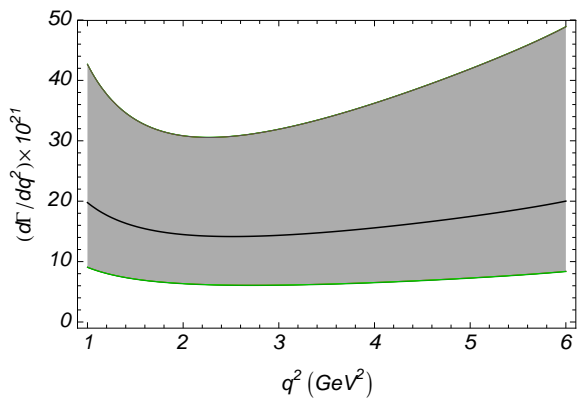}
\end{center}
\caption{SM prediction for the differential decay distribution of $\bar{B}_d \to \bar{K}^{*0} \ell^+ \ell^-$ in the $1-6\,{\mathrm{GeV}}^2$ energy range. The black line corresponds to the central value of $d\Gamma/dq^2$. The wide gray band corresponds to the uncertainties associated to $B \to K^*$ form factors (according to the parametrisation in Appendix 4 of ref. \cite{Khodjamirian:2010vf}). Hadronic (orange) and $\Lambda_{\mathrm{QCD}}/m_b$ (green) uncertainty bands are barely visible. The central value compares well with Fig.~2 in ref. \cite{Altmannshofer:2008dz} (note that the CP-averaged differential decay distribution $d(\Gamma+\bar{\Gamma})/dq^2$ is plotted there, so their central value curve is twice ours).}
\label{dGammaPlot}
\end{figure}

\begin{figure}[ht]
\centering
\subfloat[]{\label{AFBplot}
\includegraphics[width=0.47\textwidth]{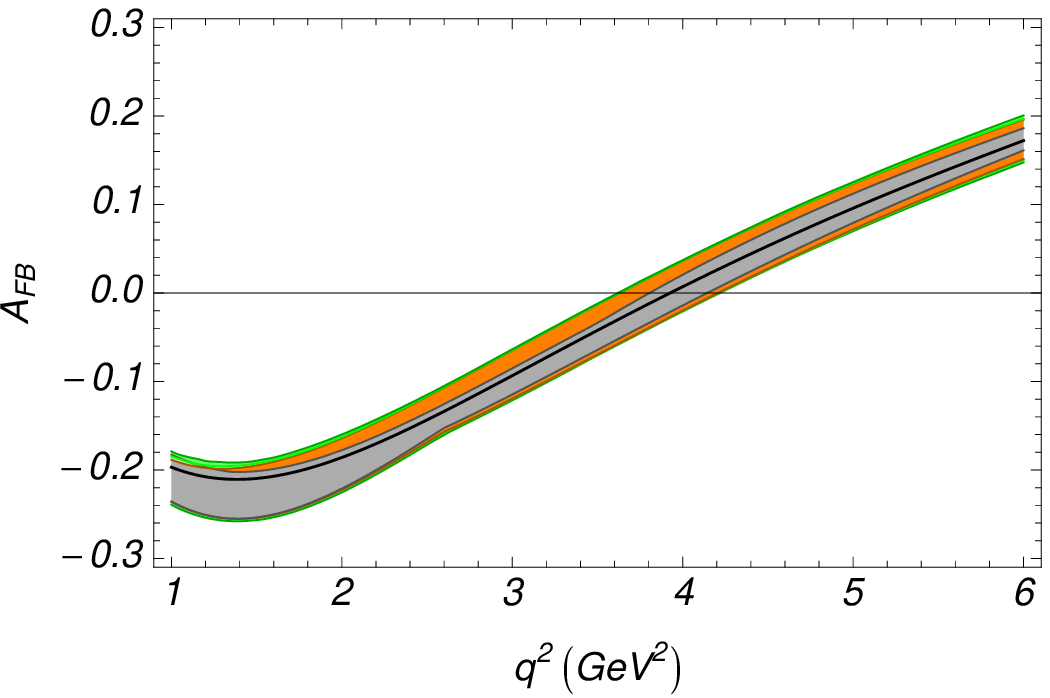}}
\subfloat[]{\label{FLplot}
\includegraphics[width=0.47\textwidth]{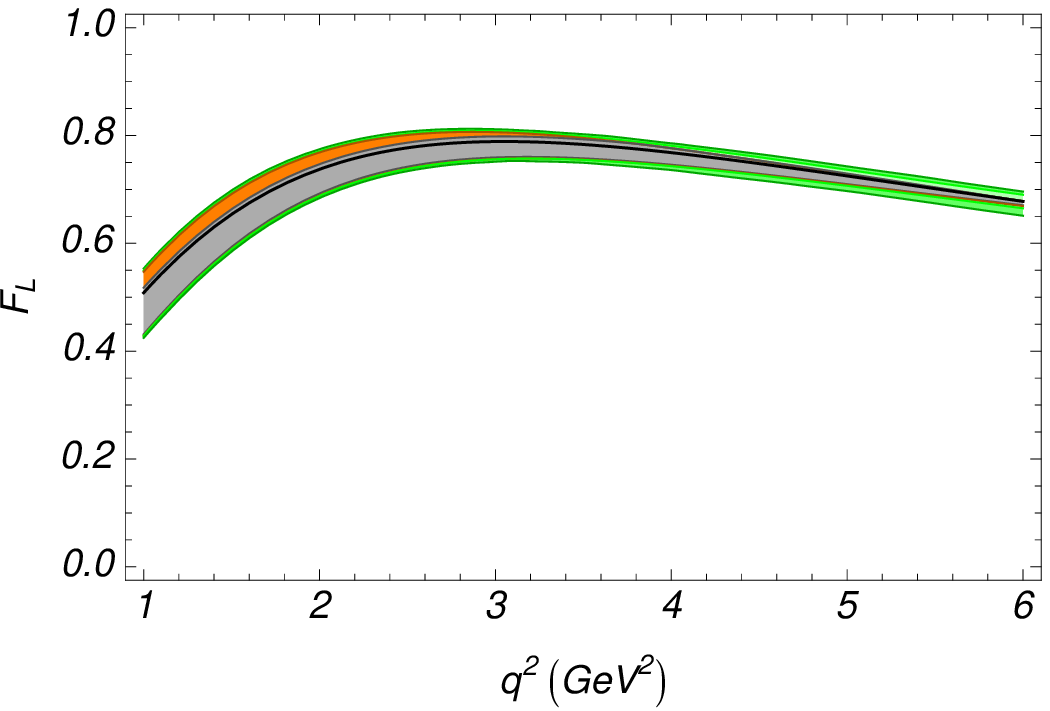}}
\caption{SM prediction for $A_{\mathrm{FB}}$ (left) and $F_{\mathrm{L}}$ (right) in the $1-6\,{\mathrm{GeV}}^2$ energy range. The color scheme used for uncertainties is the same as in Fig.~15.}
\end{figure}

The large uncertainties coming from the $B \to K^*$ form factors turn $d\Gamma / dq^2$ into a theoretically ill-controlled observable (as can be seen in Fig.~\ref{dGammaPlot}). However, since it appears only in the denominator of $A_{\mathrm{FB}}$ and $F_{\mathrm{L}}$, and the corresponding numerators display the same kind of uncertainties  correlated to those in $d\Gamma / dq^2$, $A_{\mathrm{FB}}$ and $F_{\mathrm{L}}$ become much better behaved observables (see Figs.~\ref{AFBplot} and \ref{FLplot}, and refs.~\cite{Egede:2010zc,Egede:2008uy} for an in-depth discussion of this issue). $A_{\rm T}^{(2)}$,  on the contrary, is essentially free from this problem.


As shown in refs.~\cite{Egede:2010zc, Egede:2008uy}, a full angular fit can be performed on $\bar{B}_d \to \bar{K}^{*0} \ell^+ \ell^+$ observables, but this will probably require more integrated luminosity than the one delivered by the end of the fist run of LHC \cite{Egede:2008uy, Egede:2007zz, Bharucha:2010bb}. However, we can also integrate out two of the three angles of the $K^*$ differential decay distribution to get three single-angle distributions, which, in the massless case\footnote{Massive terms are suppressed by $m_\mu^2 \simeq 0.011\,{\rm GeV}^2$ so that their impact in absence of possible large scalar/pseudoscalar or tensor/pseudotensor NP operators is negligible.} read
\begin{subequations}
  \label{eq:projections}
  \begin{eqnarray}
    \label{eq:dGammadPhi}
    \frac{1}{\Gamma^\prime}\frac{d\Gamma^\prime}{d\phi} & = & \frac{1}{2\pi}\left(
      1 + \frac{1}{2} (1 - F_{\mathrm{L}}) A_{\mathrm{T}}^{(2)} \cos 2\phi + 
      A_{\mathrm{im}} \sin 2\phi
    \right), \\
    \label{eq:dGammadThetal}
    \frac{1}{\Gamma^\prime}\frac{d\Gamma^\prime}{d\theta_l} & = & \left(
      \frac{3}{4} F_{\mathrm{L}} \sin^2\theta_l + 
      \frac{3}{8} (1-F_{\mathrm{L}}) (1+\cos^2\theta_l) +
     A_{\mathrm{FB}} \cos \theta_l 
    \right)\sin\theta_l\, , \\
    \label{eq:dGammadThetaK}
    \frac{1}{\Gamma^\prime}\frac{d\Gamma^\prime}{d\theta_K} & = &
    \frac{3}{4} \sin\theta_K \left(
      2 F_{\mathrm{L}} \cos^2\theta_K + (1-F_{\mathrm{L}}) \sin^2\theta_K
    \right),
  \end{eqnarray}
\end{subequations}
where
\begin{equation}
\Gamma^\prime \equiv \frac{d\Gamma}{dq^2} \quad {\rm and} \quad A_{\mathrm{im}} = \frac{{\mathrm{Im}}{(A_{\perp L} A_{\| L}^*)} + {\mathrm{Im}}{(A_{\perp R} A_{\| R}^*)}}{\frac{d\Gamma}{dq^2}}.
\end{equation}
Since $A_{\mathrm{FB}}$, $F_{\mathrm{L}}$ and $A_{\mathrm{T}}^{(2)}$ appear in all the expressions above, experimental data can be binned in $q^2$ and the corresponding fits performed on these bins. The value extracted from these fits is then a $\frac{d\Gamma}{dq^2}$-weighted average of each parameter. Some strategies have already been devised to perform the binning in a way that allows to increase the statistics signal for some chosen observables \cite{Alok:2010zd, Bharucha:2010bb, Lunghi:2010tr}.

\subsubsection{$\bar{B}_d \to K^{*0} \ell^+ \ell^-$ observables at leading order in the large-recoil limit}\label{sec:symmetries}

This section is devoted to the analysis of the relations existing between the different fitting functions, namely eqs.~(\ref{AT2Frel}), (\ref{AT2Grel}) and (\ref{AT2FGrel}) for $A_{\mathrm{T}}^{(2)}$, eqs.~(\ref{AFBHrel}) and (\ref{AFBIrel}) for $A_{\mathrm{FB}}$ and eqs.~(\ref{FLJrel}) for $F_{\mathrm{L}}$. The simple large recoil spin amplitudes in eqs.~(\ref{LEETeqs}) will be used to account for the existence of these relations, allowing us to proceed as we did in Sect.~6 of ref.~\cite{Egede:2010zc}. The LO large-recoil expressions are sufficient to understand these symmetries, since NLO contributions do not break the pairing of the coefficients.

The following short-hand notation will be used
\begin{eqnarray}
M_{\pm} &\equiv& 12 m_\mu^2 \pm q^2 (3+\beta_\mu^2), \qquad\quad F \equiv \frac{2 \hat{m}_b}{ \hat{s}}, \nonumber \\
P_1 &\equiv& \sqrt{2} N m_B (1-\hat{s}), \qquad\quad\quad P_2 \equiv \frac{1}{2 \sqrt{2} \hat{m}_{K^*} \sqrt{\hat{s}}} (1-\hat{s}), \nonumber
\end{eqnarray}
and
\begin{equation}
\!\!\!\!\!\!\!\!\!\!\!\!\!\!\!\!\!\!\!\!\!\!\!\!\!\!\!\!\!\!\!\!\! {\phantom{\Bigg\vert}}C_i \equiv C_i^{\mathrm{SM}}+\delta C_i, \qquad\qquad\qquad\,\, C_{i^\prime} \equiv \delta C_{i^\prime}.
\label{WCsubst}
\end{equation}
From the simplified expression of the spin amplitudes given earlier in Appendix~\ref{sffsection} we obtain:
\begin{itemize}
\item[$\bullet$] $A_{\mathrm{T}}^{(2)}$. Being built in terms of just $A_\perp$ and $A_\|$, the LO behaviour of this observable can be readily understood (see ref.~\cite{Egede:2010zc}).
\begin{equation}
A_{\mathrm{T}}^{(2)}\Big\vert_{LR} = \frac{2 \big[C_{10} C_{10^\prime} + (C_9 + C_7 F) (C_{9^\prime} + C_{7^\prime} F)\big] P_1^2\, 4 \,\xi_\perp^2}{\big[C_{10}^{\,2} + C_{10^\prime}^{\,2} + (C_9 + C_7 F)^2 + (C_{9^\prime} + C_{7^\prime} F)^2\big] P_1^2\, 4\, \xi_\perp^2},
\label{AT2LR}
\end{equation}
where $LR$ stands for "large recoil". Eq.~(\ref{AT2LR}) shows that, at LO, only the terms with primed coefficients ($\delta C_{j^\prime}$) and cross terms like $\delta C_{i} \delta C_{j^\prime}$ (with $i,j=7,9,10$) might appear in the numerator of $A_{\mathrm{T}}^{(2)}$. Neither those involving just unprimed coefficients ($\delta C_i$, with $i=7,9,10$) nor products of same chirality operators $(\delta C_{i^{(\prime)}} \delta C_{j^{(\prime)}}$) are allowed, whereas in table \ref{tab:coeffAT2-F}, the latter are also present but they come from NLO corrections and are much smaller than the LO ones\footnote{The fitting coefficients of the terms forbidden at LO are at most $4\%$ of those allowed.}.  Furthermore they are very suppressed by the corresponding NP terms in the denominator (see table~\ref{tab:coeffAT2-G}). Since both primed and unprimed coefficients enter the full expression of $A_{\mathrm{T}}^{(2)}$ at NLO in the same way, the relations on the first row of eqs.~(\ref{AT2Frel}), (\ref{AT2Grel}) and (\ref{AT2FGrel}) hold. The remaining relations can be checked trivially using eq.~(\ref{AT2LR}).
\item[$\bullet$] $d\Gamma/dq^2$. The differential decay distribution appears in the denominator of both $A_{\mathrm{FB}}$ and $F_{\mathrm{L}}$ as a sum of $I_{(i,j)}\, \delta C_i \,\delta C_j$. At LO it can be expressed as
\begin{eqnarray}
\frac{d\Gamma}{dq^2}\Bigg\vert_{LR}\!\!\!&=&\frac{P_1^2}{2q^2} \Bigg\{ M_{+} \bigg[ \Big[\big(C_9-C_{9^\prime} + (C_7 - C_{7^\prime}) F \hat{s}\big)^2 + \big(C_{10}-C_{10^\prime}\big)^2 \Big] P_2^2\, \xi_\|^2 \nonumber\\
&+& \Big[\big(C_9-C_{9^\prime} + (C_7 - C_{7^\prime}) F\big)^2 + 2 \big(C_9+C_7 F\big) \big(C_{9^\prime}+C_{7^\prime} F\big) \Big] 2\, \xi_\perp^2 \bigg]\nonumber \\
&-& M_{-} \big(C_{10}^2 + C_{10^\prime}^2\big) 2\, \xi_\perp^2  \Bigg\}.\nonumber
\label{dGammaLR}
\end{eqnarray}
Although quite cumbersome, eq.~(\ref{tab:coeffAFB-I}) allows us to understand table \ref{tab:coeffAFB-HI}, since all the coefficients there appear already at LO. In particular, we can check that the largest fitting coefficients ($I_{(0,7)}$ and $I_{(7,7)}$) are enhanced either by the square of the factor $F=(2 m_b m_B)/q^2$ (which becomes very important in the low-$q^2$ region) or by $C_9^{\mathrm{SM}}F$, whereas others are enhanced by $F$ but suppressed by $C_7^{\mathrm{SM}}$ (like $I_{(0,9)}$ and $I_{(0,9^{\prime})}$) and the remaining ones are not enhanced at all. Eq.~(\ref{tab:coeffAFB-HI}) can be used also to verify the relations in eq.~(\ref{AFBIrel}).
\item[$\bullet$] $A_{\mathrm{FB}}$. At LO and in the large recoil limit, the numerator of this observable has a structure given by
\begin{equation}
A_{\mathrm{FB}}\Big\vert_{LR} =\frac{-6 \beta_\mu P_1^2 \big[C_{10}(C_9 + C_7 F) -C_{10^\prime} (C_{9^\prime} + C_{7^\prime} F) \big] \xi_\perp^2}{d\Gamma/dq^2}.
\label{AFBnumLR}
\end{equation}

All fitting coefficients in table~\ref{tab:coeffAFB-HI} arise already at LO except for those that involve a primed coefficient, i.e. $(0,7^\prime)$, $(0,10^\prime)$ and $(7,10^\prime)$. In the case of $I_{(0,7^\prime)}$, the effect of the enhancement factor $F$ at low-$q^2$ explained above is particularly visible, while $I_{(7,10^\prime)}$, which also receives this enhancement, is suppressed by $C_7^{\mathrm{SM}}$ and $I_{(0,10^\prime)}$ is not enhanced at all.

Regarding eq.~(\ref{AFBHrel}), the first and the second equalities are LO relations due to the antisymmetric behaviour of primed and unprimed coefficients in eq.~(\ref{AFBnumLR}), whereas the last one appears only at NLO but respects the same symmetry.
\item[$\bullet$] $F_{\mathrm{L}}$. The numerator of this $F_{\mathrm{L}}$ at LO and in the large recoil limit simplifies into
\begin{equation}
F_{\mathrm{L}}\Big\vert_{LR} = \frac{2 P_1^2 P_2^2 \big[(C_{10} - C_{10^\prime})^2 + (C_9 - C_{9^\prime} + (C_7 - C_{7^\prime}) F \hat{s})^2\big] \xi_\|^2}{d\Gamma/dq^2}.
\end{equation}

Using eq.~(\ref{WCsubst}) we can expand the numerator of $F_{\mathrm{L}}$ into products of NP Wilson coefficients. This is enough to derive all relations in eq.~(\ref{FLJrel}) and to explain the enhancement of some fitting coefficients over others in the low-$q^2$ region.
\end{itemize}


\begin{thebibliography}{10}


\bibitem{Lenz:2010gu}
  A.~Lenz {\it et al.},
  Phys.\ Rev.\  D {\bf 83} (2011) 036004
  [arXiv:1008.1593 [hep-ph]].

\bibitem{Lunghi:2010gv}
  E.~Lunghi and A.~Soni,
  Phys.\ Lett.\  B {\bf 697} (2011) 323
  [arXiv:1010.6069 [hep-ph]].

\bibitem{Bevan:2010gi}
  A.~J.~Bevan {\it et al.}  [UTfit Collaboration],
  arXiv:1010.5089 [hep-ph].

\bibitem{Bhattacherjee:2010ju}
  B.~Bhattacherjee, A.~Dighe, D.~Ghosh and S.~Raychaudhuri,
  arXiv:1012.1052 [hep-ph].

\bibitem{Dighe:2011du}
  A.~Dighe, D.~Ghosh, A.~Kundu and S.~K.~Patra,
  arXiv:1105.0970 [hep-ph].


\bibitem{Belle:2009zv}
  J.~T.~Wei {\it et al.}  [BELLE Collaboration],
  Phys.\ Rev.\ Lett.\  {\bf 103}, 171801 (2009)
  [arXiv:0904.0770 [hep-ex]].

\bibitem{Gambino:2004mv}
  P.~Gambino, U.~Haisch and M.~Misiak,
  Phys.\ Rev.\ Lett.\  {\bf 94}, 061803 (2005)
  [arXiv:hep-ph/0410155].
  
\bibitem{Alok:2009tz}
  A.~K.~Alok, A.~Dighe, D.~Ghosh, D.~London, J.~Matias, M.~Nagashima and A.~Szynkman,
  JHEP {\bf 1002} (2010) 053
  [arXiv:0912.1382 [hep-ph]].
  
 \bibitem{Egede:2010zc}
  U.~Egede, T.~Hurth, J.~Matias, M.~Ramon and W.~Reece,
  JHEP {\bf 1010} (2010) 056
  [arXiv:1005.0571 [hep-ph]].
  
\bibitem{Pati:1974vw}
  J.~C.~Pati and A.~Salam,
  Phys.\ Rev.\ Lett.\  {\bf 32} (1974) 1083.

\bibitem{Mohapatra:1974gc}
  R.~N.~Mohapatra and J.~C.~Pati,
  Phys.\ Rev.\  D {\bf 11} (1975) 2558.

\bibitem{Mohapatra:1974hk}
  R.~N.~Mohapatra and J.~C.~Pati,
  Phys.\ Rev.\  D {\bf 11} (1975) 566.

\bibitem{Bernard:2006gy}
  V.~Bernard, M.~Oertel, E.~Passemar and J.~Stern,
  Phys.\ Lett.\  B {\bf 638} (2006) 480
  [arXiv:hep-ph/0603202].

\bibitem{Bernard:2007cf}
  V.~Bernard, M.~Oertel, E.~Passemar and J.~Stern,
  JHEP {\bf 0801} (2008) 015
  [arXiv:0707.4194 [hep-ph]].

\bibitem{Crivellin:2009sd}
  A.~Crivellin,
  Phys.\ Rev.\  D {\bf 81} (2010) 031301
  [arXiv:0907.2461 [hep-ph]].


\bibitem{Buras:2010pz}
  A.~J.~Buras, K.~Gemmler and G.~Isidori,
  Nucl.\ Phys.\  B {\bf 843}, 107 (2011)
  [arXiv:1007.1993 [hep-ph]].
  
\bibitem{Alok:2011gv}
  A.~K.~Alok, A.~Datta, A.~Dighe, M.~Duraisamy, D.~Ghosh and D.~London,
  arXiv:1103.5344 [hep-ph].
  
\bibitem{Bobeth:2008ij}
  C.~Bobeth, G.~Hiller and G.~Piranishvili,
  JHEP {\bf 0807}, 106 (2008)
  [arXiv:0805.2525 [hep-ph]].
  
\bibitem{Misiak:2006zs}
  M.~Misiak {\it et al.},
  Phys.\ Rev.\ Lett.\  {\bf 98} (2007) 022002
  [arXiv:hep-ph/0609232].
  
\bibitem{Lunghi:2006hc}
  E.~Lunghi and J.~Matias,
  JHEP {\bf 0704}, 058 (2007)
  [arXiv:hep-ph/0612166].
  
\bibitem{Kruger:2005ep}
  F.~Kruger and J.~Matias,
  Phys.\ Rev.\  D {\bf 71} (2005) 094009
  [arXiv:hep-ph/0502060].
  
\bibitem{Huber:2005ig}
  T.~Huber, E.~Lunghi, M.~Misiak and D.~Wyler,
  Nucl.\ Phys.\  B {\bf 740} (2006) 105
  [arXiv:hep-ph/0512066].

\bibitem{Gambino:2003zm}
  P.~Gambino, M.~Gorbahn and U.~Haisch,
  Nucl.\ Phys.\  B {\bf 673}, 238 (2003)
  [arXiv:hep-ph/0306079].
  
 \bibitem{Gorbahn:2004my}
  M.~Gorbahn and U.~Haisch,
  Nucl.\ Phys.\  B {\bf 713}, 291 (2005)
  [arXiv:hep-ph/0411071].
  
 \bibitem{Bobeth:2003at}
  C.~Bobeth, P.~Gambino, M.~Gorbahn and U.~Haisch,
  JHEP {\bf 0404}, 071 (2004)
  [arXiv:hep-ph/0312090].
  
\bibitem{Chetyrkin:1996vx}
  K.~G.~Chetyrkin, M.~Misiak and M.~Munz,
  Phys.\ Lett.\  B {\bf 400} (1997) 206
  [Erratum-ibid.\  B {\bf 425} (1998) 414]
  [arXiv:hep-ph/9612313].
  
\bibitem{Nakamura:2010zzi}
  K.~Nakamura {\it et al.}  [Particle Data Group],
  J.\ Phys.\ G {\bf 37} (2010) 075021.
 
 \bibitem{Alcaraz:2009jr}
  J.~Alcaraz,
  arXiv:0911.2604 [hep-ex], and July 2010 update on http://lepewwg.web.cern.ch/

\bibitem{ckmfitter}
  J.~Charles {\it et al.}  [CKMfitter Group],
  Eur.\ Phys.\ J.\  C {\bf 41} (2005) 1
  [arXiv:hep-ph/0406184] and Summer 2010 update on http://ckmfitter.in2p3.fr/

\bibitem{Altmannshofer:2008dz}
  W.~Altmannshofer, P.~Ball, A.~Bharucha, A.~J.~Buras, D.~M.~Straub and M.~Wick,
  JHEP {\bf 0901}, 019 (2009)
  [arXiv:0811.1214 [hep-ph]].

\bibitem{Kagan:2001zk}
  A.~L.~Kagan and M.~Neubert,
  Phys.\ Lett.\  B {\bf 539}, 227 (2002)
  [arXiv:hep-ph/0110078].
  
\bibitem{Asner:2010qj}
  D.~Asner {\it et al.}  [Heavy Flavor Averaging Group],
  arXiv:1010.1589 [hep-ex].
  
\bibitem{Aaij:2011rj}
  R.~Aaij {\it et al.}  [the LHCb Collaboration],
  arXiv:1103.2465 [hep-ex].

\bibitem{Freitas:2008vh}
  A.~Freitas and U.~Haisch,
  Phys.\ Rev.\  D {\bf 77} (2008) 093008
  [arXiv:0801.4346 [hep-ph]].
  
\bibitem{Misiak:2006ab}
  M.~Misiak and M.~Steinhauser,
  Nucl.\ Phys.\  B {\bf 764}, 62 (2007)
  [arXiv:hep-ph/0609241].

\bibitem{Misiak:2010sk}
    M.~Misiak and M.~Steinhauser,
  arXiv:1005.1173 [hep-ph].
  
\bibitem{BaBar:2009we}
  B.~Aubert {\it et al.}  [BABAR Collaboration],
  Phys.\ Rev.\ Lett.\  {\bf 103} (2009) 211802
  [arXiv:0906.2177 [hep-ex]].

\bibitem{Nakao:2004th}
  M.~Nakao {\it et al.}  [BELLE Collaboration],
  Phys.\ Rev.\  D {\bf 69}, 112001 (2004)
  [arXiv:hep-ex/0402042].

\bibitem{Feldmann:2002iw}
  T.~Feldmann and J.~Matias,
  JHEP {\bf 0301}, 074 (2003)
  [arXiv:hep-ph/0212158].

\bibitem{Beneke:2004dp}
  M.~Beneke, T.~Feldmann and D.~Seidel,
  Eur.\ Phys.\ J.\  C {\bf 41} (2005) 173
  [arXiv:hep-ph/0412400].
  
\bibitem{Grinstein:2004uu}
  B.~Grinstein, Y.~Grossman, Z.~Ligeti and D.~Pirjol,
  Phys.\ Rev.\  D {\bf 71} (2005) 011504
  [arXiv:hep-ph/0412019].
  
\bibitem{Grinstein:2005nu}
  B.~Grinstein and D.~Pirjol,
  Phys.\ Rev.\  D {\bf 73} (2006) 014013
  [arXiv:hep-ph/0510104].

\bibitem{Ball:2006cva}
  P.~Ball and R.~Zwicky,
  Phys.\ Lett.\  B {\bf 642} (2006) 478
  [arXiv:hep-ph/0609037].
  
\bibitem{Ball:2006eu}
  P.~Ball, G.~W.~Jones and R.~Zwicky,
  Phys.\ Rev.\  D {\bf 75} (2007) 054004
  [arXiv:hep-ph/0612081].
  
\bibitem{Ushiroda:2006fi}
  Y.~Ushiroda {\it et al.}  [Belle Collaboration],
  Phys.\ Rev.\  D {\bf 74} (2006) 111104
  [arXiv:hep-ex/0608017].
  
\bibitem{Aubert:2008gy}
  B.~Aubert {\it et al.}  [BABAR Collaboration],
  Phys.\ Rev.\  D {\bf 78} (2008) 071102
  [arXiv:0807.3103 [hep-ex]].
  
\bibitem{Egede:2008uy}
  U.~Egede, T.~Hurth, J.~Matias, M.~Ramon and W.~Reece,
  JHEP {\bf 0811} (2008) 032
  [arXiv:0807.2589 [hep-ph]].

 \bibitem{Beneke:2000wa}
  M.~Beneke and T.~Feldmann,
  Nucl.\ Phys.\  B {\bf 592} (2001) 3
  [arXiv:hep-ph/0008255].
  
 \bibitem{Ghinculov:2003bx}
 A.~Ghinculov, T.~Hurth, G.~Isidori, Y.~P.~Yao,
 Eur.\ Phys.\ J.\  {\bf C33 } (2004)  S288-S290.
 [hep-ph/0310187].

\bibitem{Neubert:2000ch}
 M.~Neubert,
 JHEP {\bf 0007 } (2000)  022.
 [hep-ph/0006068].

\bibitem{Bauer:2001rc}
 C.~W.~Bauer, Z.~Ligeti, M.~E.~Luke,
 Phys.\ Rev.\  {\bf D64 } (2001)  113004.
 [hep-ph/0107074].  

\bibitem{Huber:2007vv}
  T.~Huber, T.~Hurth and E.~Lunghi,
  Nucl.\ Phys.\  B {\bf 802}, 40 (2008)
  [arXiv:0712.3009 [hep-ph]].

\bibitem{Aaltonen:2011cn}
  T.~Aaltonen {\it et al.} [ CDF Collaboration ],
   Submitted to: Phys.Rev.Lett..
  [arXiv:1101.1028 [hep-ex]].

\bibitem{Alok:2010zd}
  A.~K.~Alok, A.~Datta, A.~Dighe, M.~Duraisamy, D.~Ghosh, D.~London and S.~U.~Sankar,
  arXiv:1008.2367 [hep-ph].
  
\bibitem{talk}
  U.~Egede, T.~Hurth, J.~Matias, M.~Ramon and W.~Reece,
  Acta Phys.\ Polon.\  B {\bf 3} (2010) 151
  [arXiv:0912.1339 [hep-ph]].

\bibitem{talk2}
U.~Egede, T.~Hurth, J.~Matias, M.~Ramon and W.~Reece,
  arXiv:1012.4603 [hep-ph].

\bibitem{Hurth:2008jc}
  T.~Hurth, G.~Isidori, J.~F.~Kamenik and F.~Mescia,
  Nucl.\ Phys.\  B {\bf 808} (2009) 326
  [arXiv:0807.5039 [hep-ph]].

\bibitem{Deschamps:2009rh}
  O.~Deschamps, S.~Descotes-Genon, S.~Monteil, V.~Niess, S.~T'Jampens and V.~Tisserand,
  Phys.\ Rev.\  D {\bf 82} (2010) 073012
  [arXiv:0907.5135 [hep-ph]].

\bibitem{Bobeth:2010wg}
  C.~Bobeth, G.~Hiller and D.~van Dyk,
  JHEP {\bf 1007}, 098 (2010)
  [arXiv:1006.5013 [hep-ph]].

\bibitem{Chetyrkin:1997dh}
 K.~G.~Chetyrkin,
  Phys.\ Lett.\  B {\bf 404}, 161 (1997)
  [arXiv:hep-ph/9703278].
  
\bibitem{Bauer:2004ve}
  C.~W.~Bauer, Z.~Ligeti, M.~Luke, A.~V.~Manohar and M.~Trott,
  Phys.\ Rev.\  D {\bf 70} (2004) 094017
  [arXiv:hep-ph/0408002].

\bibitem{Hoang:2000fm}
  A.~H.~Hoang,
  arXiv:hep-ph/0008102.
  
 \bibitem{Beneke:2001at}
  M.~Beneke, T.~Feldmann and D.~Seidel,
  [arXiv:hep-ph/0106067].

 \bibitem{Khodjamirian:2010vf}
  A.~Khodjamirian, T.~Mannel, A.~A.~Pivovarov and Y.~M.~Wang,
  JHEP {\bf 1009} (2010) 089
  [arXiv:1006.4945 [hep-ph]].
  
\bibitem{Seidel:2004jh}
  D.~Seidel,
  Phys.\ Rev.\  D {\bf 70} (2004) 094038
  [arXiv:hep-ph/0403185].

\bibitem{Guetta:1997fw}
  D.~Guetta and E.~Nardi,
  Phys.\ Rev.\  D {\bf 58} (1998) 012001
  [arXiv:hep-ph/9707371].

 \bibitem{Ball:2004rg}
  P.~Ball and R.~Zwicky,
  Phys.\ Rev.\  D {\bf 71} (2005) 014029
  [arXiv:hep-ph/0412079].

 \bibitem{Charles:1998dr}
  J.~Charles, A.~Le Yaouanc, L.~Oliver, O.~Pene and J.~C.~Raynal,
  Phys.\ Rev.\  D {\bf 60} (1999) 014001
  [arXiv:hep-ph/9812358].

\bibitem{Egede:2007zz}
  U.~Egede,
  \textit{Angular correlations in the $\bar{B}_d \to \bar{K}^{*0} \ell^+ \ell^+$ decay},
  \href{http://www-spires.slac.stanford.edu/spires/find/hep/www?r=CERN-LHCB-2007-038}{CERN-LHCB-2007-057}.
 
\bibitem{Bharucha:2010bb}
  A.~Bharucha and W.~Reece,
  Eur.\ Phys.\ J.\  C {\bf 69} (2010) 623
  [arXiv:1002.4310 [hep-ph]].

\bibitem{Lunghi:2010tr}
  E.~Lunghi and A.~Soni,
  JHEP {\bf 1011} (2010) 121
  [arXiv:1007.4015 [hep-ph]].


\end{thebibliography}
\end{document}